\documentclass[showpacs,nofootinbib,preprintnumbers,prd,superscriptaddress,twocolumn]{revtex4-2}

\usepackage{amsmath}
\usepackage{amsfonts}
\usepackage{amssymb}
\usepackage{graphicx}
\usepackage[titletoc]{appendix}
\usepackage{color}
\usepackage{hyperref}
\usepackage{cleveref}
\usepackage[rightcaption]{sidecap}
\usepackage{subfigure}
\usepackage{comment}
\usepackage{float}

\usepackage{dcolumn}

\usepackage{array}
\usepackage{ctable}
\usepackage{multirow}
\usepackage{siunitx}
\usepackage{longtable}
\usepackage{tabularx}
\usepackage{booktabs}

\usepackage{tensor}
\usepackage{mathtools}
\usepackage{dirtytalk}
\usepackage{siunitx}
\DeclareSIUnit{\parsec}{pc}

\graphicspath{{Graphics/}}

\def\be{\begin{equation}}
\def\ee{\end{equation}}
\def\bea{\begin{eqnarray}}
\def\eea{\end{eqnarray}}

\definecolor{vividviolet}{rgb}{0.62, 0.0, 1.0}
\definecolor{amaranth}{rgb}{0.9, 0.17, 0.31}
\definecolor{palatinateblue}{rgb}{0.15, 0.23, 0.89}
\definecolor{brightpink}{rgb}{1.0, 0.0, 0.5}
\definecolor{cornflowerblue}{rgb}{0.39, 0.58, 0.93}
\definecolor{deepcarminepink}{rgb}{0.94, 0.19, 0.22}
\definecolor{radicalred}{rgb}{1.0, 0.21, 0.37}
\hypersetup{ linktoc=all,
    colorlinks, linkcolor={palatinateblue},
    citecolor={brightpink}, urlcolor={amaranth}
}

\begin{document}

\title{Linear and nonlinear clusterings of Horndeski-inspired dark energy models with fast transition}

\author{Orlando Luongo}
\email{orlando.luongo@unicam.it}
\affiliation{Universit\`a di Camerino, Divisione di Fisica, Via Madonna delle Carceri, 9a, 62032 Camerino MC, Italy.}
\affiliation{Istituto Nazionale di Fisica Nucleare (INFN), Sezione di Perugia, Via Alessandro Pascoli, 23c, 06123 Perugia PG, Italy.}
\affiliation{Department of Mathematics and Physics, SUNY Polytechnic Institute, Utica, NY 13502, USA.}
\affiliation{Istituto Nazionale di Astrofisica (INAF), Osservatorio Astronomico di Brera, Via Brera, 28, 20121 Milano MI, Italy}
\affiliation{Al-Farabi Kazakh National University, Almaty, 050040, Kazakhstan.}

\author{Francesco Pace}
\email{francesco.pace@unito.it}
\affiliation{Dipartimento di Fisica, Universit\`a degli Studi di Torino, Via Pietro Giuria, 1, 10125, Torino TO, Italy.}
\affiliation{Istituto Nazionale di Fisica Nucleare (INFN), Sezione di Torino, Via Pietro Giuria, 1, 10125 Torino TO, Italy.}
\affiliation{Istituto Nazionale di Astrofisica (INAF), Osservatorio Astrofisico di Torino, Via Osservatorio, 20, 10025 Pino Torinese TO, Italy}.

\author{Sebastiano Tomasi}
\email{sebastiano.tomasi@studenti.unicam.it}
\affiliation{Universit\`a di Camerino, Divisione di Fisica, Via Madonna delle Carceri, 9a, 62032 Camerino MC, Italy.}
\affiliation{Istituto Nazionale di Fisica Nucleare (INFN), Sezione di Perugia, Via Alessandro Pascoli, 23c, 06123 Perugia PG, Italy.}

\date{\today}

\begin{abstract}
We analyze time-dependent dark energy equations of state through linear and nonlinear structure formation and their quintessence potentials,  characterized by fast, recent transitions, inspired by parameter space studies of selected classes of the more general Horndeski models. The influence of dark energy on structures comes from modifications to the background expansion rate and from perturbations as well. In order to compute the structures growth, we employ a generalization of the \emph{spherical collapse} formalism that includes perturbations of fluids with pressure. We numerically solve the equations of motion for the perturbations and the field. Our analysis suggests that a true Heaviside step transition is a good approximation for most of the considered models, since most of the quantities weakly depend on the transition speed. We find that transitions occurring at redshifts $z_{\rm t}\gtrsim 2$ cannot be distinguished from the $\Lambda$CDM model if dark energy is freezing, i.e, the corresponding equation of state tends to $-1$. For fast, recent transitions, the redshift at which the properties of dark energy have the most significant effect is $z=0.6\pm 0.2$. We also find that in the freezing regime, the $\sigma_8$ values can be lowered by about $8\%$, suggesting that those models could relieve the $\sigma_8$-tension. Additionally, freezing models generally predict faster late-time merging rates but a lower number of massive galaxies at $z=0$. Finally, the nonlinear matter power spectrum for smooth dark energy shows a valley centered in $k\approx1\,h\,{\rm Mpc}^{-1}$ which in the clustering case is replaced by a sharp increase for $k\gtrsim 0.2\,h\,{\rm Mpc}^{-1}$ and a peak at $k\approx 2\,h\,{\rm Mpc}^{-1}$.
\end{abstract}



\maketitle

\section{Introduction}

Experimental evidences favor a late-time cosmic acceleration \cite{SupernovaSearchTeam:1998bnz, SupernovaCosmologyProject:1998vns, SupernovaSearchTeam:2004lze} and suggest the existence of an exotic fluid, exhibiting a negative equation of state (EOS), called dark energy (DE). In the standard cosmological puzzle, DE is intimately related to the cosmological constant, $\Lambda$, arising from vacuum energy quantum fluctuations \cite{Martin:2012bt,Luongo:2018lgy}. 

Generally, two different possibilities are commonly developed to explain the fundamental nature of DE. 
The first is framing out the EOS, $w(z)$, through parametric functions of redshift and observationally constraining its evolution\footnote{For the sake of completeness, determining the EOS as a function of redshift may not directly reveal the nature of DE. However, it clearly plays a significant role in distinguishing between different DE models \cite{Liberato:2006un,Bamba:2012cp,Muccino:2020gqt,Capozziello:2022jbw,Luongo:2020aqw}.} \cite{Aviles:2016wel,Aviles:2012ay}. The second is deriving the DE EOS from first principles, making \emph{a priori} hypotheses on its microphysics \cite{Capozziello:2019cav,Sharma:2021ayk,Sharma:2021ivo}.

In this respect, the use of field theories appears of utmost importance to model DE in terms of its intrinsic constituent\footnote{Extended theories of gravity are often conformally equivalent to scalar fields, for example \cite{Capozziello:2011et,Afonso:2018hyj}. Investigating scalar theories can, therefore, provide hints on how to depart from Einstein's gravity.}. The most studied models, also due to their relative simplicity, are quintessence, phantom, $k$-essence and $f(R)$ gravity models. $k$-essence models represent a generalization of both quintessence ($w>-1$) and phantom models ($w<-1$), while $f(R)$ gravity can be transformed into Brans-Dicke models via a conformal transformation. All these models are specific subclasses of the most general scalar-tensor theory of gravitation in four dimensions, namely the Horndeski theory \cite{Kobayashi2019hrl}. Given a model Lagrangian, different functional forms for the EOS arise, according to the specific class considered. Therefore, one can see that there is a direct link between the scalar field and the EOS. Hence, describing DE within the Horndeski framework implies that the EOS reconstruction with observational data leads to constraints on the corresponding scalar field potential\footnote{For a recent discussion about the background properties of the most studied scalar field models we refer to \cite{Battye:2016alw}.}.

However, there is an important caveat in this. A generic Horndeski model, even after the constraints due to the detection of gravitational waves \cite{LigoVirgo2017,LigoVirgoIntegral2017}, is too general to be used, unless a particular subclass is considered. Hence, albeit later we focus only on minimally coupled models (quintessence, phantom and $k$-essence), we consider Horndeski models the framework where to embed our predictions. To make a concrete example of this difficulty, we refer to \cite{Battye:2016alw} where the authors considered the evolution of the scalar field for different models. When considering $k$-essence, they needed to specify a functional form for the kinetic term to write the relation between the EOS and the scalar field potential $V(\phi)$. This reinforces the necessity of considering a particular subclass of Horndeski to proceed and obtain results.

At small redshifts, the majority of DE models fully degenerates \cite{Luongo:2015zgq}, indicating that late-time constraints appear less predictive in clarifying how DE evolves throughout the Universe evolution \cite{Luongo:2022bju,Luongo:2021pjs}. Nevertheless, DE exerts its influence through its impact on cosmic perturbations. Accordingly, it significantly affects the rate of formation and growth of virialized structures, such as galaxies, galaxy clusters, and more. Indeed, the cosmic expansion, if driven by DE, acts to decelerate the gravitational collapse of overdense regions through the \emph{Hubble drag}. As DE becomes dominant, overdense regions experience slower growth, and the matter gravitational collapse may even reverse on scales comparable to the Hubble horizon. 

Particularly, if DE is not the cosmological constant, it could exhibit fluctuations in both space and time. Consequently, DE is not only influenced by matter overdensities but also gives rise to its own overdensities, that, in turn, exert a nonlinear influence on matter overdensities.

Motivated by the aforementioned points, in this work we analyze how effective DE fluid models, characterized by an EOS with fast transitions, affect the background evolution and structure formation. This class of paradigms is still under study and far from being rejected as suitable candidate for DE \cite{de_eos_1, de_eos_2, de_eos_3, de_eos_5, de_eos_6}. While in previous efforts a given parametrization has been investigated, in this work we analyze six different parameterizations, four of which belong to the same fast transition class, in order to understand if the models can be distinguished from one another and from the $\Lambda$CDM scenario. Accordingly, we determine their effects on structures' growth by employing a generalization of the spherical collapse formalism \cite{Lee:2009qq,Chang:2017vhs} that includes fluids with pressure, adopting the formalism developed in \cite{Lima:1996at,Abramo:2007iu}. We then numerically solve the equations of motion (EOM) of the perturbation and field and investigate linear and nonlinear features of our models. As a key output, our analysis seems to suggest that a true Heaviside step transition works fairly well as an alternative to the cosmological constant and provide a good approximation to fast transition models. Even though promising, we emphasize that transitions happening at $z\geq 2$ cannot be distinguished from the standard cosmological model, if DE is \emph{freezing}\footnote{A freezing DE model is one where the DE EOS is approaching the cosmological constant, $w=-1$, over time.}. Particularly, in the freezing regime, slight improvements on the $\sigma_8$ tension can be achieved, roughly indicating that such field models may in principle heal this cosmic tension. Finally, the nonlinear matter power spectrum for smooth dark energy shows a valley centered in $k\approx1\,h\,{\rm Mpc}^{-1}$ which in the clustering case is replaced by a sharp increase for $k\gtrsim 0.2\,h\,{\rm Mpc}^{-1}$ and a peak at $k\approx 2\,h\,{\rm Mpc}^{-1}$.

The paper is structured as follows. In Sec.~\ref{sec:stt} we work out how DE can be described in a scalar tensor theory and introduce the framework we work in. In Sec.~\ref{subsec:effectiveFluidDescription}, we specialize to the effective fluid description of quintessence theory, for the reasons discussed above. In Sec.~\ref{sec:non_linear_perturbations} we set the perturbation framework, introducing the spherical collapse formalism, the pseudo-Newtonian approximation to gravitational interaction. Section~\ref{sec:fast_trans_eos} is devoted to the analysis of the fast transition models. After a brief introduction to the models, we begin the analysis computing numerically the quintessence potential and field that correspond to each EOS. We then compute the linear matter perturbations and make extensive use of numerical simulations to compute the virialization overdensity and the linearly extrapolated density contrast at collapse for the matter perturbations. These quantities allow for a semiquantitative analysis of the models and are needed to compute the halo mass function (HMF) in Sec.~\ref{subsec:hmf}. The HMF is quite an important quantity as it can be directly compared to results of large $N$-body cosmological simulations and will be hopefully measured accurately by the \textit{Euclid} survey \cite{Castro:2016jmw}. Conclusions and perspectives are discussed in Sec.~\ref{sec:discussion}. Throughout this article, we will assume $c=1$.

\section{Scalar field theories}\label{sec:stt}

While general relativity is the most successful theory for explaining gravity, alternatives have been introduced and severely investigated. Scalar-tensor theories represent a powerful instrument to produce self-consistent modified theories of gravity. According to Lovelock's theorem, Einstein equations are the only possible second-order Euler-Lagrange equations derived from a Lagrangian scalar density in $3+1$ dimensions that is constructed solely from the metric, $\mathcal{L} = \mathcal{L}(g\indices{_\mu_\nu})$ \cite{Navarro:2010zm}. To extend Einstein's theory of gravity, we can relax this hypothesis of Lovelock's theorem. A simple way to do this, is to add another degree of freedom different from the metric, e.g. a scalar field. In doing so, the EOM contains higher-order derivatives, possibly leading to Ostrogradsky instabilities \cite{Kobayashi2019hrl}.

The Horndeski theory is also known, for historical reasons, as the \emph{generalized covariant Galileon} \cite{Kobayashi:2019hrl}. The procedure to obtain the Horndeski theory is to start from the most general scalar field theory on a Minkowsky spacetime, which yields second-order equations, under the assumptions that the Lagrangian contains at most second-order derivatives of the field, behaving as polynomial in $\partial_{\mu}\partial_{\nu}\phi$.

One then promotes the theory to be covariant, replacing partial derivatives with covariant derivatives, and adds appropriate unique counterterm to retain second-order field equations. The overall procedure can be realized in any dimension and, particularly, in four dimensions, the Horndeski theory is given by \cite{Kobayashi2019hrl,kase2019dark,Charmousis:2011bf} 
\begin{equation}
    \begin{aligned}\label{eq:HorndeskiTheory}
        \mathcal{L}_{\mathrm{H}} = &\, G_2(\phi,X)-G_3(\phi,X)\Box \phi+G_4(\phi,X)R+\\
        &\, G_{4X}\left[\left(\Box \phi\right)^2-\phi\indices{^\mu^\nu}\phi\indices{_\mu_\nu}\right]   +G_5(\phi,X)G\indices{^\mu^\nu}\phi\indices{_\mu_\nu}\\
        &\, -\frac{G_{5X}}{6}X\left[\left(\Box \phi\right)^3-3\Box\phi\phi\indices{^\mu^\nu}\phi\indices{_\mu_\nu}+2\phi\indices{_\mu_\nu}\phi\indices{^\nu^\lambda}\phi\indices{^\mu_\lambda} \right]\,,
    \end{aligned}
\end{equation}
where $\phi_{\mu}\equiv\nabla_{\mu}\phi$, $\phi\indices{_\mu_\nu}\equiv\nabla_{\mu}\nabla_{\nu}\phi$,  $X=-g\indices{^\mu^\nu}\phi_{\mu}\phi_{\nu}/2$, and $G_2$, $G_3$, $G_4$ and $G_5$ are arbitrary functions, playing the role of superpotentials, and depend both on $\phi$ and $X$ and, finally, $f_X=\partial f/\partial X$. Eq.~\eqref{eq:HorndeskiTheory} represents the most general scalar-tensor theory exhibiting second-order field equations in four dimensions.

Thanks to the constraints posed by the gravitational wave detection \cite{LigoVirgo2017,LigoVirgoIntegral2017}, the previous Lagrangian can be simplified to be in agreement with them by setting $G_5=0$ and $G_4=G_4(\phi)$. However, as we discussed previously in the Introduction, the resulting Lagrangian is still too generic and to move forward we need to specify a particular class of models. This is the reason why, in the following, we consider only quintessence and phantom models.

These models are described by the following Lagrangian\footnote{For an alternative perspective, we refer the reader to \cite{Luongo:2018lgy,Belfiglio:2023eqi,Belfiglio:2023rxb,Luongo:2023aaq}.}
\begin{equation}\label{eq:quint}
 \mathcal{L} = -\frac{1}{2}\eta\nabla_{\mu}\phi\nabla^{\mu}\phi - V(\phi)\,,
\end{equation}
where $V(\phi)$ is the scalar field potential and $\eta$ is a constant whose value is $+1$ for quintessence and $-1$ for phantom models.
Consequently, the Lagrangian density includes a further term, contributing to the energy-momentum tensor, within the Einstein-Hilbert action
\begin{equation*}
 S = \int_{\mathcal{D}}\mathrm{d}^4x\,\sqrt{-g}\,(\mathcal{L}_{\rm EH}+\mathcal{L}_{\phi})\,,
\end{equation*}
where $g$ stands for the determinant of the metric, while $\mathcal{L}_{\rm EH}$ is the Einstein-Hilbert Lagrangian and $\mathcal{D}$ the integration domain.

The usual Klein-Gordon equation of motion (EOM) for the field holds, and reduces, in a Friedmann-Robertson-Walker (FRW) Universe, to
\begin{equation}\label{eq:eomScalarField}
    \Ddot{\phi}+3H\Dot{\phi}+\eta\frac{\mathrm{d} V}{\mathrm{d} \phi}=0\,.
\end{equation}
In the next subsection we describe more in detail the properties of quintessence and phantom models.

\subsection{Quintessence as effective fluid}\label{subsec:effectiveFluidDescription}
Quintessence and phantom models are a subclass of the Horndeski theories and are obtained by setting $G_3 = G_5 = 0$, $G_2 = \eta X-V(\phi)$ and $G_4=1/(16\pi G)$.

We now want to bridge the gap between the fluid description of DE and its quintessence/phantom interpretation. By the effective fluid description, we mean to describe DE through its EOS only. In doing so, there are two main approaches:

\begin{itemize}
    \item[-] We can specify a potential $V(\phi)$ and then derive the EOS, $w$.
    \item[-] We can alternatively fix the EOS and calculate the corresponding quintessence/phantom potential.
\end{itemize}  

We will follow the latter approach. Thus, we suppose that the EOS is a given function $w(a)$ \cite{Battye:2016alw}. The contact point between fluid and quintessence descriptions is the EOS $w(a)$. We, therefore, start from its expression in terms of the field 
\begin{equation*}
    w(\phi,\Dot{\phi}) = \frac{P_{\phi}}{\rho_{\phi}} = \frac{\eta\frac{\Dot{\phi}^2}{2}-V(\phi)}{\eta\frac{\Dot{\phi}^2}{2}+V(\phi)}\,,
\end{equation*}
which allows us to analyze both quintessence ($\eta=+1$) and phantom models ($\eta=-1$).

Thus we have to design the potential in a way that this expression gives the desired $w(a)$.
The field satisfies the EOM, that is equivalent to the continuity equation $\Dot{\rho}_{\phi}+3H\rho_{\phi}(1+w)=0$. By changing the differentiation variable from $t$ to $a$ and integrating, we obtain 
\begin{equation*}
    \rho_{\phi} = \rho_{0\phi}e^{-3\int_1^a\frac{1+w(x)}{x}\mathrm{d}x}\coloneqq \rho_{0\phi}g(a)\,.
\end{equation*}

Consequently, the potential reads
\begin{equation}
    V(a)=\frac{\rho_{0\phi}}{2}g(a)[1-w(a)]\,,
\end{equation}
and, so, to convert $V(a)$ into $V(\phi)$, we need to compute $\phi(a)$, which is given by
\begin{equation}
    \phi(a)-\phi_0 = \pm\sqrt{\frac{3}{8\pi G}}\int_1^a \mathrm{d}x\,\frac{\sqrt{\Omega^0_{\phi} \eta g(x)\left[1+w(x)\right]}}{x\sqrt{\frac{\Omega^0_{\rm m}}{x^3}+\Omega^0_\phi g(x)}}\,,
\end{equation}
where $\phi_0$ is the integration constant. We will select it to ensure that $\phi(a_{\rm min})=0$. In this context, $a_{\rm min}$ denotes the smallest scale parameter value that we employ in our numerical integration.

There are very few cases in which the solution for $g(a)$ is analytical, but if we have an analytic $g(a)$, there are some cases in which $\phi(a)$ can be computed. Even though this can be possible, generally it is not trivial to invert $\phi(a)$ to obtain $a(\phi)$ and, therefore, this task should be computed numerically. 

By defining
\begin{equation}
 \tilde{V}\equiv \frac{16\pi G}{3H_0^2}V\,, \quad\quad \tilde{\phi}\equiv  \sqrt{\frac{8\pi G}{3}}\phi\,,
\end{equation}
we have the dimensionless quantities 
\begin{align}
 \tilde{V}(a) = &\,\Omega^0_{\phi}g(a)[1-w(a)]\,, \label{eq:dimensionlessPotField}\\
 \tilde{\phi}(a) = &\,\pm\int_1^a \frac{\sqrt{\Omega^0_{\phi}\eta g(x)\left[1+w(x)\right]}}{x\sqrt{\frac{\Omega^0_{\rm m}}{x^3}+\Omega^0_\phi g(x)}}\mathrm{d}x\,, \label{eq:dimensionlessField}
\end{align}
that can easily be integrated through numerical methods.

\section{Perturbations framework}\label{sec:non_linear_perturbations}

In this section we set the framework to study both linear and nonlinear perturbations in DE models described by a given EOS $w(a)$. Because linear equations can be obtained by linearizing the full nonlinear equations, we will not repeat the derivation twice, but we will start directly from the nonlinear equations and then obtain, in the appropriate limit, the linear ones. In addition, we will consider two rather different cases regarding the evolution of perturbations in DE models: one where the DE fluid affects the background only (smooth models) and the other where also the DE component can have perturbations (clustering models). In the latter case, we also need to consider the effects of pressure perturbations.

Investigating perturbations beyond the linear approximation poses significant mathematical difficulties. To tackle these caveats, large numerical simulations are often considered and the spherical collapse model is one of the most used approximation. The model describes the dynamics of an isolated, spherical region of matter, that is collapsing under its own gravitational pull. This model provides a simple, yet effective framework for analyzing the growth of density perturbations in the early Universe and their eventual evolution into bound structures such as galaxies and galaxy clusters. 

Heuristically, what is requested in the spherical collapse model is to pop into existence an overdense sphere, very early in the cosmic history. Since the Universe is expanding, initially the sphere expands with the Hubble flow. However, the overdensity causes an inward gravitational pull that resists the Hubble flow. At a certain point, dubbed \emph{turnaround}, the sphere radius stops growing. After turnaround, the sphere starts collapsing on itself until it reaches zero radius and infinite density. In the physical world, the shrinking stops before the collapse due to nonspherical symmetry, peculiar velocities of the particles, friction, pressure and other effects that are ignored in the most basic formulation of the spherical collapse model. To overcome some of those shortcomings, it is assumed that the sphere reaches a time in which it becomes stable and gets into virial equilibrium.

\subsection{Spherical collapse model}\label{subsec:sphcolmodel}

To grasp the development of nonlinear perturbations in a realistic two-component Universe that contains DE, fully-characterized by $w$, and matter, one can extend the Newtonian theory to incorporate DE, as an additional term in the acceleration equation \cite{Abramo:2007iu,Pace:2010,Lima:1996at}. 

To do so, we define $\delta_{i} \equiv (\rho_{i} - \Bar{\rho}_{i})/\Bar{\rho}_{i}$ as the density contrast of the $i$th fluid, where $\rho_i(\boldsymbol{x},t)$ represents the perturbed density and $\Bar{\rho}_i(t)$ denotes the background density. Furthermore, we define $\theta_i \equiv \nabla \cdot \mathbf{v}_i$ as the divergence of the peculiar velocity $\mathbf{v}_i$. While, in principle, we can have a peculiar velocity for each perturbed species, in our setting this does not happen as the Euler equation is the same for all the species, in our case dark matter and DE.

With respect to standard dark matter perturbations, we need to take into account the fact that DE perturbations involve both density and pressure perturbations. This introduces, therefore, an additional degree of freedom to the model, which can only be set once one knows the microphysics of the fluid via its Lagrangian formulation. As from a given EOS it is not possible to infer the Lagrangian, unless one specifies the type of DE model \emph{a priori}, i.e, quintessence or $k$-essence, we take the common Ansatz of relating pressure perturbations to density perturbations via an effective sound speed, $c_{\rm eff}^2 \equiv \delta P / \delta\rho \geq 0$ \cite{Pace:2016klp}. For quintessence models, $c_{\rm eff}^2 = 1$ and the DE component can be considered effectively smooth, as perturbations propagate at the speed of light. Conversely, $k$-essence models can have a small sound speed and this increases the level of clustering. In the limit of $c_{\rm eff}^2 = 0$, DE can cluster similarly to dark matter.

We note that for adiabatic fluids, $P=P(\rho)$, the relation $P = w\rho$ holds both at the background and at the perturbation level, implying $c_{\rm eff}^2 = w$. However, this is not the case here, as the fluid is not adiabatic and having a negative sound speed will lead to instabilities in the perturbations.

We start by considering smooth DE models, where the only perturbed fluid is the dark matter one. In this case, it is easy to write the EOM of the matter density contrast
\begin{equation}
\label{eq:mperturb}
 \delta''_{\rm m}+ \frac{3}{2a}(1-w_{\rm{eff}})\delta'_{\rm m}-\frac{4}{3}\frac{(\delta'_{\rm m})^2}{1+\delta_{\rm m}} = \\
 \frac{3}{2a^2}\Omega_{\rm m}(1+\delta_{\rm m})\delta_{\rm m}\,,
\end{equation}
obtained perturbing a homogeneous and isotropic Universe, modeled by the FRW spacetime, with $a(t)$ the scale factor. The primes denote derivatives with respect to $a$, whereas $w_{\rm{eff}}$ is the effective EOS for the background, defined by
\begin{equation}
     w_{\rm{eff}}=\frac{\Bar{P}}{\Bar{\rho}}=-\left(1+\frac{2a}{3}\frac{H'}{H}\right)
     =\frac{w_{\rm de}\tilde{\Omega}_{\rm de}}{\tilde{\Omega}_{\rm m}+\tilde{\Omega}_{\rm de}}\,,
     \label{eq:effectiveEos}
\end{equation}
where $\tilde{\Omega}_i=\Omega_i^0g_i(a)$ is the density evolution of the fluid $i$, with the function $g_i(a)$ given by 
\begin{equation}
    g_i(a)=e^{-3\int_1^a\frac{1+w_i(x)}{x}\mathrm{d}x}\,,
    \label{eq:densityEvolution}
\end{equation}
and $w_i(a)$ the EOS for the $i$th fluid.

These equations are valid only well within the horizon, meaning that the radius of the overdensity, $R$, satisfies $R\ll 1/H$, where $H$ is the Hubble function.

From the full nonlinear equation, it is easy to derive the equation describing the evolution of linear perturbations. The resulting expression is the so-called growth-factor equation and it reads
\begin{equation}\label{eq:linearDensityContrast}
 \delta''_{\rm m} + \frac{3}{2a}(1-w_{\rm{eff}})\delta'_{\rm m} - \frac{3}{2a^2}\Omega_{\rm m}\delta_{\rm m} = 0\,.
\end{equation}

To find the initial conditions of Eq.~(\ref{eq:linearDensityContrast}), we use the Ansatz that at early-times $\delta_{\rm m}\propto a^n$, which reduces the differential equation to an algebraic one for the exponent $n$. The solution found in this way is valid only at early-times, in a suitable range of scale parameter values, where the coefficients are approximately constant. By neglecting the decaying mode, the solution is given by $\delta_{\rm m}^{\rm{lin}} = c_1 a^n$, where $c_1$ is an arbitrary constant and
\begin{equation}
 n=\frac{1}{4}\left(-1+\sqrt{24\Omega_{\rm m}+(1-3w_{\rm{eff}})^2}+3w_{\rm{eff}}\right)\,.
 \label{eq:deltaMatterLinear}
\end{equation}
The solution is usually denoted as $D_+(a)$. For a Universe dominated by matter (baryons and cold dark matter) at early times, we recover the standard Einstein--de Sitter (EdS) solution and $n=1$ \cite{Haude:2019qms}.

We now consider the case of clustering DE. Because the fluid is not adiabatic, the EOS of DE perturbations inside the sphere is different from that outside represented by the background value. Suppose that $w_{\rm de}$ is the EOS for the background DE while $w_{\rm de}^{\rm s}$ is the EOS inside the sphere. Since $c_{\rm{eff}}^2\coloneqq\delta P/\delta \rho$, we have
\begin{equation}
 \delta w \equiv w_{\rm de}^{\rm s} - w_{\rm de} = (c_{\rm{eff}}^2 - w_{\rm de})\frac{\delta_{\rm{de}}}{1+\delta_{\rm{de}}}\,,
    \label{eq:effectiveSoundSpeed}
\end{equation}
where $\delta_{\rm de}$ represents the density perturbations of the DE fluid. When perturbations are small ($\delta_{\rm de}\ll 1$), the perturbed EOS is approximately equal to the background one, while when $\delta_{\rm de}\gg 1$ then $w_{\rm de}^{\rm s} \approx c_{\rm{eff}}^2$. This shows that DE can behave similarly to matter in terms of perturbations when $c_{\rm{eff}}^2\ll 1$.

To derive the equations of the spherical collapse when DE perturbations need to be taken into account, we can use a generalized approach based on the pseudo-Newtonian approximation to gravitational interactions \cite{Abramo:2007iu,Pace:2017qxv}. Perturbing the continuity, Euler and Poisson equations for both fluids leads to
\begin{equation}\label{eq:first_oredr_system}
 \begin{aligned}
  & \delta_{\rm m}' + \left(1+\delta_{\rm m}\right)\frac{\tilde{\theta}}{a} = 0 \,,\\
  & \delta'_{\rm de} + \frac{3}{a}(c_{\rm{eff}}^2 - w_{\rm de})\delta_{\rm{de}} + \left[1+w_{\rm de} + (1+c_{\rm{eff}}^2)\delta_{\rm{de}}\right] \frac{\tilde{\theta}}{a} = 0\,,\\
  & \tilde{\theta}' + \frac{1}{2a}(1-3w_{\rm{eff}})\tilde{\theta} + \frac{\tilde{\theta}^2}{3a} + \\
  & \frac{3}{2a}\left[\Omega_{\rm m}\delta_{\rm m} + (1+3c_{\rm{eff}}^2)\Omega_{\rm{de}}\delta_{\rm{de}}\right] = 0\,,
 \end{aligned}
\end{equation}
where $\tilde{\theta} = \theta/H$, with $\theta$ the divergence of the peculiar velocity.

As done before, we can linearize the previous equations and determine how the growth factor equation is affected by DE perturbations and impose the appropriate initial conditions for a model with arbitrary sound speed to both linear and nonlinear equations.

In a Universe containing DE, the exponent $n$ is normally very close to unity with the exception of early DE models. In those cases, at early times matter is not completely dominant and therefore $\Omega_{\rm m}$ can be quite different from $1$. For what concerns the linearized DE perturbation equation, by neglecting the decaying mode and using the solution in Eq.~(\ref{eq:deltaMatterLinear}) found for the matter contrast, we obtain 
\begin{equation}
    \delta_{\rm{de}}^{\rm{lin}} = \frac{3\Omega_{\rm m}(1 + w_{\rm de})}{n(1+2n-3w_{\rm{eff}})}\delta_{\rm m}^{\rm{lin}}+c_2\,.
    \label{eq:deltaDELinear}
\end{equation}
Because of the adiabatic condition $\delta_{\rm{de}} = (1+w_{\rm de})\delta_{\rm m}$, we set $c_2=0$. When nonadiabatic initial conditions are present, pressure imbalances might eventually equalize, causing any nonadiabatic component to dissipate during the evolution of the perturbations \cite{Abramo:2007iu}, only leaving nonvanishing conventional adiabatic fluctuations.

So the early-time solutions, Eqs.~(\ref{eq:deltaMatterLinear}) and (\ref{eq:deltaDELinear}), can be used to set the initial conditions for both linear and nonlinear equations, by evaluating them at the scale factor from which we want to start the numerical integration, hereafter $a_{\rm min}$.

For completeness, we report the equations to be solved to determine the growth factor in clustering dark energy models. However, note that we report a second order equation only for matter perturbations, as to avoid derivatives of the sound speed in the equation for DE
\begin{equation}\label{eq:growth_system}
 \begin{aligned}
  & \delta_{\rm m}'' + \frac{3}{2a}\left(1-w_{\rm eff}\right)\delta_{\rm m}' - \frac{3}{2a^2}\left[\Omega_{\rm m}\delta_{\rm m} + (1+3c_{\rm{eff}}^2)\Omega_{\rm{de}}\delta_{\rm{de}}\right] = 0 \,,\\
  &\delta'_{\rm de} + \frac{3}{a}(c_{\rm{eff}}^2 - w_{\rm de})\delta_{\rm{de}} + \left(1+w_{\rm de}\right) \frac{\tilde{\theta}}{a} = 0\,,\\
  &\tilde{\theta}' + \frac{1}{2a}(1-3w_{\rm{eff}})\tilde{\theta} + 
  \frac{3}{2a}\left[\Omega_{\rm m}\delta_{\rm m} + (1+3c_{\rm{eff}}^2)\Omega_{\rm{de}}\delta_{\rm{de}}\right] = 0\,.
 \end{aligned}
\end{equation}
These equations show that matter perturbations are affected and sourced by DE perturbations.

To determine the initial conditions, we need to assume that the perturbation collapses at a given scale factor $a_{\rm c}$ (or equivalently redshift $z_{\rm c}$). At this time, the matter density contrast diverges.

Since the nonlinear equations can only be solved numerically, to define the collapse we must specify a value for the numerical infinity, indicated by $\delta_{\infty}$. To find the initial conditions that lead to the collapse at $a_{\rm c}$, we integrate the nonlinear equations (\ref{eq:first_oredr_system}) in the range $a\in \left[a_{\rm min},a_c\right]$ with initial conditions
\begin{subequations}
 \begin{align}
    \delta_{\rm m}(a_{\rm min}) = &\, \delta_{\rm c}^{*} \left(\frac{a_{\rm min}}{a_{\rm c}}\right)^n\,, \label{eq:numericalInitialConditions}\\
    \delta_{\rm{de}}(a_{\rm min}) = &\, \frac{n(1 + w_{\rm de})}{n+3(c^2_{\rm{eff}}-w_{\rm de})}\delta_{\rm m}(a_{\rm min})\,,\\
    \tilde{\theta}(a_{\rm min}) = &\, -n\delta_{\rm m}(a_{\rm min})\,,
 \end{align}
\end{subequations}
where $n$ is defined in Eq.~(\ref{eq:deltaMatterLinear}). We choose initial conditions with that form because in the EdS Universe the quantity $\delta_{\rm c}^{*}$ is exactly the linearly extrapolated density contrast at collapse. In the EdS case, the solution of Eq.~(\ref{eq:linearDensityContrast}) is indeed\footnote{Eq.~(\ref{eq:linearDensityContrast}) can be solved exactly in the EdS case, since the coefficients are constant.}
\begin{equation}
 \delta_{\rm m}^{\rm{lin}} = \delta_{\rm c}^{*}\left(\frac{a}{a_{\rm c}}\right)\,,
\end{equation}
and therefore $\delta_{\rm m}^{\rm{lin}}(a_{\rm c}) = \delta_{\rm c}^{*}$.

If the Universe is not an EdS, the linear matter density contrast should be numerically integrated. Thus, in a more general case, the value $\delta_{\rm c}^{*}$ does not represent the linearly extrapolated density contrast at collapse, but it fully specifies the initial conditions in Eq.~(\ref{eq:numericalInitialConditions}).

The numerical algorithm that we set up searches for the zero of the auxiliary function
\begin{equation}
    f(\delta_{\rm c}^{*}) = \delta_{\rm m}(\delta_{\rm c}^{*},a_{\rm c}) - \delta_{\infty}\,,
    \label{eq:bisection_function}
\end{equation}
with the bisection method, while the system in Eqs.~(\ref{eq:first_oredr_system}) is integrated with a Runge-Kutta 4th-order-integrator at each iteration of the bisection, to obtain $\delta_{\rm m}$. 

It is clear that when the function in Eq.~(\ref{eq:bisection_function}) is zero, the nonlinear density contrast at collapse is equal to the chosen numerical infinity. The meaning of the functional dependence of $\delta_{\rm m}$ in Eq.~(\ref{eq:bisection_function}) on the variables is that the nonlinear density contrast depends on the value of the initial condition $\delta_{\rm c}^{*}$, considered as the independent variable, and on the chosen collapse scale parameter $a_{\rm c}$, which is fixed. Hence, the bisection algorithm returns the value $\delta_{\rm c}^{*}$. Knowing the correct value of $\delta_{\rm c}^{*}$ fully specifies the initial conditions. Then, the linearized version of Eq.~(\ref{eq:first_oredr_system}) is integrated to obtain $\delta_{\rm m}^{\rm{lin}}(a_{\rm c})$, that is the linearly extrapolated matter density contrast at collapse.

Another quantity that we shall compute is the turnaround scale factor, $a_{\rm{ta}}$. To determine it numerically, it is sufficient to find the value of the scale factor that maximizes the sphere radius, proportional to
\begin{equation}
    x(\delta)=\zeta_{\rm{ta}}^{1/3}\frac{y}{(1+\delta)^{1/3}}\,,
    \label{eq:sphereRadiusOfDelta}
\end{equation} 
where the subscript ``$\rm{ta}$'' stands for turnaround, $y=a/a_{\rm{ta}}$, $x=R/R_{\rm{ta}}$ with $R$ the radius of the sphere and $\zeta_{\rm ta} = \delta_{\rm m,ta}+1$ the overdensity at turnaround. To compute the scale factor at virialization, $a_{\rm{vir}}$, we use the result in Ref.~\cite{Wang:1998gt} for the radius of the sphere at virialization
\begin{equation}
    x_{\rm{vir}} = \frac{1-\eta_{\rm v}/2}{2+\eta_{\rm t} - 3\eta_{\rm v}/2}\,,
\end{equation}
where $\eta_{\rm t} = 2\zeta_{\rm{ta}}^{-1}\Omega_{\rm{de}}(a_{\rm{ta}})/\Omega_{m}(a_{\rm{ta}})$ and $\eta_{\rm v} = 2\zeta_{\rm{ta}}^{-1}(a_{\rm{ta}}/a_{\rm c})^3\Omega_{\rm{de}}(a_{\rm c})/\Omega_{\rm m}(a_{\rm c})$.

Hence, to compute the virialization scale factor, we solve the equation $x_{\rm{vir}}-\zeta_{\rm{ta}}^{\frac{1}{3}}\frac{a}{a_{\rm{ta}}(1+\delta_{\rm m}(a))^{\frac{1}{3}}}=0$, that is, we search for the scale factor that equates the radius of the sphere and the virialization radius.

It is known through the analytical analysis for the EdS Universe that a more reasonable value for the matter overdensity at virialization is obtained by computing the numerator of the overdensity $\zeta = \zeta_{\rm{ta}}\frac{y^3}{x^3}$ at collapse and the denominator at virialization. We name this value\footnote{Another way to obtain the same is reported in Ref. \cite{Pace:2017qxv}, by solving the first order system, Eq.~(\ref{eq:first_oredr_system}), and using the new variable $f\equiv 1/\delta$, leading to a considerable noise reduction.} $\zeta_{\rm{vir}}^*$, being $\zeta_{\rm{vir}}^{*}=\zeta_{\rm{ta}}\left(\frac{a_{\rm c}}{a_{\rm{vir}}x_{\rm{vir}}}\right)^3$.

Our Python code is designed to solve the background equations (\texttt{friedman\_solver}) and the nonlinear perturbations (\texttt{nonlinear\_perturbations\_solver}) and, as we will clarify later in the text, the quintessence potential and field (\texttt{quintessence\_solver}). The complete code is freely available in the following GitHub Repository \cite{software}.

\section{Fast transition models analysis}\label{sec:fast_trans_eos}
In this section, we introduce the six fast transition models that we analyze. We then delve into the quintessence description of the dark fluid by computing the scalar field and potential corresponding to each model, focusing specifically on recent and fast transitions of the EOS. We then see how linear and nonlinear matter perturbations grow in a Universe containing the dark fluids just mentioned. We consider perturbations in two cases: 
\begin{itemize}
    \item[-] Smooth DE evolution.
    \item[-] Clustering of DE.
\end{itemize} 
We mostly focus on models with no phantom behavior \cite{Lonappan:2017lzt}, but one of them, namely $w_6$, is in the phantom regime. 

\subsection{Introduction to fast transition EOS}\label{subsec:eos_intro}

Recently, there has been some evidence of a tension between the standard cosmological model, the $\Lambda$CDM paradigm, and modern observations \cite{zhao2012examining,Ding:2015vpa,Sahni:2014ooa}. Thus, evolving DE scenarios are to date acquiring a renewed importance to heal cosmic tensions. 

Here we explore the consequences of some parameterizations, characterized by a steplike transition, shaped by four parameters, $w_{\rm i}$, $w_{\rm f}$, $z_{\rm t}$ and $\Gamma$, where
\begin{itemize}
    \item[-] $w_{\rm i}$ is the initial EOS value. This is the value that DE had for most of the Universe lifetime.
    \item[-] $w_{\rm f}$ is the final EOS value, constrained to $w_{\rm f}<-1/3$ since the value of the deceleration parameter at the present time must be negative.
    \item[-] $z_{\rm t}$ is the parameter fixing the transition; specifically, it is referred to as the \emph{transition redshift}.
    \item[-] $\Gamma$ is the parameter that regulates the speed or steepness of the transition.
\end{itemize}
We consider six DE models, described by the following functional forms 

\begin{subequations}\label{eq:de_eos_combined}
    \begin{align}
        w_1(z) = &\, \frac{1}{2}(w_{\rm i} + w_{\rm f}) - \nonumber \\
                 &\, \frac{1}{2}(w_{\rm i} - w_{\rm f})\tanh{\left[\Gamma\ln\left(\frac{1+z_t}{1+z}\right)\right]}\,, \quad  \label{eq:de_eos_1} \\
        w_2(z) = &\, w_{\rm f} + (w_{\rm i} - w_{\rm f})\frac{\left(\frac{z}{z_{\rm t}}\right)^\Gamma}{1+\left(\frac{z}{z_{\rm t}}\right)^\Gamma}\,, \quad\quad\quad\quad  \label{eq:de_eos_2} \\
        w_3(z) = &\, w_{\rm i} + \frac{w_{\rm f} - w_{\rm i}}{1+e^{\frac{z-z_{\rm t}}{\Gamma}}}\,, \quad\quad\quad\quad\quad\quad\quad\quad\;\; 
        \label{eq:de_eos_3} \\
        w_4(a) = &\, w_{\rm f} + \nonumber\\ 
        &\, (w_{\rm i} - w_{\rm f})\frac{1+e^{\frac{a_{\rm t}}{\Gamma}}}{1+e^{-\frac{a-a_{\rm t}}{\Gamma}}}\frac{1-e^{-\frac{a-1}{\Gamma}}}{1-e^{\frac{1}{\Gamma}}}\,. \quad\quad\!\! \label{eq:de_eos_4}
    \end{align}
\end{subequations}
In the same order of the above equations, those models are described in the works \cite{de_eos_1,de_eos_2,de_eos_3,Pace:2010}.
In Eq.~(\ref{eq:de_eos_4}), $a_{\rm t}=1/(1+z_{\rm t})$. The four models above are characterized by the four parameters previously discussed. We furthermore consider
\begin{subequations}\label{eq:de_eos_combined2}
    \begin{align}
        w_5(z) & = -1-\frac{\Gamma}{3\ln(10)}\bigg\{ 1 + \nonumber\\
        & \tanh{\left[\Gamma\log\left(\frac{1+z}{1+z_{\rm t}}\right)\right]}\bigg\}\,, \,\,\,\qquad \quad\label{eq:de_eos_5} \\
        w_6(z) & = -1-\frac{\left\{1+\tan{[\log(1+z)]}\right\}}{3\ln(10)}\,. \quad  \label{eq:de_eos_6}
    \end{align}
\end{subequations}
In the same order of the above equations, those two models are described in the papers \cite{de_eos_5,de_eos_6}.
The model in Eq.~(\ref{eq:de_eos_5}) has two free parameters while the last one, Eq.~(\ref{eq:de_eos_6}),  has no free parameters.

\begin{figure}
    \centering
    \includegraphics[width=1\columnwidth]{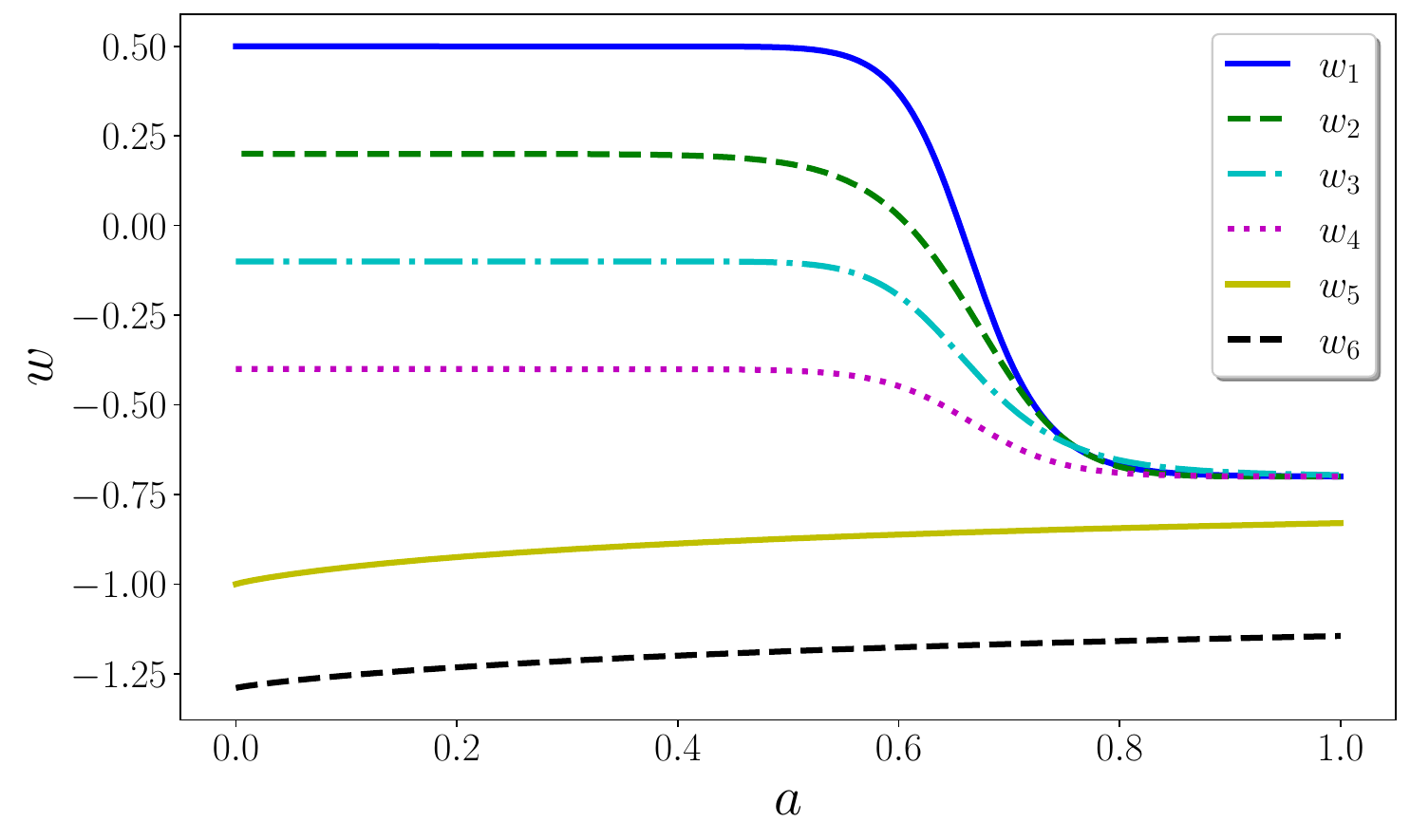}
    \caption{Models of EOS as presented in Eqs.~(\ref{eq:de_eos_1})--(\ref{eq:de_eos_6}), using the same order, with parameter values chosen to make the curves clear and not overlapping. With the exception of the last two models, all the others exhibit similar properties.}
    \label{fig:all_eos}
\end{figure}

Figure ~\ref{fig:all_eos} shows all the forms of the EOS employed in the paper. The first five models can be interpreted as nonphantom, with suitable parameters choices, while the last is in the phantom regime. The first four models can mimic a fast transition. The difference between them is in the exact shape of the transition, while they are identical far away from it. 

We now provide a brief description of the main properties of some of the fast transition models. Referring to Eq.~(\ref{eq:de_eos_1}), for $\Gamma\rightarrow 0$ it reduces to $w_1=(w_{\rm i}+w_{\rm f})/2$ which is the average between the initial and final EOS values. The average value is also reached when $z=z_{\rm t}$, thus at $z_{\rm t}$ the transition is half way through. For $\Gamma\rightarrow \infty$ it tends to the Heaviside step function centered at $z_{\rm t}$. The transition speed can be related to the redshift interval $\Delta z$ around $z_{\rm t}$ in which the transition takes place. For Eq.~(\ref{eq:de_eos_1}) it can be computed to be $\Delta z = 2(1+z_{\rm t})\sinh{(2\Gamma^{-1})}$. To obtain this, we define $\Delta z$ as the interval between the redshifts where Eq.~(\ref{eq:de_eos_1}) takes the values $(w_{\rm i} + w_{\rm f})/2 \pm \Delta w \tanh(2)/2$, where $\tanh(2)\approx 0.96$. For this model, the transition speed depends linearly on the transition redshift and thus those two parameters are not decoupled. In Eq.~(\ref{eq:de_eos_3}), $\Delta z$ depends only on $\Gamma$ and thus when the transition happens it does not influence the transition speed.

Most of the DE models have four free parameters that must be constrained. The first constraint we apply is considering fast and recent transitions. In fact recent developments \cite{de_eos_2,de_eos_3} demonstrate that the parametrization in Eq.~(\ref{eq:de_eos_2}) performs optimally when there is a recent and rapid transition. Furthermore, in Ref. \cite{de_eos_1} an early-time transition is indistinguishable from a genuine Chevallier-Polarski-Linder parametrization \cite{Chevallier:2000qy,Linder:2002et}. Finally, we also demonstrate that only a recent transition has observable effects. Therefore, these points justify why we focus on fast and recent transitions. We choose as ``default'' the values reported in the top part of Table~\ref{tab:params}. We are thus considering mostly freezing models with the exception of Eq.~(\ref{eq:de_eos_6}) which is thawing\footnote{Thawing models have a value of $w$ which begins near $-1$ and increases with time.}. This choice also allows us to interpret the models as quintessence theories, as the EOS is always above the phantom divide line $w=-1$. Throughout this paper, we consistently utilize the background parameters listed in the bottom part of Table~\ref{tab:params}.

\begin{table}
    \caption{Top: default parameters used in the models. $\Gamma$ is not shown since it is highly dependent on the model. Bottom: background cosmological parameters. We have approximated the values obtained by \cite{Planck:2018vyg}.}
    \label{tab:params}
    \renewcommand{\arraystretch}{1.5} 
    \begin{ruledtabular}
        \begin{tabular}{|>{\centering\arraybackslash}m{2cm}|>{\centering\arraybackslash}m{2cm}|>{\centering\arraybackslash}m{2cm}|>{\centering\arraybackslash}m{2cm}|}
            $w_{\rm i}$  & $w_{\rm f}$ & $z_{\rm t}$ & $\Gamma$ \\
            \hline
            $-0.4$ & $-1$  & $0.5$ & $-$ \\
            \hline\hline\hline
            $\Omega_{\rm{m}}$  & $\Omega_{\rm{de}}$ & $H_0$ & $z_{\rm{ls}}$ \\
            \hline
            $0.3$ & $0.7$  & $67$ & $1090$ \\
        \end{tabular}
    \end{ruledtabular}
\end{table}

\subsection{Quintessence potential and field}\label{subsec:quintessencePotentials}

In view of the discussion above, we numerically compute the quintessence potential and field for the DE models reported in Sec.~\ref{subsec:eos_intro}. To do so, we assume the Lagrangian in Eq.~(\ref{eq:quint}).

To perform the numerical integration, it is more convenient to work with the dimensionless quantities $\tilde{\phi}$ and $\tilde{V}$, as defined by Eqs.~(\ref{eq:dimensionlessPotField}) and (\ref{eq:dimensionlessField}). As noted in Sec.~\ref{subsec:effectiveFluidDescription}, we have the freedom to choose the integration constant, $\tilde{\phi}_0$, of the scalar field. For simplicity, we set $\tilde{\phi}(a_{\rm min})=0$, where $a_{\rm min}$ is the lower integration limit.

\begin{figure}
  \centering
  \includegraphics[width=1\columnwidth,clip]{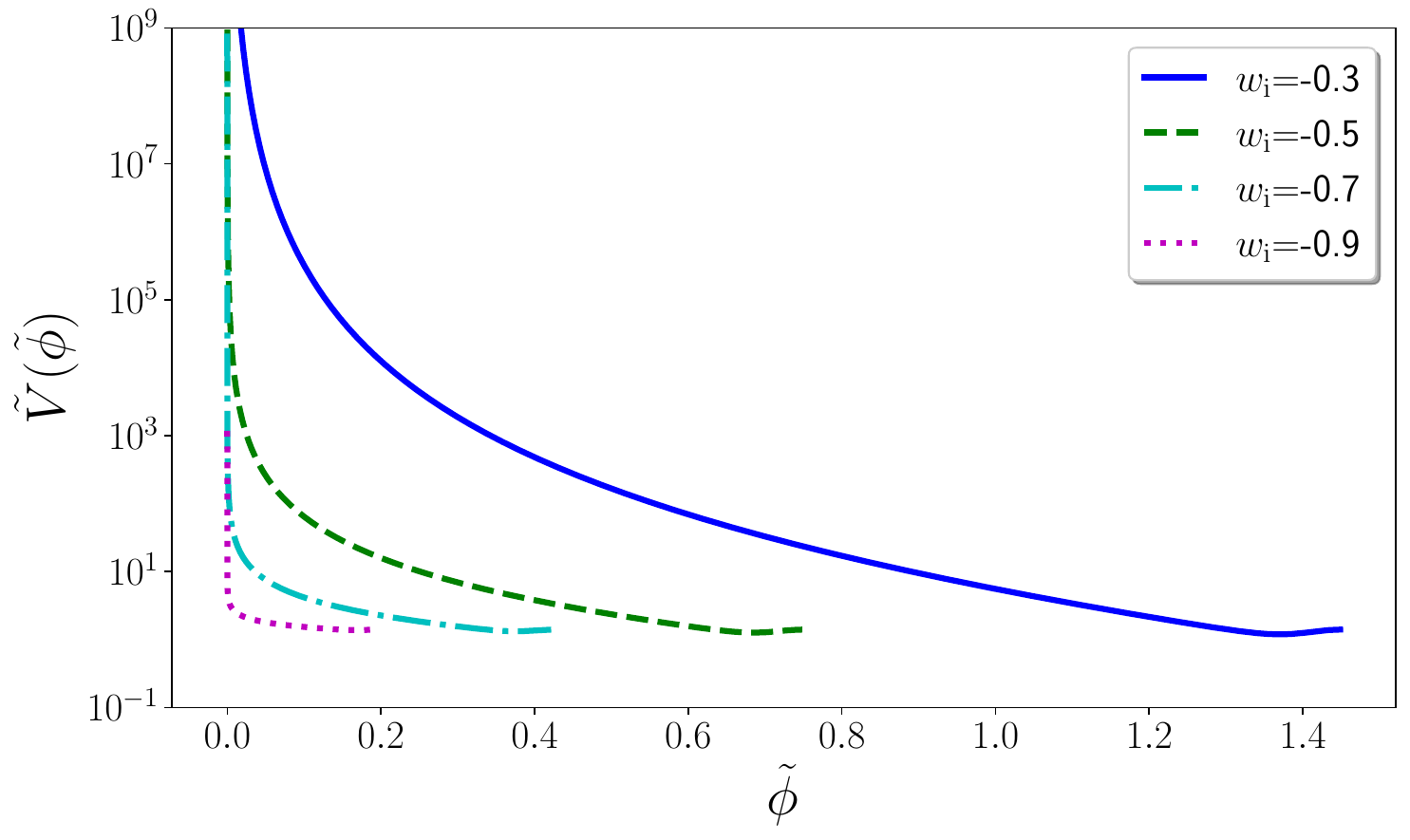}
  \includegraphics[width=1\columnwidth,clip]{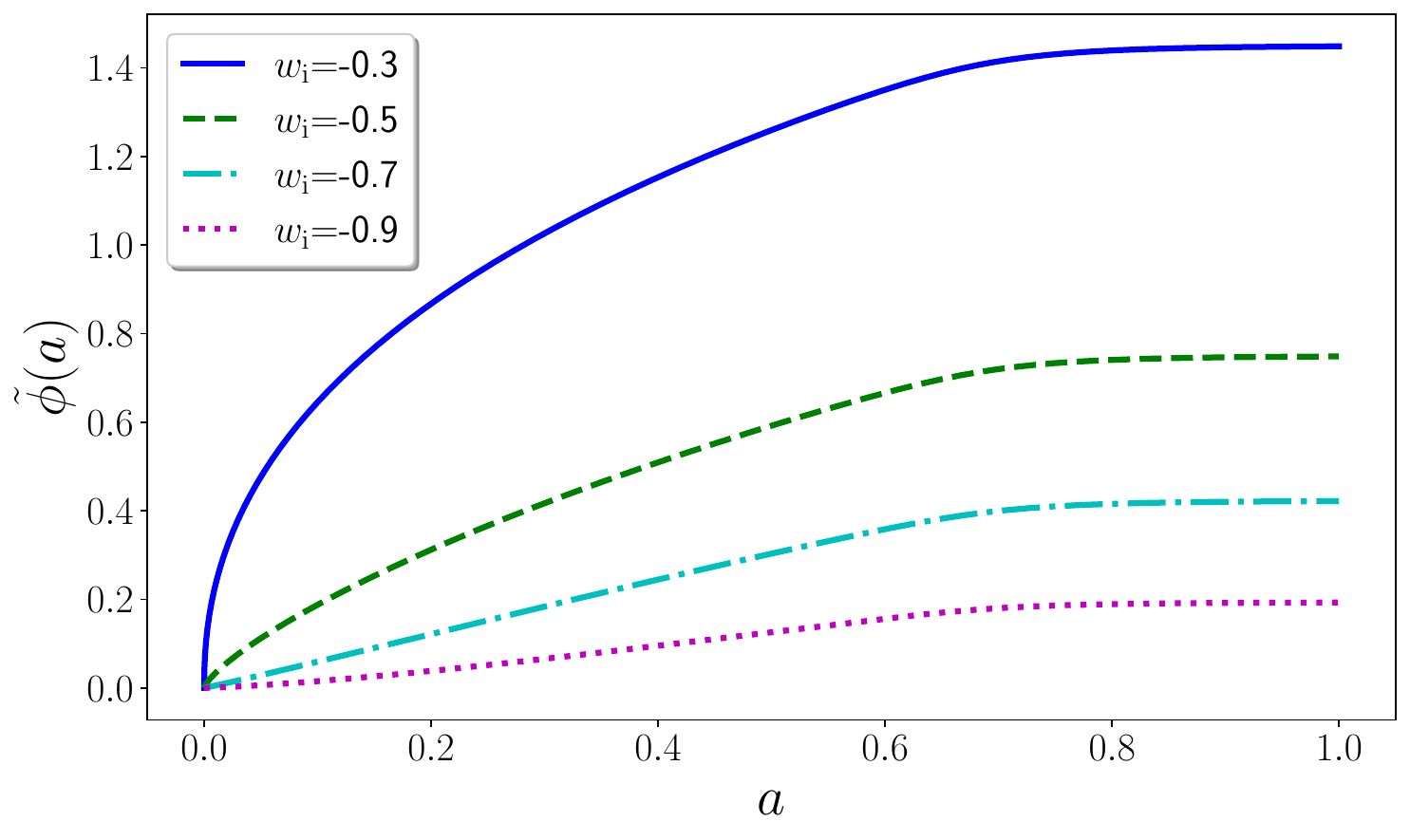}
  \caption{Model with the EOS $w_1$ with $w_{\rm f}=-1$, $\Gamma=10$, $z_{\rm t}=0.5$. Top panel: quintessence potential as a function of the scalar field. The closer is $w_{\rm i}$ to zero, the faster is the divergence for $\tilde{\phi}\rightarrow0$. Bottom panel: time evolution of the scalar field. After the transition to $w_{\rm f}$, the field flattens as it enters a slow-rolling period.}
  \label{fig:quint_1_w_i}
\end{figure}

In Fig.~\ref{fig:quint_1_w_i}, we present the evolution of the scalar field and its potential for the first model, varying the $w_{\rm i}$ parameter. We immediately notice that as $w_{\rm i}$ approaches $-1$, $\tilde{\phi}$ decreases, resulting in a smaller domain for the potential. In the limit where both $w_{\rm i}$ and $w_{\rm f}$ equal $-1$, the model reduces to the cosmological constant, for which the potential does not depend on $\tilde{\phi}$. Numerically, this degeneracy results in the potential becoming the single point $(\tilde{\phi}_0, 2\Omega_{\rm{de}}^0)$. As $w_{\rm i}$ approaches zero, DE tends to a matter-like fluid. For any value of $w_{\rm i}$ inside the interval $(-1,0]$, the potential is divergent in zero and its slope is shallower when $w_{\rm i}$ increases and becomes less negative. After the transition to $w_{\rm f}=-1$, both the field and the potential become approximately constant, as in this regime DE behaves as a cosmological constant.

\begin{figure}
    \centering
    \includegraphics[width=1\columnwidth,clip]{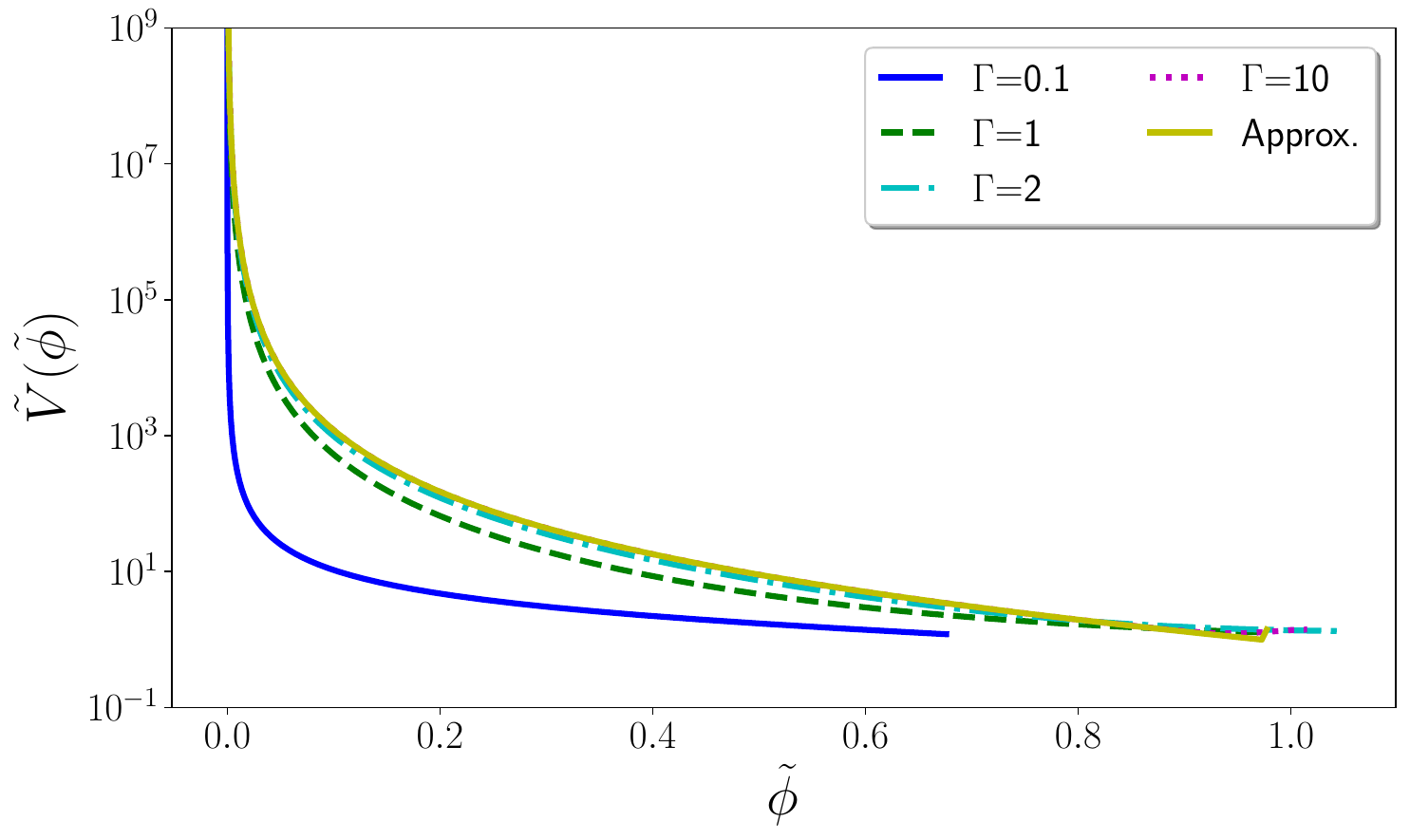}
    \includegraphics[width=1\columnwidth,clip]{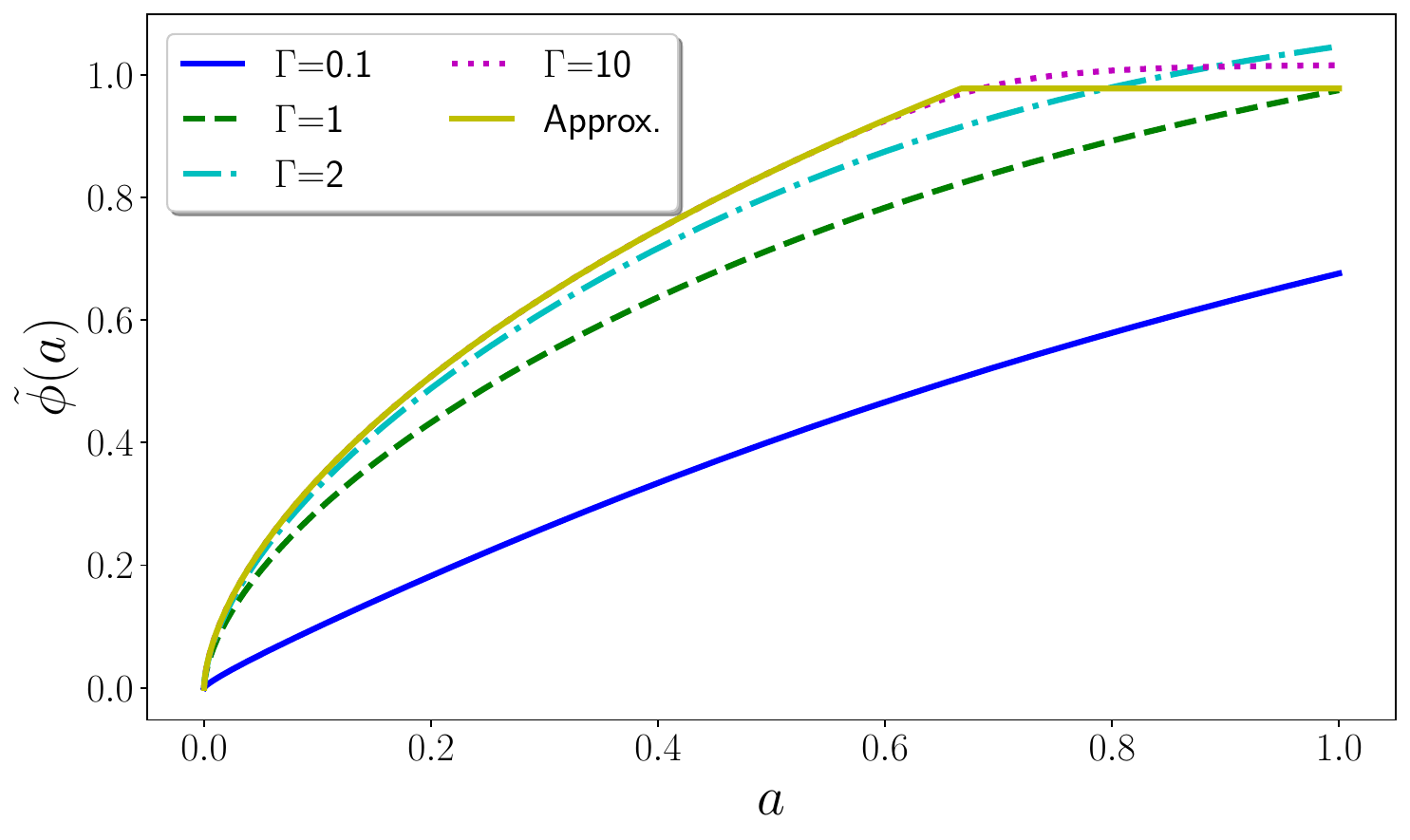}
    \caption{Plots for model $w_1$ with $w_{\rm i}=-0.4$, $w_{\rm f}=-1$, $z_{\rm t}=0.5$. The solid yellow line is computed using a true step transition, which in this model corresponds to the limit $\Gamma\rightarrow\infty$. Top panel: corresponding quintessence potential. Bottom panel: corresponding scalar field. A very fast transition results in $w_{\rm f}=-1$ right after $z_{\rm t}$, while a smoother transition needs additional time to reach the asymptotic value of $w_{\rm f}=-1$, corresponding to the slowly rolling field.}
    \label{fig:quint_1_trans_steepness}
\end{figure}

Figure ~\ref{fig:quint_1_trans_steepness} shows the shape of the potential and the evolution of the scalar field obtained varying $\Gamma$. The dependence on this parameter is relatively weak, with only one substantially different curve corresponding to $\Gamma=0.1$. With this choice of parameter, the steplike character of the transition is lost. Apart from this limiting value, the potential and the field do not depend strongly on how fast the transition is. Thus, a true step transition provides a good approximation for fast transition models, and it has the benefit of having analytical $\tilde{V}(a)$ and $\tilde{\phi}(a)$.

The other models we examined, Eqs.~(\ref{eq:de_eos_2})--(\ref{eq:de_eos_4}), are quite similar to Eq.~(\ref{eq:de_eos_1}) as we can choose their parameter values in a way that they can mimic Eq.~(\ref{eq:de_eos_1}).

Thus, we shift our attention to the fifth model, Eq.~(\ref{eq:de_eos_5}), which is markedly distinct. This functional form contains only two free parameters, $z_{\rm t}$ and $\Gamma$, which influence the final result in a nontrivial manner, making it an intriguing subject. Moreover, all other missing degrees of freedom are completely dependent on these two free parameters, resulting in a restricted range of possible behaviors compared to the other models. 

Notably, this model does not exhibit a slow-roll period, since the EOS is always different from $-1$. Additionally, Eq.~(\ref{eq:de_eos_5}) cannot be regarded as a fast transition. In fact, while the model starts to behave as a fast transition for large negative values of $\Gamma$, this results in the final value $w(z=0)>-1/3$, which is inadmissible as it would not lead to an accelerated expansion. 
Although the parameters influence each other, the scalar field and the potential show a weak dependence on the transition speed and the final value of the EOS, compared to the other parameters. Thus, even if the parameter values depend on each other, the general behavior is mostly dictated by the parameter $z_{\rm t}$. Eq.~(\ref{eq:de_eos_6}) has no free parameters and is classified as a phantom model since it has an EOS $w$ that is always smaller than $-1$.

For additional plots of the background quantities, please refer to Appendix \ref{sec:appendix}.

Before moving on with the analysis of the effects that fast transition induces on linear and nonlinear perturbations, we want to comment more about the results obtained in this section. As extensively discussed above, here we are following the designer approach, which implies fixing the background EOS $w(a)$ and then reconstructing the scalar field and its potential once the functional form of the Lagrangian has been fixed. For simplicity, we considered results only for the quintessence/phantom case as the Lagrangian is unequivocally defined, however we could have performed the same exercise considering a $k$-essence Lagrangian. In this case though, as illustrated in \cite{Battye:2016alw}, we can distinguish at least three different functional forms and to solve for the potential we need not only fix the Lagrangian, but also the functional form of the kinetic term. This will make the investigation extremely complex. Nevertheless, the analysis in \cite{Battye:2016alw} shows that the potentials in the different cases, even though derived for a constant EOS, do not differ qualitatively, but only quantitatively. This reinforces our choice to consider only quintessence models.

Being based in the Horndeski framework, we can move beyond the simplest models and consider other subclasses, such as KGB, Gauss-Bonnet and $f(R)$. While the first two effectively involve scalar fields, the latter will lead to the determination of the functional form for $f(R)$ thanks to which one can study the evolution of the cosmological model if it satisfies certain viability criteria. In all the cases, the analysis will not be trivial and not directly comparable with what done here. We, therefore, prefer to be more conservative and discuss models for which we can provide an immediate physical interpretation.

When studying perturbations, we would face the same problem, even considering the fluid description. For quintessence/phantom models, perturbations are negligible, but this is not necessary the case for $k$-essence models. To have an exact value for the sound speed of perturbations, we would need, once again, the precise Lagrangian we started with. Hence, we simply consider the case where dark energy, in its broader sense, fully clusters, as we discuss in the next section.

\subsection{Linear regime}\label{subsec:linearRegime}
In this section, we analyze how the linear matter density contrast is affected by the fast transition models presented in Sec.~\ref{subsec:eos_intro}. We solve the differential equations given by Eq.~(\ref{eq:first_oredr_system}) with initial conditions obtained from the procedure outlined in Sec.~\ref{subsec:sphcolmodel}, with $a_{\rm min}=a_{\rm{ls}}$. When we include DE perturbations, we assume that the DE EOS is not the same inside or outside the collapsing regions. The difference is encoded in the effective sound speed, that we will fix to $c_{\rm{eff}}=0$ since we lack the underlying model\footnote{If one assumes quintessence,  $c_{\rm{eff}}=1$. Different generalized versions of quintessence, under the form of quasi-quintessence fields can be seen in Ref. \cite{DAgostino:2022fcx}.} from which the parametrizations emerge. This is the minimum allowed effective sound speed value and it maximizes DE perturbations.

\begin{figure}
    \centering
    \includegraphics[width=1\columnwidth,clip]{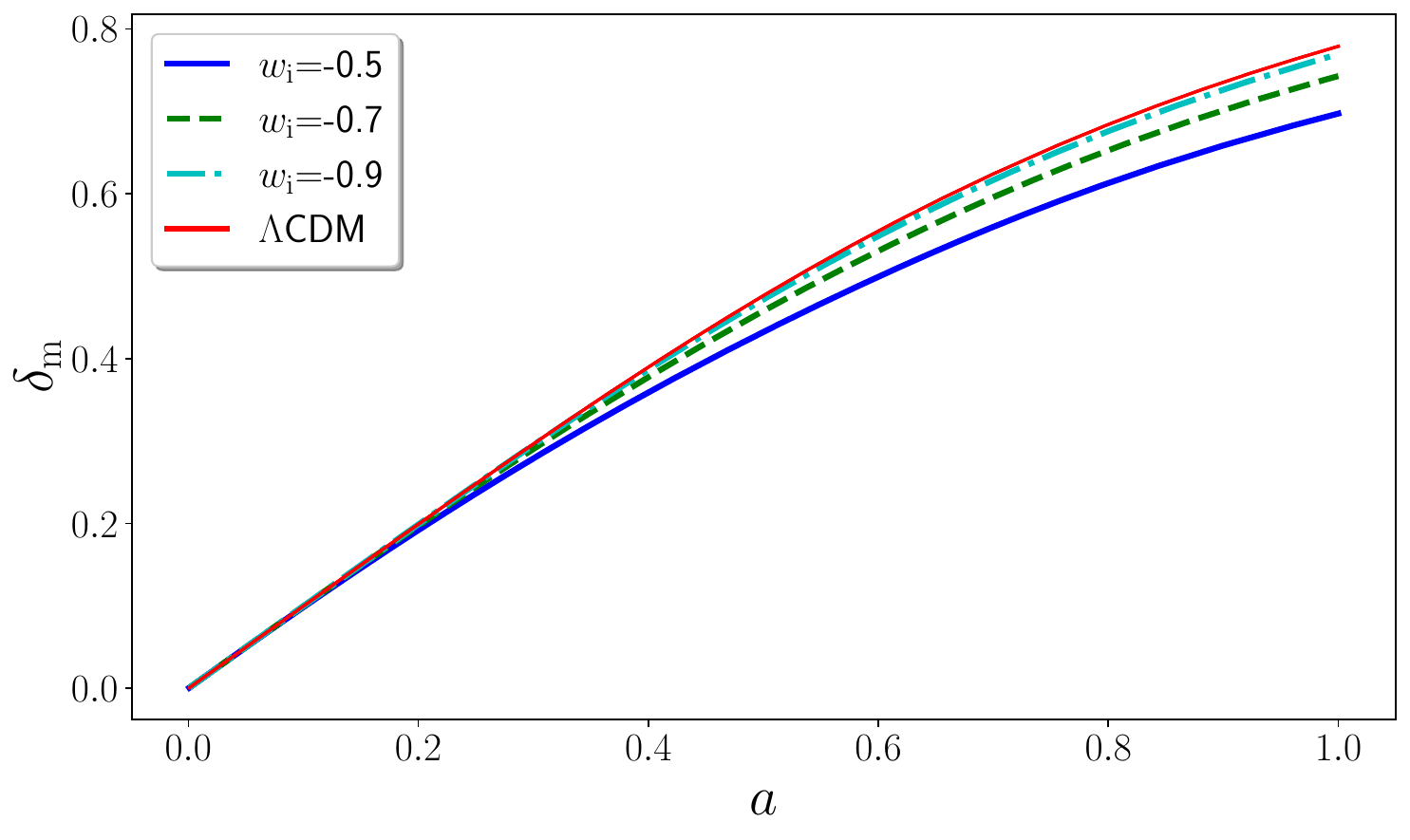}
    \caption{Evolution of the linear matter density contrast for model $w_1$ with $w_{\rm f}=-1$, $\Gamma=10$, $z_{\rm t}=0.5$, when $\delta_{\rm{de}}=0$. Corresponding background quantities are reported in Fig.~\ref{fig:unpert_1_w_i_background}.}
    \label{fig:unpert_1_w_i}
\end{figure}

We start by analyzing the first model given by Eq.~(\ref{eq:de_eos_1}). Figure ~\ref{fig:unpert_1_w_i} illustrates how linear perturbations depend on the $w_{\rm i}$ parameter in the absence of DE perturbations. Linear perturbations increase as $w_{\rm i}$ decreases, as in this case the model becomes more similar to the $\Lambda$CDM one. In fact, even if lower values of $w$ correspond to higher late-time accelerations\footnote{In the $w$CDM model, $\Ddot{a}_0=-0.5\left[\Omega_{\rm m}^0+(1+3w)\Omega_{\rm de}^0\right]$.}, they also correspond to lower Hubble functions for $0<a<1$, as shown in Fig.~\ref{fig:hubble_function}. Due to the Hubble drag, we expect that lower Hubble functions result in higher matter density contrast. Using these considerations, we can explain our results by taking into account the amount of matter at initial conditions. At early times, as there is a small amount of DE, perturbations grow more slowly than in a $\Lambda$CDM model as the rate of expansion is higher in the past. This leads to a smaller value of perturbations today as shown in Fig.~\ref{fig:unpert_1_w_i}. We remark that, even though DE perturbations are often neglected in the analysis of perturbations, this is strictly true only for $w_{\rm de}=-1$, as evident also from Eq.~(\ref{eq:first_oredr_system}).

\begin{figure}
  \centering
  \includegraphics[width=1\columnwidth,clip]{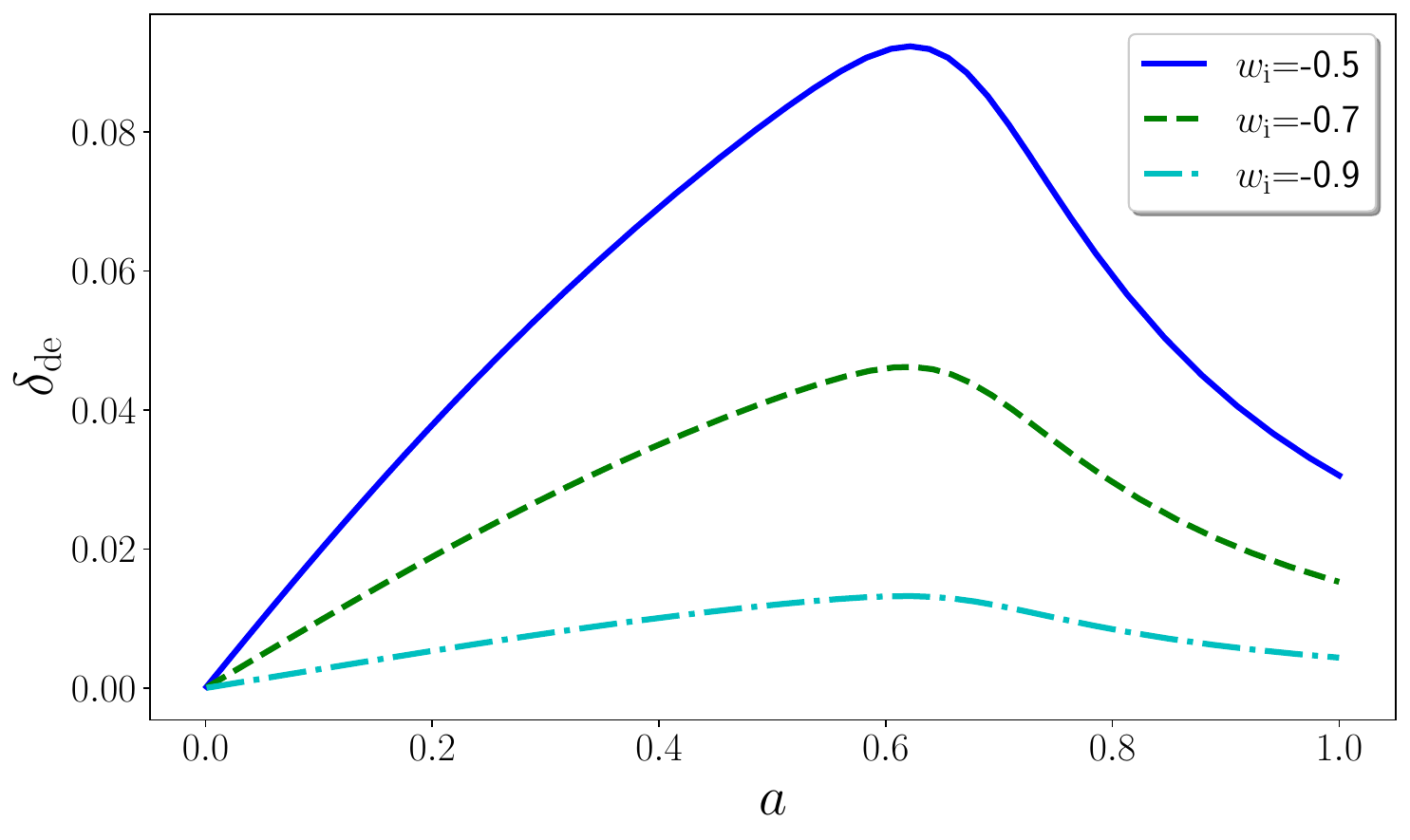}
  \includegraphics[width=1\columnwidth,clip]{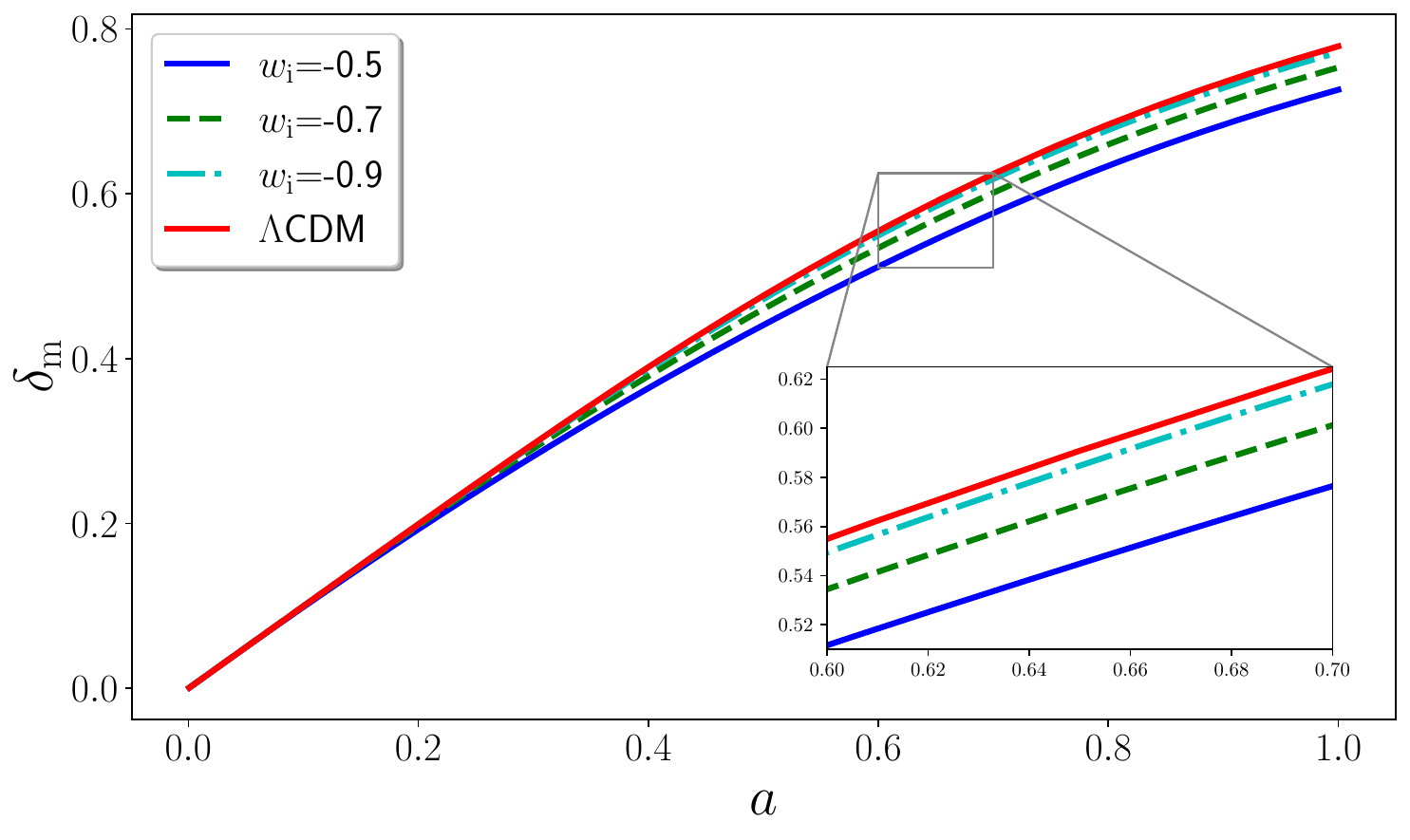}
  \caption{Plots for model $w_1$ with $w_{\rm f}=-1$, $\Gamma=10$, $z_{\rm t}=0.5$. Top panel: DE perturbations. After the transition, perturbations start to decay since the cosmological constant cannot cluster and the already formed overdensities are diluted by the expansion. Bottom panel: linear matter perturbations with $\delta_{\rm{de}}\neq 0$.}
  \label{fig:pert_1_w_i_2}
\end{figure}

What happens if we include DE perturbations is shown in Fig.~\ref{fig:pert_1_w_i_2}. When $\delta_{\rm{de}}\neq 0$ the overall behavior remains qualitatively similar, but with some notable differences. The presence of DE perturbations results in a reduced sensitivity of matter perturbations to the parameters value. Furthermore, if the DE has an EOS with $w_{\rm de}<-1/3$, it tends to dilute matter perturbations, whereas if $w_{\rm de}>-1/3$, it tends to amplify them. This can be deduced by inspecting the source term of Eq.~(\ref{eq:growth_system}). As the transition occurs, DE perturbations cease to grow, and the already present ones gradually decay due to the expansion. As a result, the feedback mechanism weakens, and matter perturbations gradually decouple from DE perturbations. In agreement with general expectations, dark energy perturbations are subdominant and are about an order of magnitude smaller than matter perturbations. Note that here we just consider the effects of $\delta_{\rm de}$ on $\delta_{\rm m}$, without attempting to discuss an eventually more general definition of matter perturbations based on the combination of the two.

\begin{figure}
    \centering
    \includegraphics[width=1\columnwidth,clip]{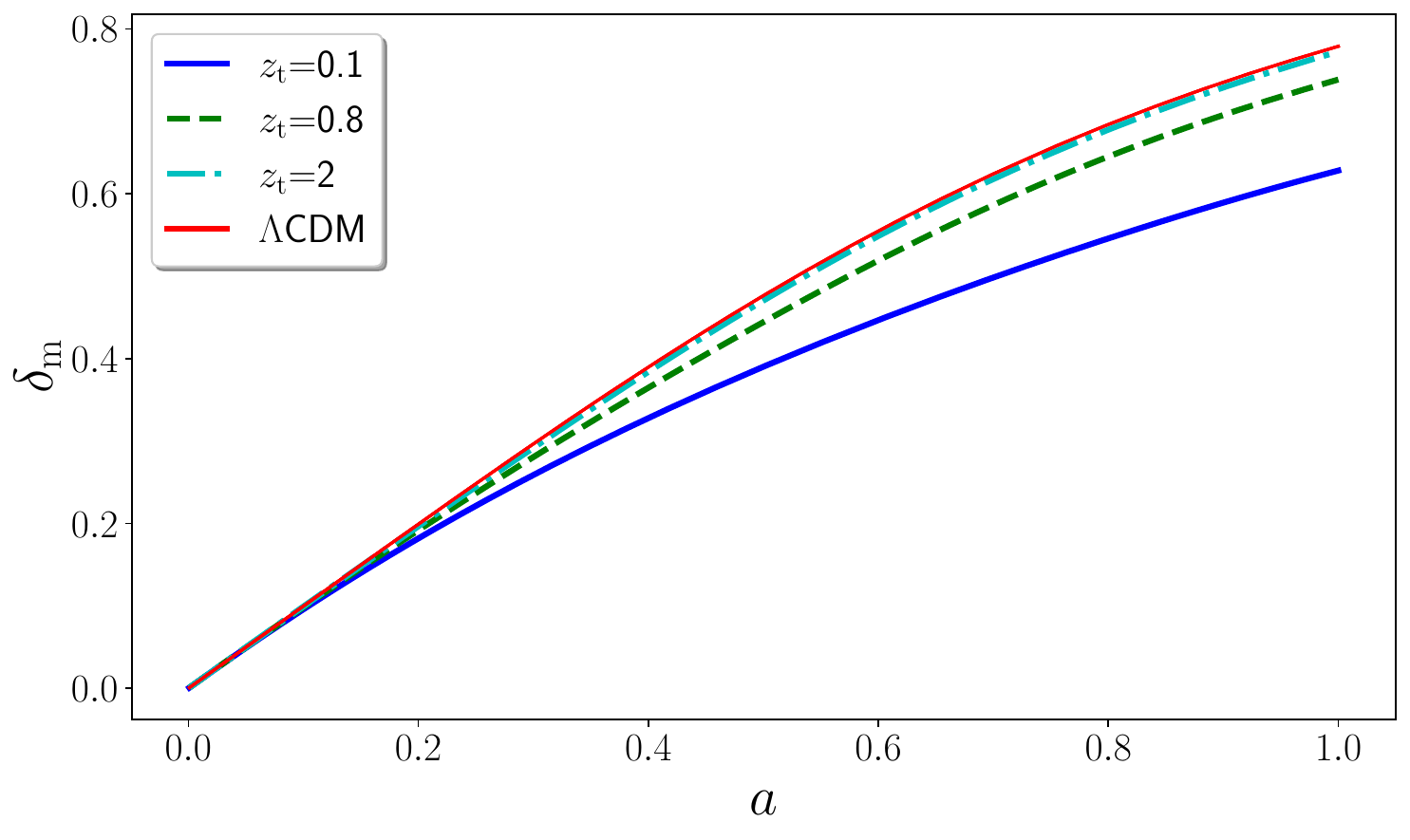}
    \caption{Linear perturbations with $\delta_{\rm{de}}\neq 0$, for model $w_1$ with $w_{\rm i}=-0.4$, $w_{\rm f}=-1$, $\Gamma=10$. For $z_{\rm t}>2$ the curves become almost indistinguishable from the $\Lambda$CDM model.}
    \label{fig:pert_1_z_t_1}
\end{figure}

Figure ~\ref{fig:pert_1_z_t_1} presents the results obtained by varying the $z_{\rm t}$ parameter. As one could expect, for $z_{\rm t}>2$, the curves become nearly indistinguishable from the $\Lambda$CDM model due to the sub-dominant role of DE at early times. Consequently, an early transition to $w_{\rm f}=-1$ makes the model very similar to $\Lambda$CDM. This observation has led us to concentrate on recent transitions and to set $z_{\rm t}=0.5$, while varying the other parameters. On the other hand, if the transition occurs during DE domination, it significantly impacts matter perturbations. We chose the initial EOS value $w_{\rm i}=-0.4$, which is quite distant from $-1$. This choice amplifies the difference between a late-time and an early-time transition. In fact, for a recent transition, this value represents what the EOS had for most of the Universe’s evolution, making the difference with the $\Lambda$CDM model even more pronounced.

\begin{figure}
    \centering
    \includegraphics[width=1\columnwidth,clip]{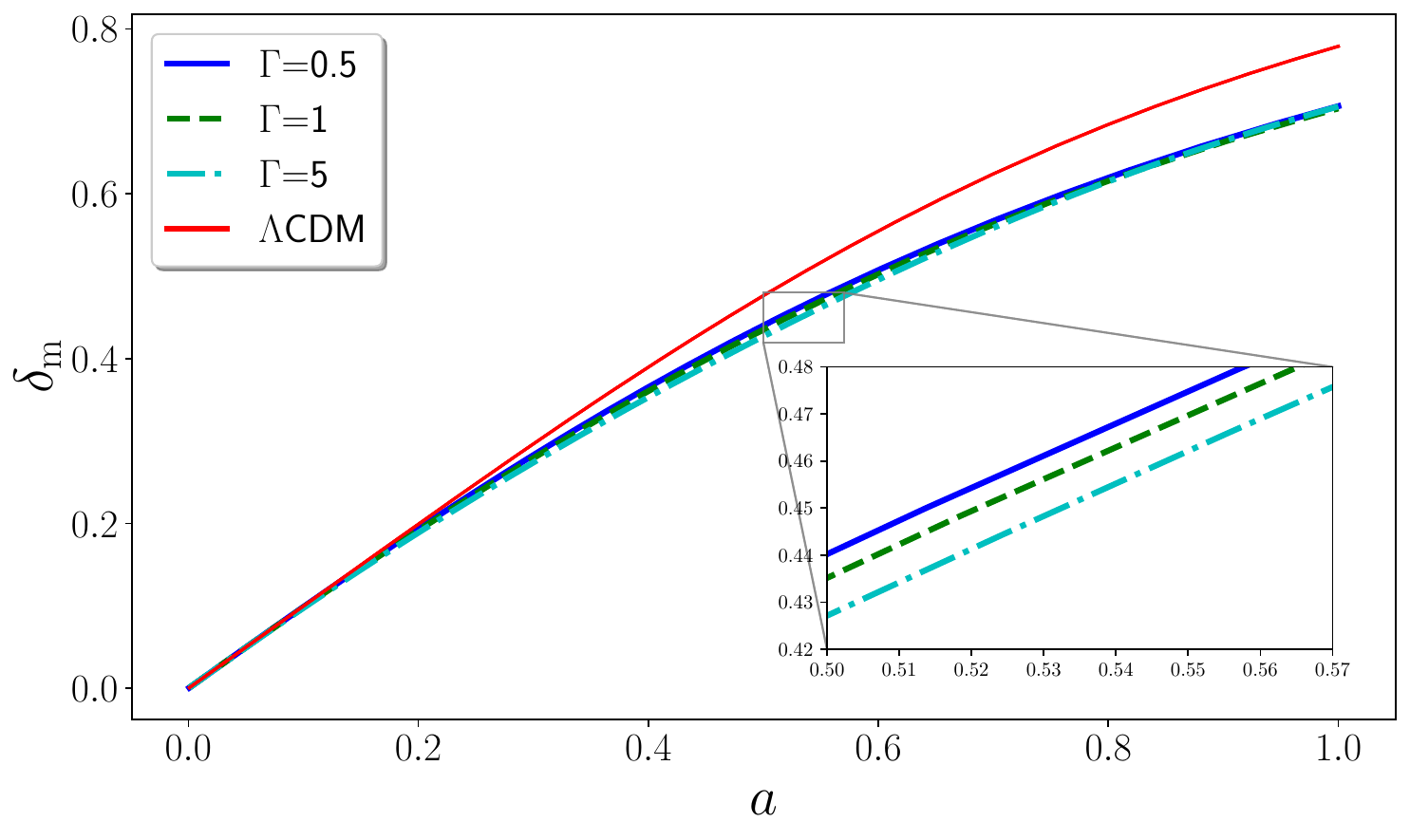}
    \caption{Linear perturbations with $\delta_{\rm{de}}\neq 0$, for model $w_1$ with $w_{\rm i}=-0.4$, $w_{\rm f}=-1$, $z_{\rm t}=0.5$. Despite the extreme variations of the EOS, the curves corresponding to the DE model are almost indistinguishable.}
    \label{fig:pert_1_trans_steepness_2}
\end{figure}

In Fig.~\ref{fig:pert_1_trans_steepness_2}, the results obtained by varying the $\Gamma$ parameter are shown. Even though we have considered a wide range of $\Gamma$ values, leading to significant changes in the EOS, it is evident that the linear perturbations are relatively insensitive to this parameter. Thus, a step transition can approximate a smooth transition quite well at the perturbative level.

\subsection{Nonlinear regime}\label{subsec:nonlinear_regime}

In this section, we look at nonlinear perturbations when DE is described by the models reported in Sec.~\ref{subsec:eos_intro}. In particular, we analyze the behavior of the virialization overdensity and the linearly extrapolated matter density contrast at collapse. The equations to be solved are Eqs.~(\ref{eq:first_oredr_system}) with $c_{\rm{eff}}=0$ for clustering DE and Eq.~(\ref{eq:mperturb}) for smooth models.

\begin{figure}
    \centering
    \includegraphics[width=1\columnwidth,clip]{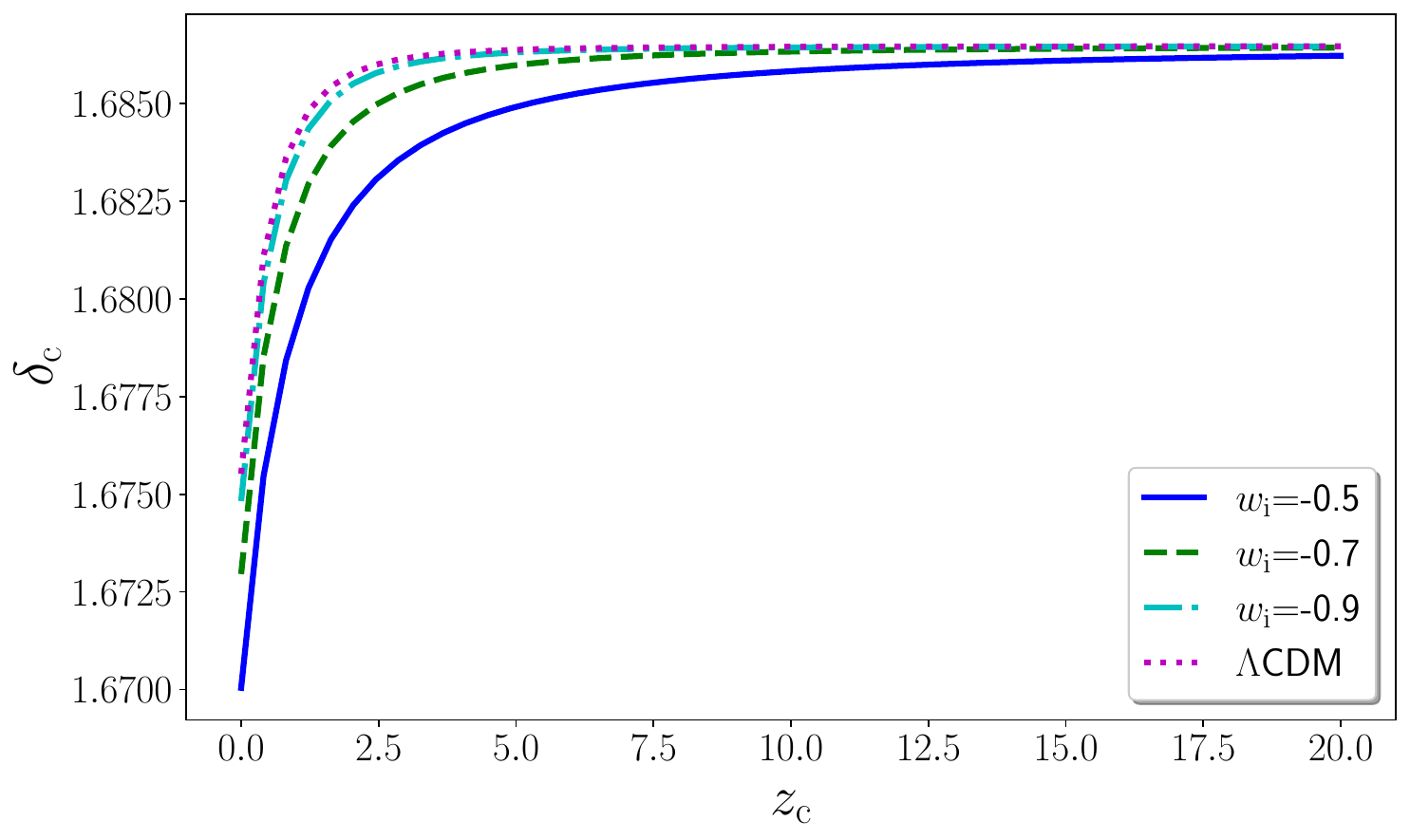}
    \includegraphics[width=1\columnwidth,clip]{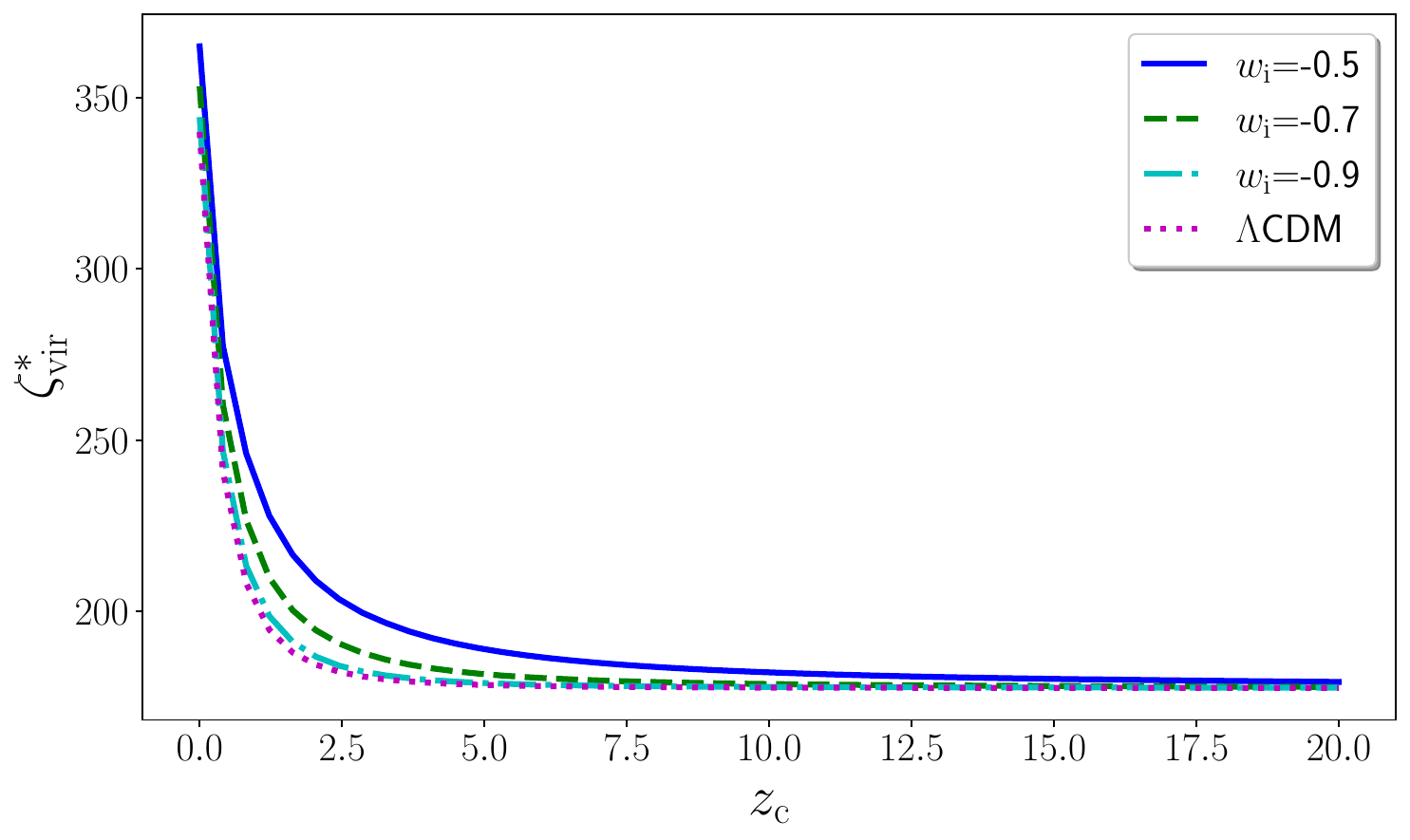}
    \caption{Plot for model $w_1$ with $w_{\rm f}=-1$, $\Gamma=10$, $z_{\rm t}=0.5$ and $\delta_{\rm{de}}=0$. Top panel: $\delta_{\rm c}$ as a function of the collapse redshift $z_{\rm c}$. Bottom panel: $\zeta_{\rm vir}^{*}$ as a function of the collapse redshift $z_{\rm c}$. The $w_{\rm i}=-0.5$ curve has the highest virialization overdensity.}
    \label{fig:nonlin_unpert_2_w_i}
\end{figure}

In Fig.~\ref{fig:nonlin_unpert_2_w_i} we show the results for nonlinear matter perturbations with smooth DE. The curves corresponding to $w_{\rm i}=-0.5$ exhibit significant deviation from the $\Lambda$CDM model. This is because, during the evolution of the Universe, the EOS, $w_{\rm de}$, had for most of the time the value $-0.5$, which is far from the cosmological constant EOS value. As $w_{\rm i}=-0.5$ is relatively close to zero, DE starts to dominate earlier than models with $w_{\rm i}<-0.5$. Thus, the corresponding curves deviate more from the $\Lambda$CDM case, compared to others. The fact that the curve corresponding to $w_{\rm i}=-0.5$ displays the highest value across all redshifts for $\zeta_{\rm vir}^{*}$ can be attributed to the same effect of the Hubble function that was discussed in the linear perturbation analysis (for more details, see Sec.~\ref{subsec:linearRegime}).

\begin{figure}
    \centering
    \includegraphics[width=1\columnwidth,clip]{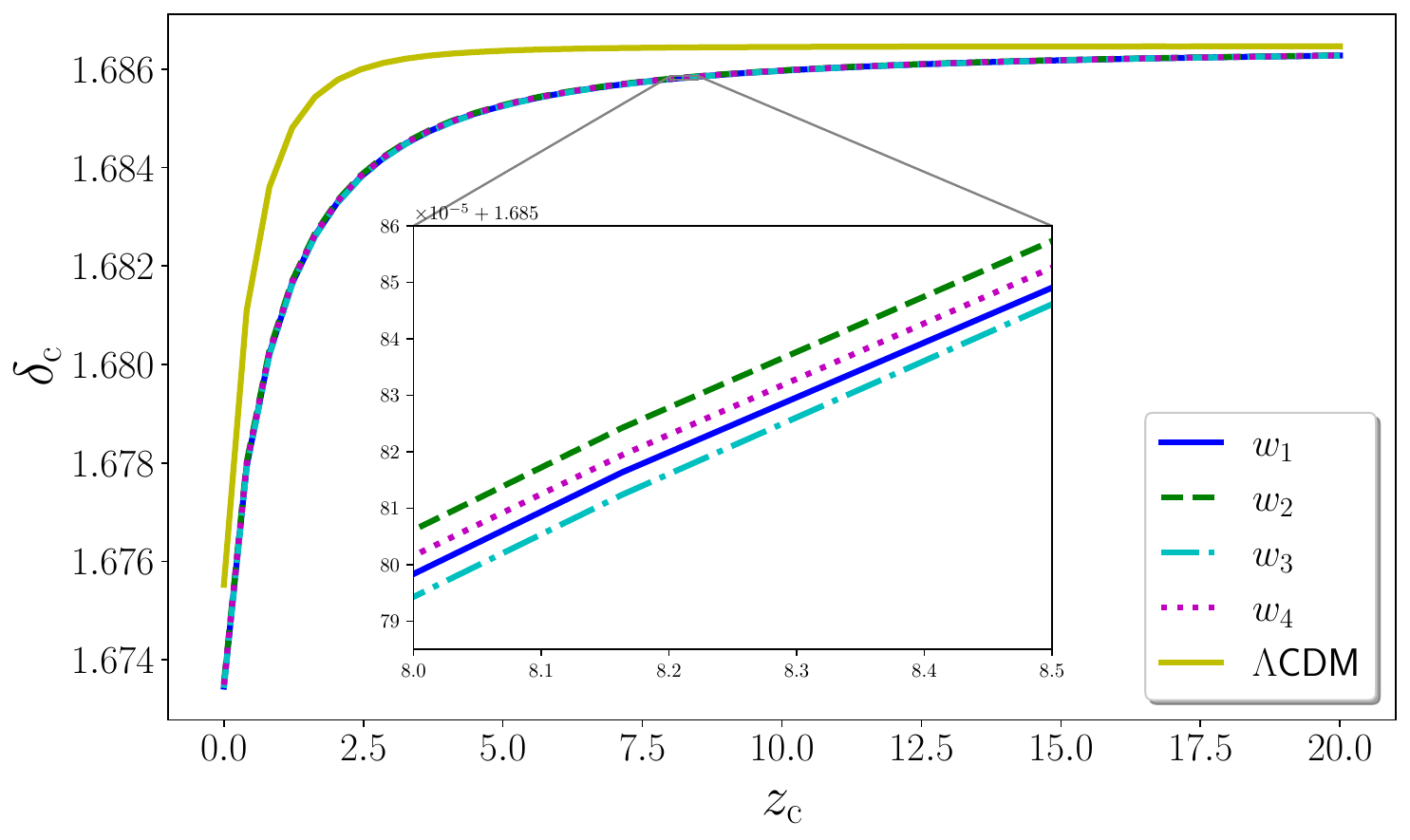}
    \caption{Plot for models $w_1$, $w_2$, $w_3$, $w_4$ respectively, with $w_{\rm i}=-0.5$, $w_{\rm f}=-1$, $\Gamma_1=10$, $\Gamma_2=6.5$, $\Gamma_3=0.07$, $\Gamma_4=0.033$, showing the matter density contrast at collapse. As we can see, the differences between these four models are negligible.}
    \label{fig:nonli_model_diff}
\end{figure}

Since the models defined by Eqs.~(\ref{eq:de_eos_1})--(\ref{eq:de_eos_4}) can be made indistinguishable by appropriately choosing the transition speeds, we do not show results for all of them. The differences between the first four models are shown in Fig.~\ref{fig:nonli_model_diff}. The maximum difference between the corresponding $\delta_{\rm c}$ is less than $0.001\%$, making them completely indistinguishable from each other, but with appreciable differences from the $\Lambda$CDM model.

For additional plots and a more detailed explanation of the observed behaviors, please refer to Appendix \ref{sec:appendix}.

\subsection{Matter power spectrum}\label{subsec:matter_Pk}

Given that our primary objective is to compare our results with simulations and observations, we compute the halo mass function (HMF) for different fast transition models. Understanding the dependency of the HMF on the DE model is crucial, especially in anticipation of precise measurements in upcoming surveys such as \textit{Euclid} \cite{Castro:2016jmw}. The results presented in this section are partially obtained using the {\scriptsize CLASS} code \cite{blas2011cosmic}, which is used to compute the linear matter power spectra, while for the nonlinear matter power spectrum we used the R{\scriptsize E}A{\scriptsize CT} code \cite{Bose:2020wch} based on the halo model reaction \cite{Cataneo:2018cic}. 

In the halo model reaction, the nonlinear matter power spectrum can be evaluated using the following relation:
\begin{equation}
    P_{\rm NL}(k,z) = P_{\rm NL}^{\rm pseudo}(k,z)\,\mathcal{R}(k,z)\,,
\end{equation}
where the pseudo matter power spectrum is such that the nonlinear physics is given in a general relativistic framework and the initial conditions are adjusted to mimic the modified linear clustering and the required redshift. The function $\mathcal{R}(k,z)$ is the reaction term and is the corrected ratio of target-to-pseudo halo model spectra
\begin{equation}
\mathcal{R}(k,z) =  \frac{\{[1-\mathcal{E}(z)]e^{-k/k_\star(z)} + \mathcal{E}(z)\} P_{\rm 2H}(k,z)  +  P_{\rm 1H}(k,z)}{P_{\rm hm}^{\rm pseudo}(k,z)}\,, \label{eq:reaction}
\end{equation}
whose components are
\begin{align}
  P_{\rm hm}^{\rm pseudo}(k,z) = &   P_{\rm 2H} (k,z) + P_{\rm 1H}^{\rm pseudo}(k,z)\,, \label{Pk-halos} \\ 
  \mathcal{E}(z) = & \lim_{k\rightarrow 0} \frac{ P_{\rm 1H}^{\rm }(k,z)}{ P_{\rm 1H}^{\rm pseudo}(k,z)} \,, \label{mathcale} \\ 
   k_{\rm \star}(z) = & - \bar{k} \left\{\ln \left[ 
    \frac{A(\bar{k},z)}{P_{\rm 2H}(\bar{k},z)} - \mathcal{E}(z) \right] - \ln\left[1-\mathcal{E}(z) \right]\right\}^{-1}\,, \label{kstar}
\end{align}
and
\begin{align}
    A(k,z) = & \frac{P_{\rm 1-loop}(k,z)+ P_{\rm 1H}(k,z)}
    {P^{\rm pseudo}_{\text{1-loop}}(k,z)+ P_{\rm 1H}^{\rm pseudo}(k,z)}  P_{\rm hm}^{\rm pseudo}(k,z) \\
    & -  P_{\rm 1H}(k,z)\,. \nonumber
\end{align}
In the expressions above, $P_{\rm L}(k,z)$ is the linear matter power spectrum, while the 1-halo terms with and without corrections to the standard spherical collapse are $P_{1\rm H}(k,z)$ and $P_{1\rm H}^{\rm pseudo}(k,z)$, respectively. Finally, $P_{2\rm H}(k,z)$ is the 2-halo term while $P_{\text{1-loop}}(k,z)$ and $P_{\text{1-loop}}^{\rm pseudo} (k,z)$ are the 1-loop predictions with and without nonlinear modifications to the $\Lambda$CDM model, respectively.

Both codes have been suitably modified to evolve the perturbations for the models considered in this work. Subsequently, we calculate the HMF using the results from Sec.~\ref{subsec:nonlinear_regime}. In this work, we assume a spectral index $n_{\rm s}=0.96$ and use the background parameters provided in Table~\ref{tab:params}. Unless stated otherwise, if $\delta_{\rm{de}}=0$, we also set $c_{\rm{eff}}^2=1$. Conversely, if $\delta_{\rm{de}}\neq0$, we set $c_{\rm{eff}}^2=0$. This means that if DE cannot cluster, we set the effective sound speed to unity in the {\scriptsize CLASS} code\footnote{Formally, the {\scriptsize CLASS} code always evolves a set of equations in which the fluid perturbations are included. However, setting $c_{\rm{eff}}^2=1$ makes sure that perturbations are effectively suppressed and relevant only on the largest scales.}. However, if DE can cluster, we maximize its perturbations by setting the sound speed to zero.
In Table~\ref{tab:sigma8} we present the values of $\sigma_8(z=0)$ for the model described by Eq.~(\ref{eq:de_eos_1}). Allowing DE to cluster consistently increases the $\sigma_8$ value, thereby enhancing matter clustering.

\begin{table}
    \caption{ Present values of $\sigma_8$, for model $w_1$. The superscript $0$ correspond to $\delta_{\rm{de}}=0$, $c^2_{\rm{eff}}=1$, while $1$ to $\delta_{\rm{de}}\neq0$, $c^2_{\rm{eff}}=0$. }
    \label{tab:sigma8}
    \begin{ruledtabular}
        \begin{tabular}{|>{\centering\arraybackslash}m{2cm}|>{\centering\arraybackslash}m{2cm}>{\centering\arraybackslash}m{2cm}>{\centering\arraybackslash}m{2cm}|}
            $w_{\rm i}$  & $-0.5$ & $-0.7$ & $-0.9$ \\
            \hline
            $\sigma^{0}_8$ &$0.71$&	$0.754$&	$0.754$	 \\
            $\sigma^{1}_8$ &$0.729$&	$0.765$&	$0.784$	 \\
            \hline
            \hline
            $w_{\rm f}$  & $-0.5$ & $-0.7$ & $-0.9$ \\
            \hline
            $\sigma^{0}_8$ &$0.636 $&	$0.635$&	$0.663$\\
            $\sigma^{1}_8$ &$0.705$&	$0.711$&	$0.724$	\\	
            \hline
            \hline
            $z_{\rm t}$  & $0.1$ & $0.8$ & $2$ \\
            \hline
            $\sigma^{0}_8$ &$0.626$ &	$0.713$ &	$0.771$\\
            $\sigma^{1}_8$& $0.679$ &	$0.755$ &	$0.788$	\\
            \hline
            \hline
            $\Gamma$  & $0.5$ & $1$ & $5$ \\
            \hline
            $\sigma^{0}_8$ &$0.704$ &	$0.707$ &	$0.677$\\
            $\sigma^{1}_8$& $0.743$ &	$0.743$ &	$0.718$	\\
        \end{tabular}
    \end{ruledtabular}
\end{table}

For freezing models, DE can cluster until the transition occurs, since, according to Eq.~(\ref{eq:first_oredr_system}), only DE with $w\neq -1$ can cluster. This mechanism could result in a decrease of $\sigma_8$ by about $8\%$, without significantly impacting the background evolution, for example the maximum deviation of the age of the Universe is less then $3\%$.

\begin{figure}
   \centering
   \includegraphics[width=1\columnwidth,clip]{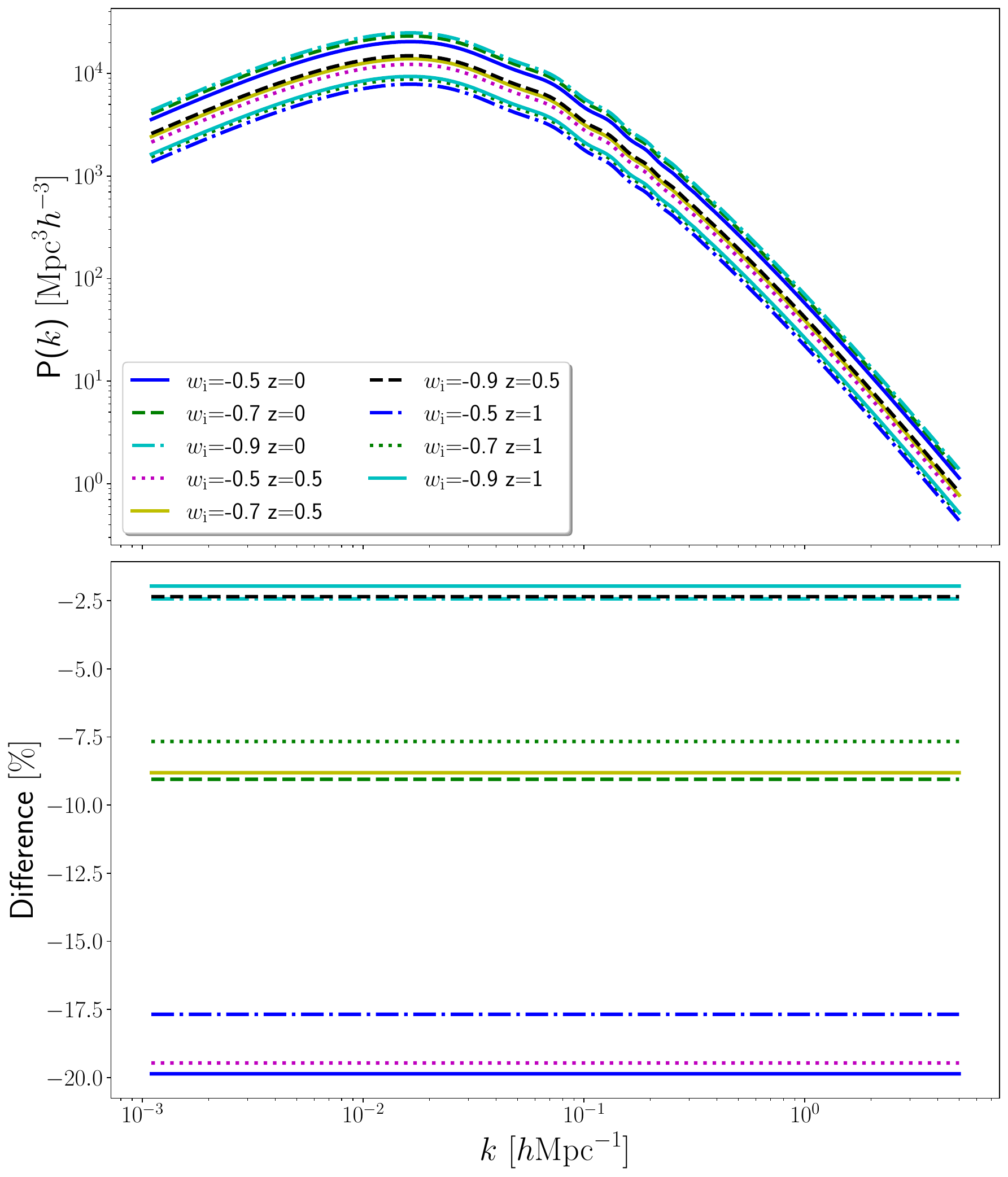}
   \includegraphics[width=1\columnwidth,clip]{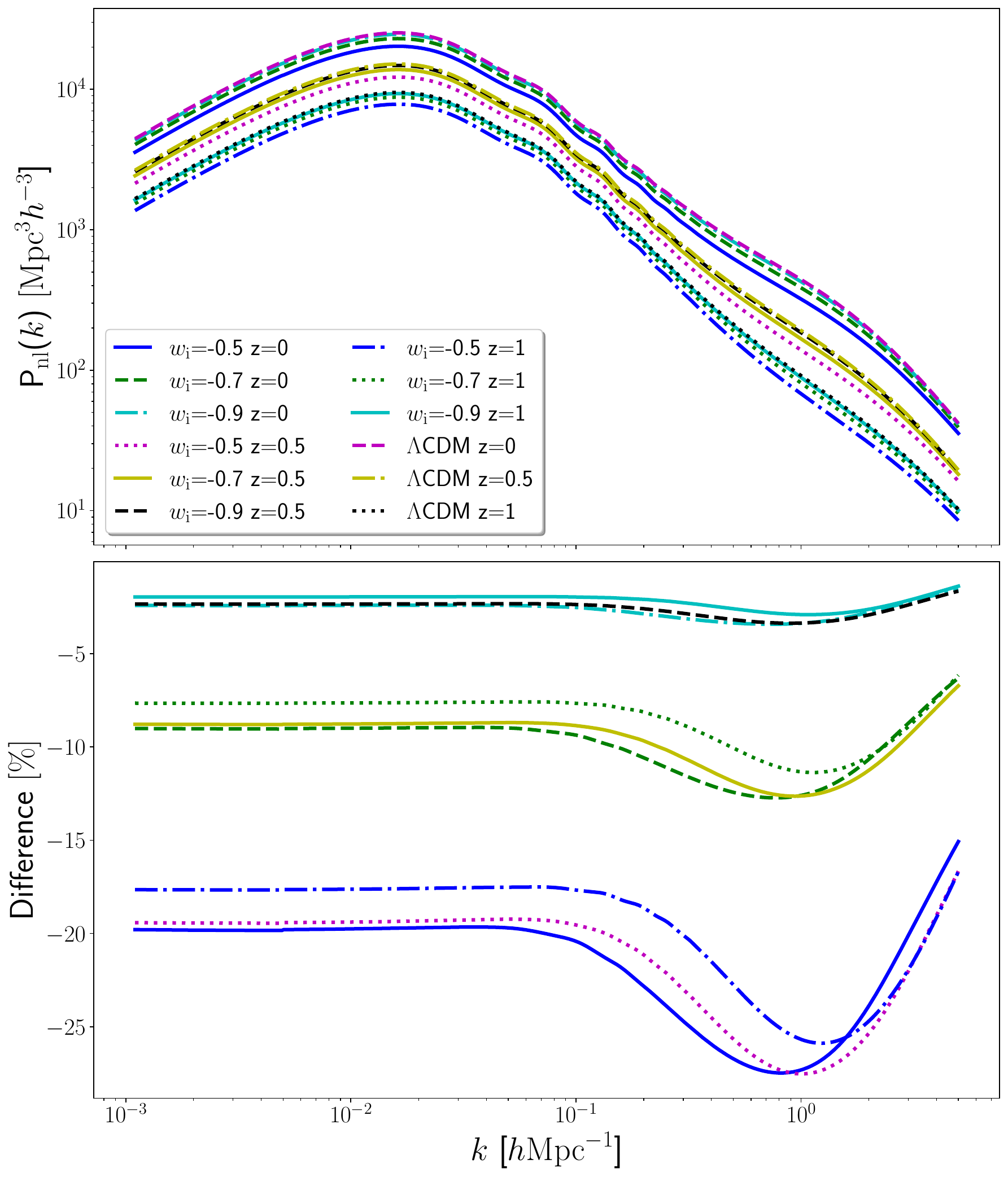}
   \caption{Plots for model $w_1$ with $w_{\rm f}=-1$, $\Gamma=10$, $z_{\rm t}=0.5$, $\delta_{\rm{de}}=0$. \textit{Top panel}: linear matter power spectrum. \textit{Bottom panel}: nonlinear matter power spectrum.}
    \label{fig:pk_w_i_1}
\end{figure}

Figure ~\ref{fig:pk_w_i_1} presents our results about the linear (top panel) and nonlinear (bottom panel) matter power spectra with the corresponding percentage difference with respect to $\Lambda$CDM for the model described by Eq.~(\ref{eq:de_eos_1}) with $\delta_{\rm{de}}=0$.

\begin{figure}
   \centering
   \includegraphics[width=1\columnwidth,clip]{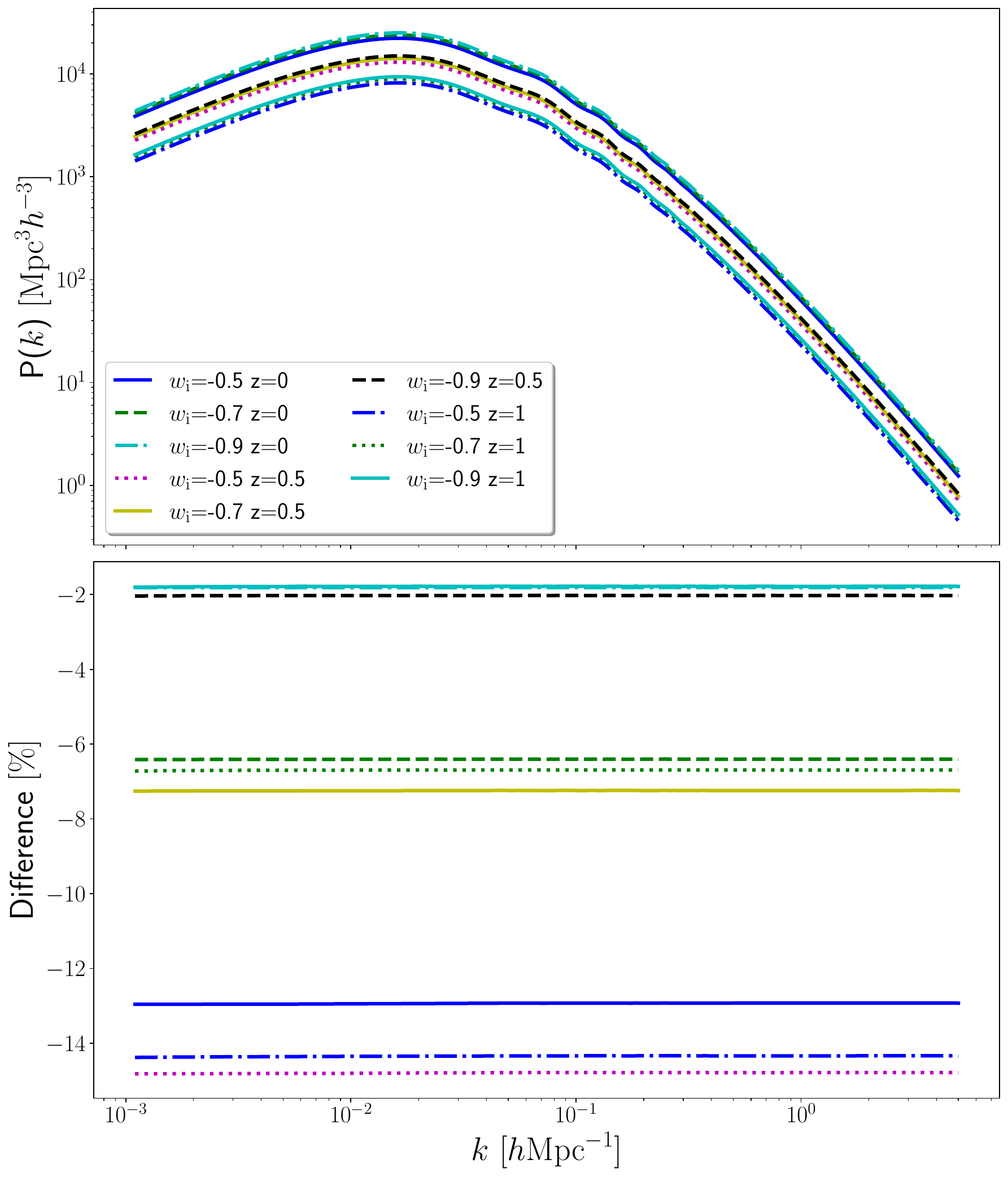}
   \includegraphics[width=1\columnwidth,clip]{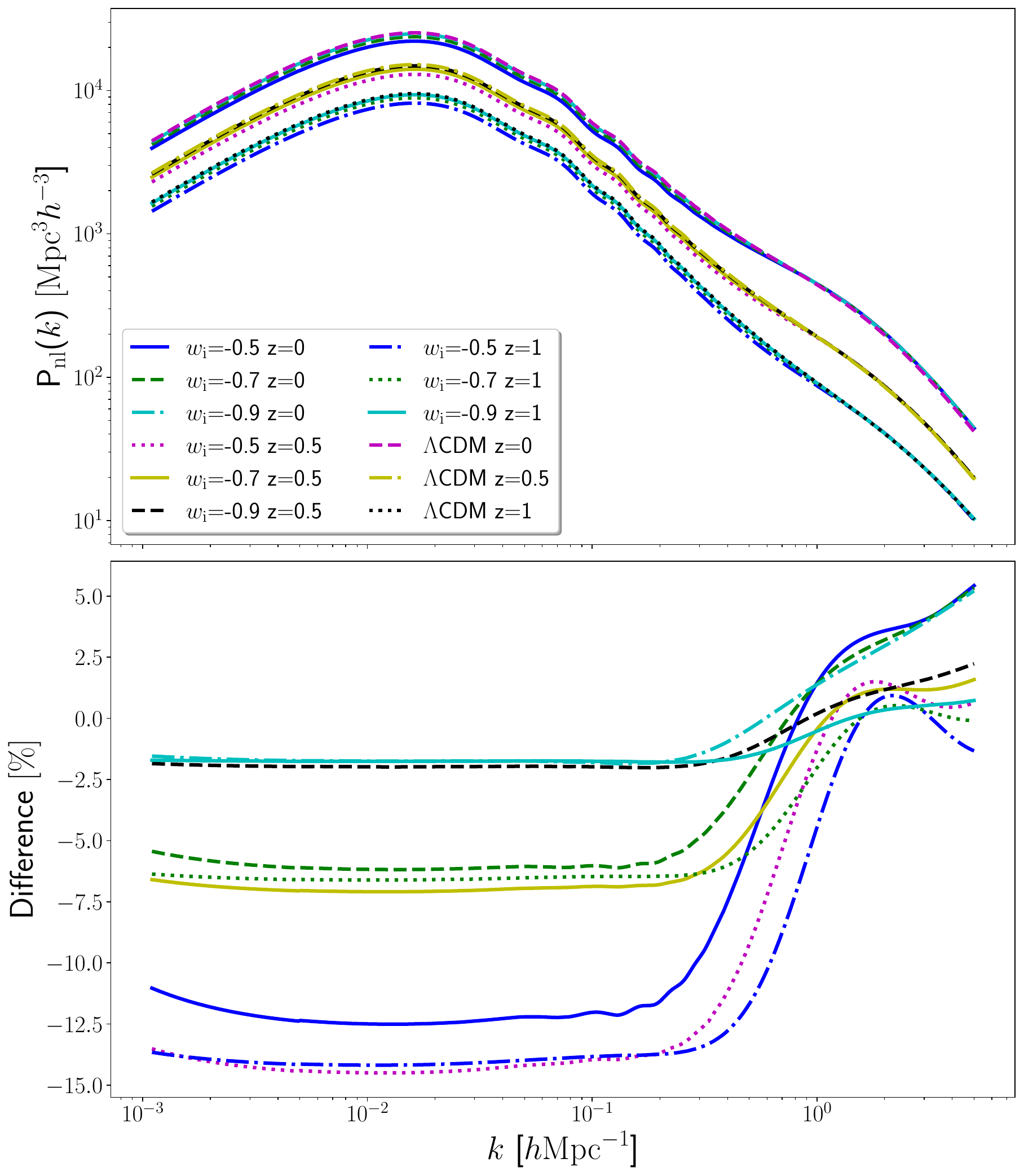}
   \caption{ Plots for model $w_1$ with $w_{\rm f}=-1$, $\Gamma=10$, $z_{\rm t}=0.5$, $\delta_{\rm{de}}\neq0$. \textit{Top panel}: linear matter power spectrum. \textit{Bottom panel}: nonlinear matter power spectrum.}
    \label{fig:pk_w_i_2}
\end{figure}

Note that the positions of the throats in the ratio, shown in the bottom panel, are not due to the transition, as altering $z_{\rm t}$ does not shift them, but are due to the exact scale where the spectrum starts to become nonlinear. By comparing with Fig.~\ref{fig:pk_w_i_2}, which shows similar results for clustering DE, we note that the models behave very similarly on small scales. At larger scales, we see that in the clustering case, peaks replace the throats observed in the smooth case. We also compared our results with those obtained by deriving the nonlinear matter power spectrum using the halo model approach. We found qualitatively similar results, with differences of the order of $5\%$ between the two approaches. In particular, with the R{\scriptsize E}A{\scriptsize CT} code, the nonlinear effects are stronger.

We notice that, however, to understand the exact level of accuracy of our predictions, we need to rely on full $N$-body simulations, which solve the full equations consistently. These simulations would then permit a direct comparison with our results and allow us to establish the exact scales to which we can trust the R{\scriptsize E}A{\scriptsize CT} code or any other formalism used. For clustering dark energy, an available framework is provided by the \texttt{\textit{k}-evolution} code \citep{Farbod2019,Farbod2020} which also evolves the equations of clustering dark energy. The authors showed that for large values of the sound speed, nonlinear effects of the dark energy component are either negligible or very small and results from the {\scriptsize CLASS} code are very accurate, while this is not the case for very small values of the sound speed. The \texttt{\textit{k}-evolution} code solves the equations assuming that the dark energy component is represented by a scalar field, while in this work we assumed it to be represented by a fluid. Hence, the formulation proposed by \cite{Blot2023}, which is based on the integration of the continuity and Euler equations is closer to what we have done here, even though, in their formulation, additional terms are considered. It is, therefore, important, to compare the different codes, establish their level of agreement and then improve the accuracy of our results.

\subsection{The halo mass function}\label{subsec:hmf}

One of the most useful quantities that can be computed in order to test a model is the HMF. This function gives the number of virialized objects with mass $M\in[M,M+\mathrm{d}M]$ per unitary comoving volume, at a given redshift. Calling $n$ the number density of virialized object, the mass function is $\mathrm{d}n/\mathrm{d}M$. 

The most consolidate treatment to obtain an analytical expression for the mass function is the Press and Schechter (PS) HMF \cite{1974ApJ...187..425P}. To obtain the corresponding expression, we assume that the density field is Gaussian and if $\delta_M$ is the density field filtered on a scale $M=4/3\pi R^3 \Bar{\rho}_{\rm m,0}$, then we have
\begin{equation}\label{eq:gauss_delta}
    p(\delta_M)=\frac{1}{\sqrt{2\pi}\sigma(M)}e^{-\frac{\delta_M^2}{2\sigma^2(M)}}\,,
\end{equation}
where $\sigma(M)$ is the variance of the filtered contrast. Then the probability that the density contrast is larger than a critical value $\delta_{\rm c}$ can be computed as
\begin{equation}
    P(\delta_M>\delta_{\rm c})=\int_{\delta_{\rm c}}^{\infty}p(\delta_M)\mathrm{d}\delta_M=\mathrm{Erf}\left(\frac{\delta_M}{\sqrt{2}\sigma(M)}\right)^{\infty}_{\delta_{\rm c}}\,,
\end{equation}
with $\mathrm{Erf}$ the complementary error function. This quantity is approximately proportional to the number of structures with masses larger than $M$. Thus, we have that 
\begin{equation}
    2\Bar{\rho}_{\rm m,0}\frac{\mathrm{d}P(M)}{\mathrm{d}M} = M\frac{\mathrm{d}n}{\mathrm{d}M}\,,
\end{equation}
yielding the standard formula
\begin{equation}\label{eq:press_schechter_mf}
    \frac{\mathrm{d}n}{\mathrm{d}\ln{M}}=\frac{\Bar{\rho}_{\rm m,0}}{M}f(\nu)\left|\frac{\mathrm{d\ln{\sigma}}}{\mathrm{d}\ln{M}}\right |\,,
\end{equation}
where $\Bar{\rho}_{\rm m,0}=\Bar{\rho}_{\rm m}(a=1)$. The multiplicity function is $f(\nu)=\sqrt{\frac{2}{\pi}}\nu e^{-\nu^2/2}$, with $\nu=\delta_{\rm c}/\sigma$. Usually, the critical value $\delta_{\rm c}$ is taken to be the linearly extrapolated density contrast at collapse as computed in the spherical collapse model, since this value should correspond to fully collapsed objects. The redshift dependence of the HMF comes from both $\delta_{\rm c}(z)$ and the fact that $\sigma(M,z)=D_+(z)\sigma(M)$ where $D_+(z)$ is the growth factor.

The Sheth–Tormen (ST) \cite{Sheth:1999mn,Sheth:1999su} HMF extends the PS formalism by assuming that halos are ellipsoidal and not perfectly spherical. The result is that the function $f$ of Eq.~(\ref{eq:press_schechter_mf}) changes to
\begin{equation}
 f(\nu')=\sqrt{\frac{2}{\pi}}A\left[1+\frac{1}{\left(\nu'\right)^{2q}}\right]\nu' e^{-\left(\nu'\right)^{2}/2}\,,
\end{equation}
with $\nu'=\sqrt{a}\nu$ and fiducial values for the parameters  $q=0.3$, $A= 0.322$, $a= 0.707$, coming from fits of numerical simulations data.

\begin{figure}
\includegraphics[width=1\columnwidth,clip]{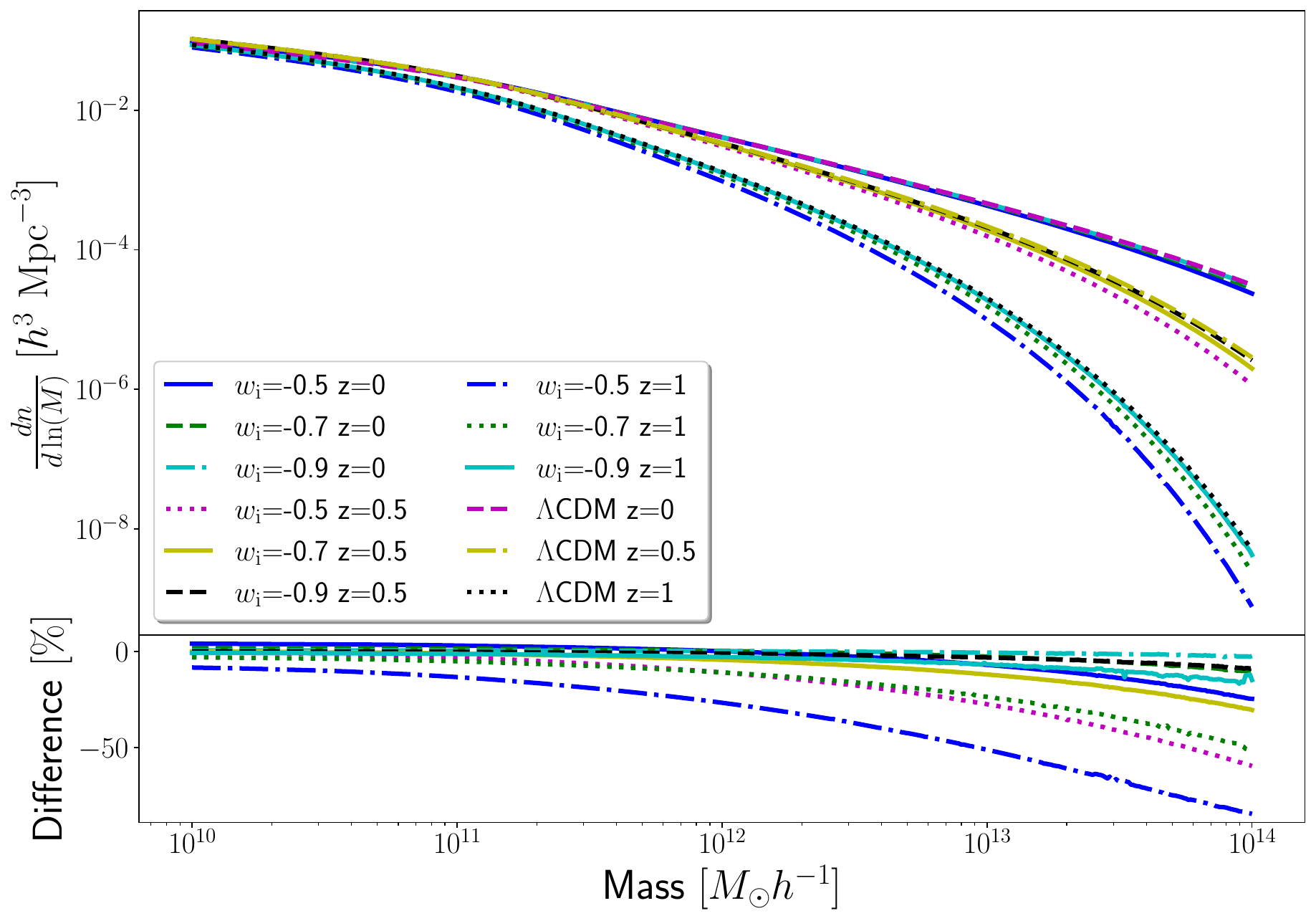}
 \caption{Plots for model $w_1$ with $w_{\rm f}=-1$, $\Gamma=10$, $z_{\rm t}=0.5$, $\delta_{\rm{de}}=0$. Top panel: ST mass function for various values of $w_{\rm i}$ and redshift. Bottom panel: percentage difference with respect to the $\Lambda$CDM model defined through the relation $\Lambda$CDM $(1+\%)$.}
  \label{fig:mf_w_i_1}
\end{figure}

In Fig.~\ref{fig:mf_w_i_1}, we present the Sheth-Tormen (ST) HMF for smooth DE. For a galaxy of average size, such as the Milky Way, which has a mass of approximately $7\times 10^{11} M_{\odot}h^{-1}$, we predict the deviations outlined in Table~\ref{tab:mf_w_i_deviations}. As anticipated from the analysis in Sec.~\ref{subsec:nonlinear_regime}, we expect the difference with respect to the $\Lambda$CDM model to peak at $z\approx 0.8$. The deviations reported in Table~\ref{tab:mf_w_i_deviations}, even if the considered redshift range is too narrow to draw a definitive conclusion, suggest this behavior as well for the HMF.

\begin{table}
    \caption{ Percentage deviations of the ST HMF from the $\Lambda$CDM model, for an average size ($7\cdot 10^{11} M_{\odot}h^{-1}$) galaxy, as extracted from Fig.~\ref{fig:mf_w_i_1}. }
    \label{tab:mf_w_i_deviations}
    \begin{ruledtabular}
        \begin{tabular}{|>{\centering\arraybackslash}m{2cm}|>{\centering\arraybackslash}m{2cm}>{\centering\arraybackslash}m{2cm}>{\centering\arraybackslash}m{2cm}|}
             $\rm Redshift$ & $w_{\rm i}=-0.5$ & $w_{\rm i}=-0.7$ & $w_{\rm i}=-0.9$ \\
            \hline
            $\,z=0$ & $ +0.57\%$ & $+0.39\%$ & $+0.14\%$ \\
            \hline
            $\quad z=0.5$ & $-9.7\%$ & $ -3.8\%$ & $-1\%$ \\
            \hline
            $\,z=1$ & $-24\%$ & $-10\%$ & $-2.7\%$ \\
        \end{tabular}
    \end{ruledtabular}
\end{table}

\begin{figure}
\centering
\includegraphics[width=1\columnwidth,clip]{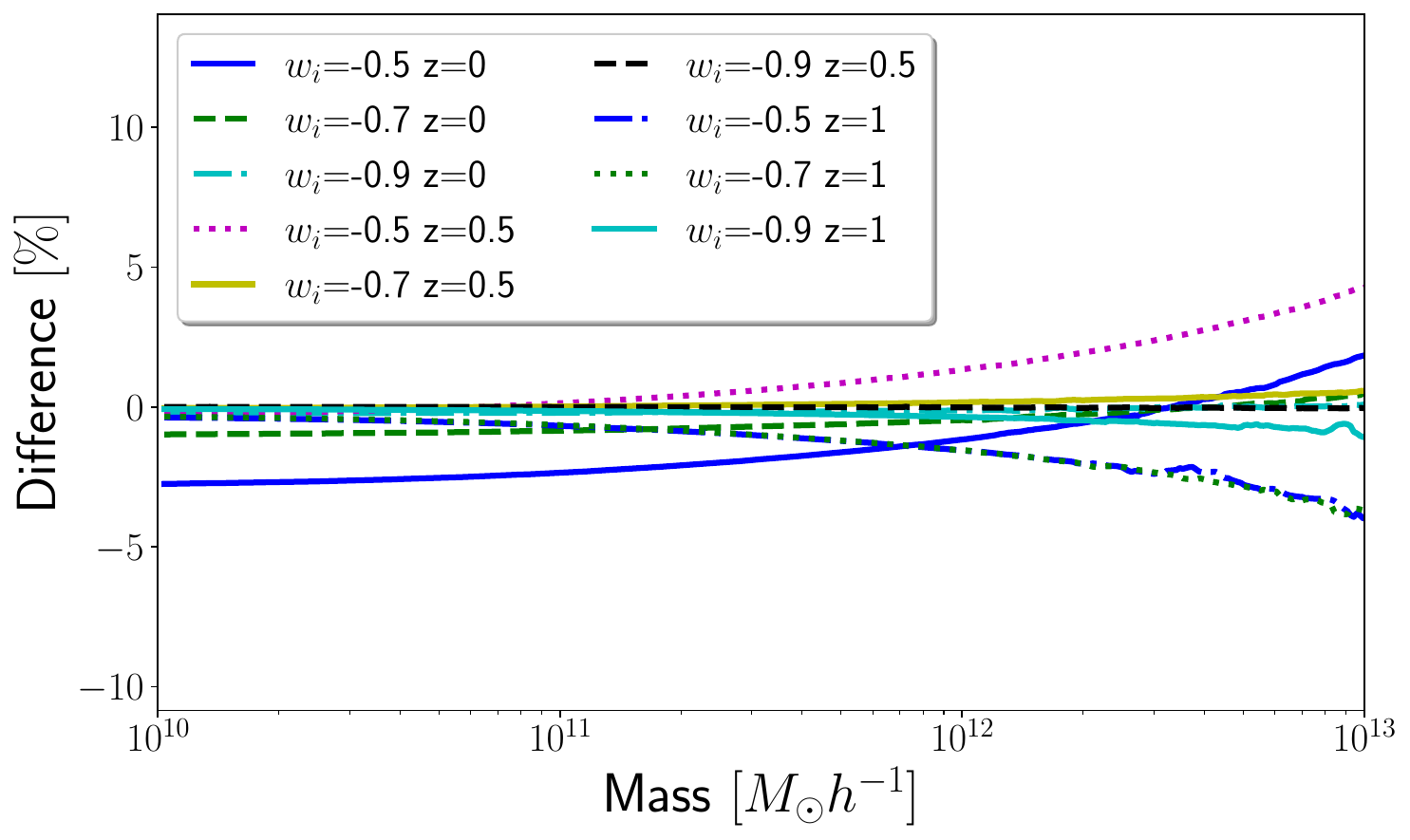}
\includegraphics[width=1\columnwidth,clip]{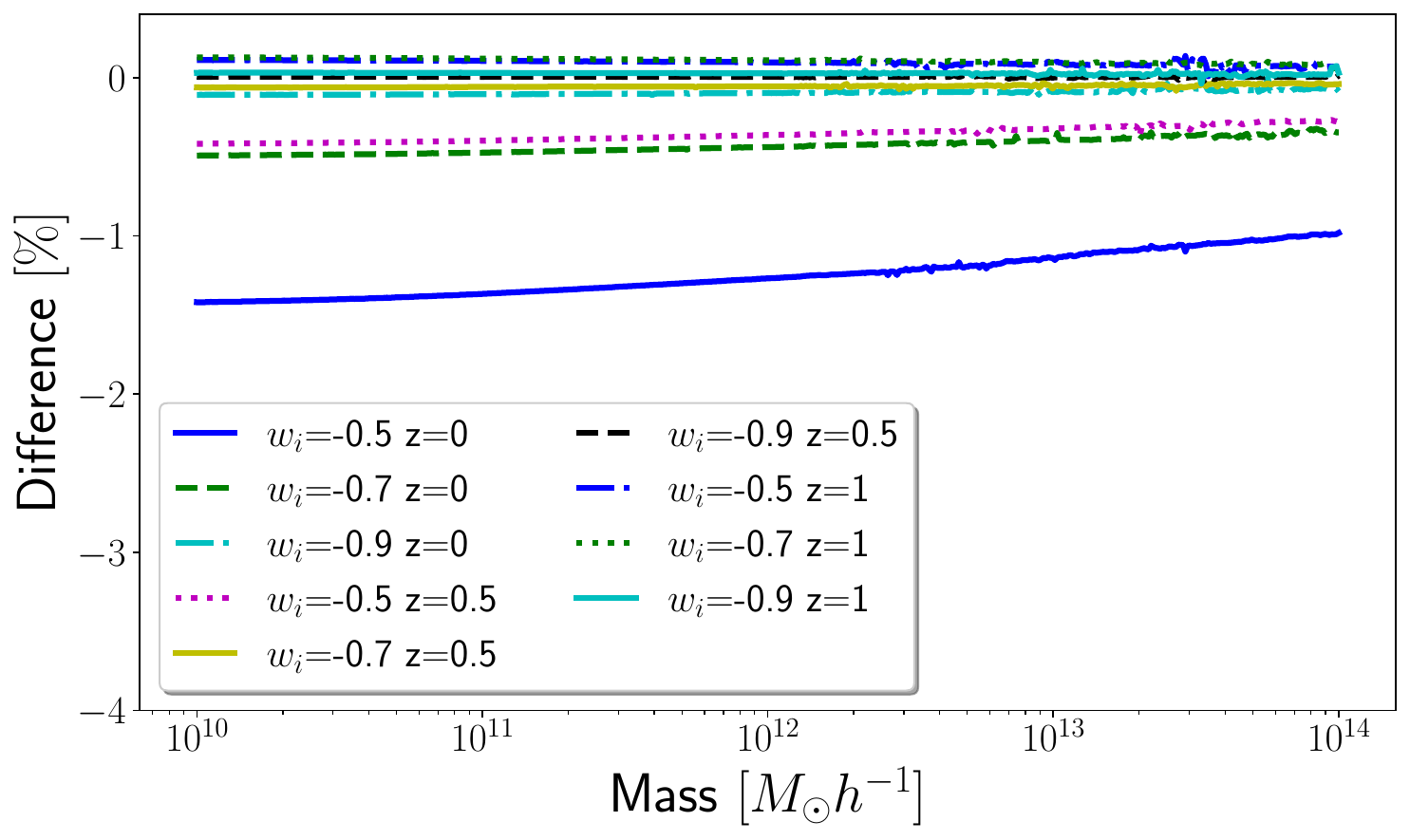}
 \caption{Plots for model $w_1$ with $w_{\rm f}=-1$, $\Gamma=10$, $z_{\rm t}=0.5$. Top panel: the plot shows $[{\rm PS}(c_{\rm{s}}=0)/{\rm PS}(c_{\rm{s}}=1)-1]\times 100$, where the PS functions ratio is used to scale the ST mass function as prescribed by Eq.~(\ref{eq:ps_scaling}). Bottom panel: percentage difference between the scaled ST mass function and the ST mass function with $c_{\rm{s}}=0$.}
  \label{fig:ps_scaling_w_i_1}
\end{figure}

The ST formalism offers a possible description of the HMF in clustering DE models. A prescription by \cite{Creminelli:2009mu} implies a \emph{rescaling} of the ST mass function via the PS mass function \cite{1974ApJ...187..425P} according to the following recipe
\begin{equation}\label{eq:ps_scaling}
    \frac{\mathrm{d}n}{\mathrm{d}\ln{M}} = \frac{\mathrm{d}n_{\rm{ST}}(c_{\rm eff}^2=1)}{\mathrm{d}\ln{M}} \frac{\mathrm{d}n_{\rm{PS}}(c_{\rm eff}^2=0)/\mathrm{d}\ln{M}}{\mathrm{d}n_{\rm{PS}}(c_{\rm eff}^2=1)/\mathrm{d}\ln{M}}\,.
\end{equation}
The main idea is that the errors in the PS approach will partially cancel in doing the ratio. As we can see in the top panel of Fig.~\ref{fig:ps_scaling_w_i_1}, the effect is greater at higher masses where the recipe tells us that we can expect deviations of about $5\%$ around $M\approx 10^{13}\,M_{\odot}\,h^{-1}$. In the bottom panel of Fig.~\ref{fig:ps_scaling_w_i_1}, we show the percentage difference between the scaled mass function computed through Eq.~(\ref{eq:ps_scaling}) and the ST mass function with clustering DE. In this case the deviations are around $1\%$ and almost constant across the mass range. This confirms that taking the ratio of the PS HMFs partially removes its shortcomings. The top panel in Fig.~\ref{fig:ps_scaling_w_i_1} also shows the difference between clustering and smooth DE on halo formation. With the exception of models with $w_{\rm i}=-0.5$, the difference in the number density of galaxies with average mass, between the clustering and smooth cases, is of the order of $1\%$ which is not particularly significant. Only at higher masses and for higher initial EOS values the effect becomes more significant.

\begin{figure}
\centering
\includegraphics[width=1\columnwidth,clip]{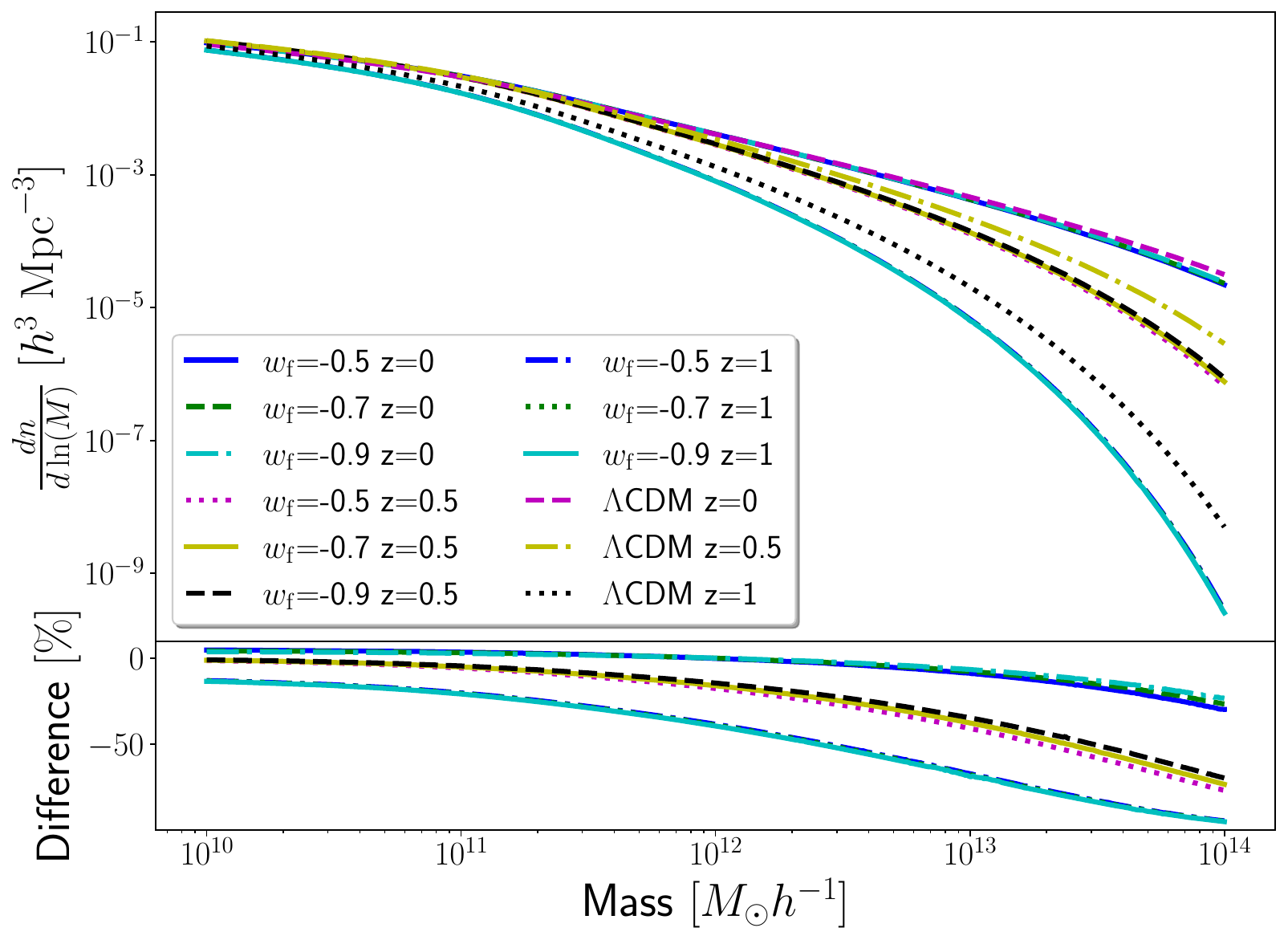}
 \caption{Plots for model $w_1$ with $w_{\rm i}=-0.4$, $\Gamma=10$, $z_{\rm t}=0.5$, $\delta_{\rm{de}}\neq0$. Top panel: ST mass function for various values of $w_{\rm f}$ and redshift. Bottom panel: percentage difference with respect to the $\Lambda$CDM model defined through the relation $\Lambda$CDM $(1+\%)$.}
  \label{fig:mf_w_f_1}
\end{figure}

In Fig.~\ref{fig:mf_w_f_1}, we illustrate the HMF, taking into account variations in the final value of the EOS. Note that deviations are expected to increase with redshift up to $z\approx 0.8$. The curves are grouped according to their corresponding redshifts. Despite the high sensitivity of the virialization overdensity to $w_{\rm f}$ variations, the HMF's sensitivity primarily hinges on redshift rather than the specific value of $w_{\rm f}$. This is anticipated since the HMF is a function of the growth factor and our analysis showed that it is much less sensitive to $w_{\rm f}$ with respect to the nonlinear regime. At $z=1$, the curves become indistinguishable due to $z=1$ being prior to both matter-DE equality and the DE transition. Therefore, the deviation between curves for clusters with the same redshift of the standard model can be attributed to the initial value of the EOS.

\begin{figure}
\centering
\includegraphics[width=1\columnwidth,clip]{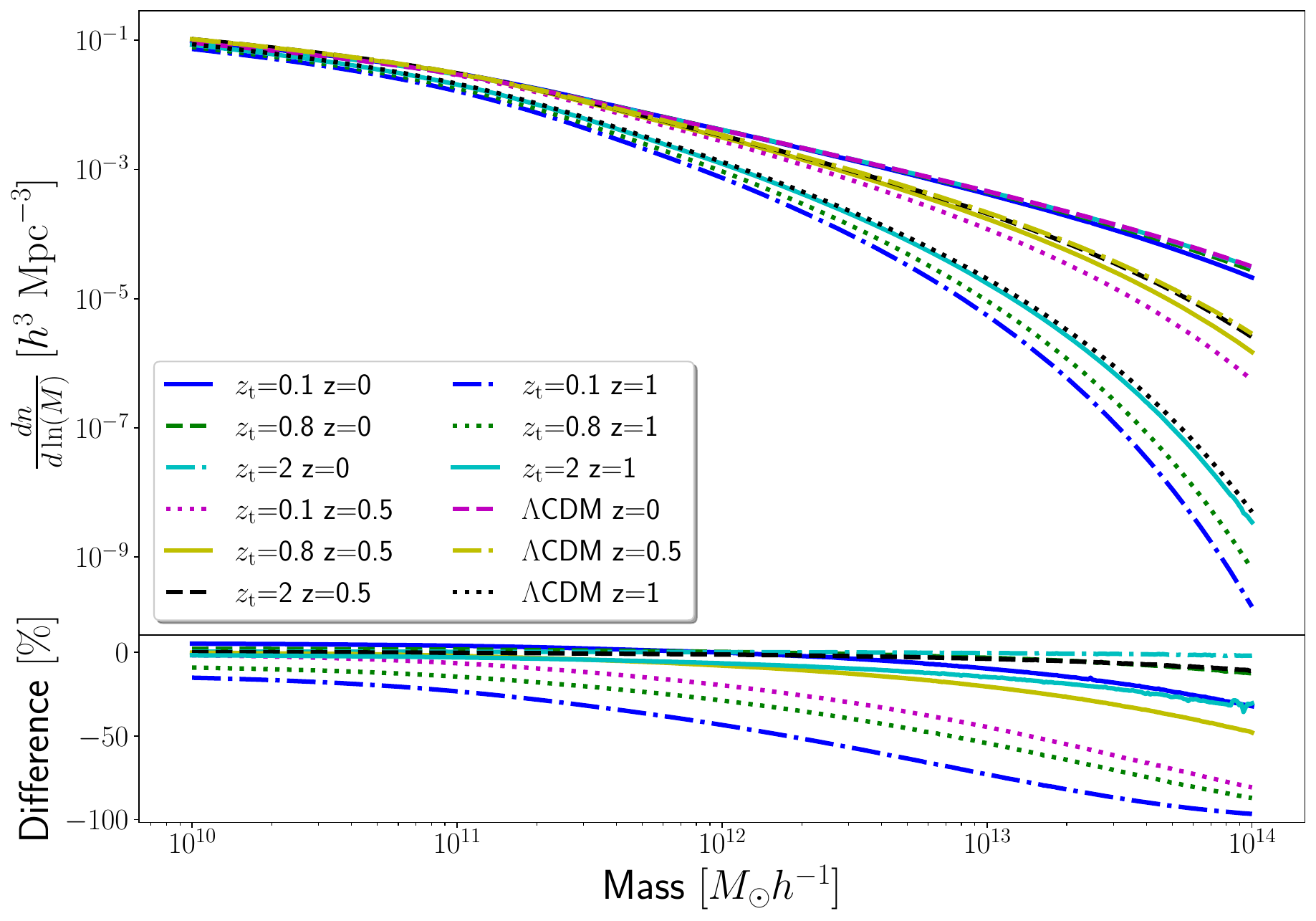}
 \caption{Model $w_1$ with $w_{\rm i}=-0.4$, $w_{\rm f}=-1$, $\Gamma=10$, $\delta_{\rm{de}}\neq0$ for several values of $z_{\rm t}$.}
  \label{fig:mf_z_t_1}
\end{figure}

In Fig.~\ref{fig:mf_z_t_1}, we display the HMF, adjusting the transition redshift parameter. In this scenario, we anticipate the differences to reach their peak at $z\approx 2$. If the transition is quite recent (as in the case of $z_{\rm t}=0.1$), the model behaves similarly to a $w$CDM with $w=w_{\rm i}=-0.4$, resulting in the maximum difference compared to the $\Lambda$CDM paradigm. However, if the transition to $w_{\rm f}=-1$ occurs before matter-DE equality, the model behaves like the standard one at $z=0$, but shows deviations at higher redshifts.

\begin{figure}
\includegraphics[width=1\columnwidth,clip]{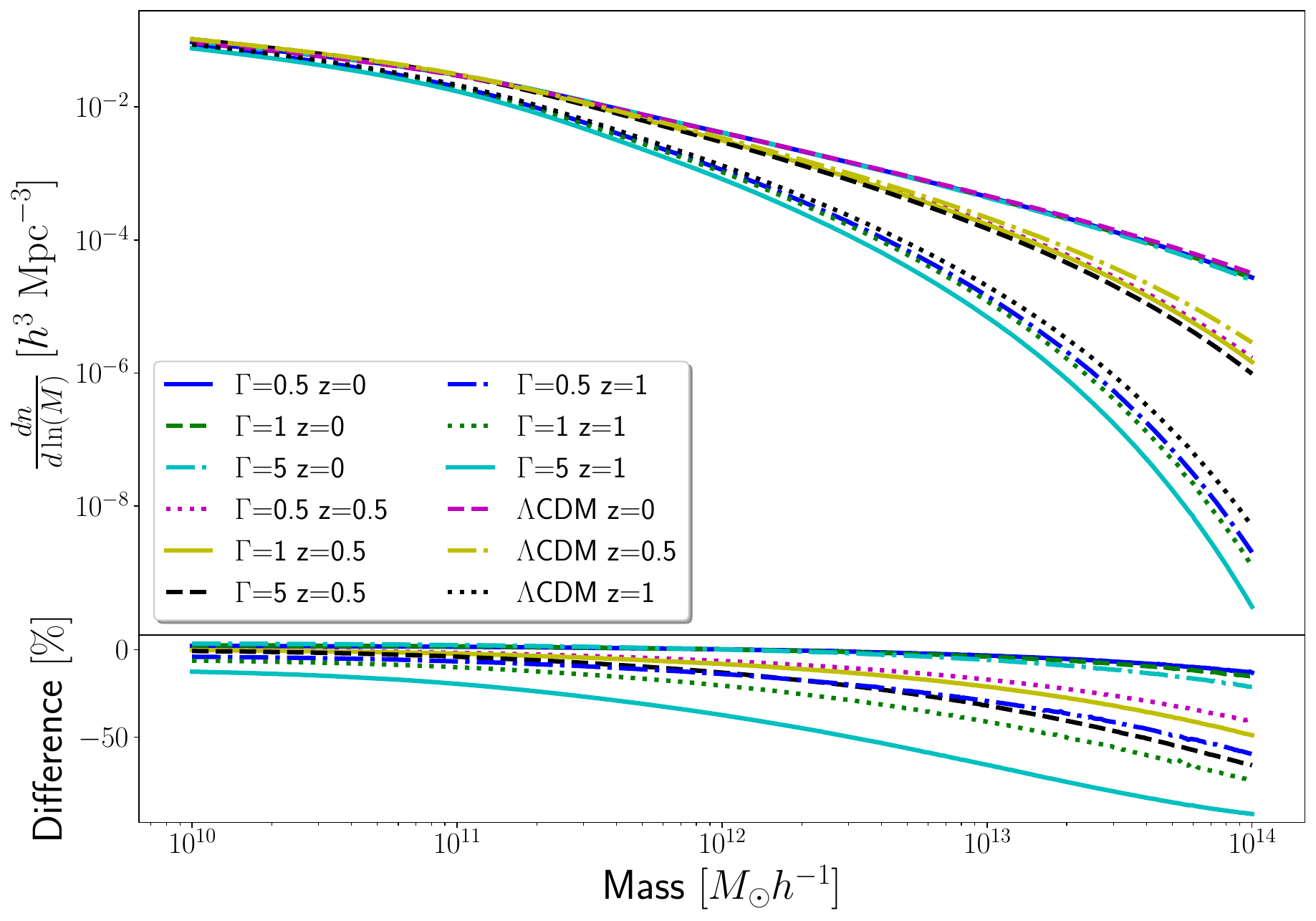}
 \caption{Plot for model $w_1$ with $w_{\rm i}=-0.4$, $w_{\rm f}=-1$, $z_{\rm t}=0.5$, $\delta_{\rm{de}}\neq0$ for various values of $\Gamma$ and redshift.}
  \label{fig:mf_gamma_1}
\end{figure}

In Fig.~\ref{fig:mf_gamma_1} the results for the HMF when the transition speed is varied are shown. The effects of the transition speed on galaxy formation are quite low at $z=0$: the differences between the curves are $\approx 0.1\%$ for an average mass galaxy. The differences become more prominent at higher redshifts: curves with $z=1$ can differ by $22\%$, while at $z=0.5$ can differ by $6\%$. Those differences at high redshifts are amplified by the choice $w_{\rm i}=-0.4$. An initial value closer to $-1$ dampens those differences. Despite that, only measurements at $z\gtrapprox 0.5$ could possibly detect the transition speed effects. 

\begin{figure}
\centering
\includegraphics[width=1\columnwidth,clip]{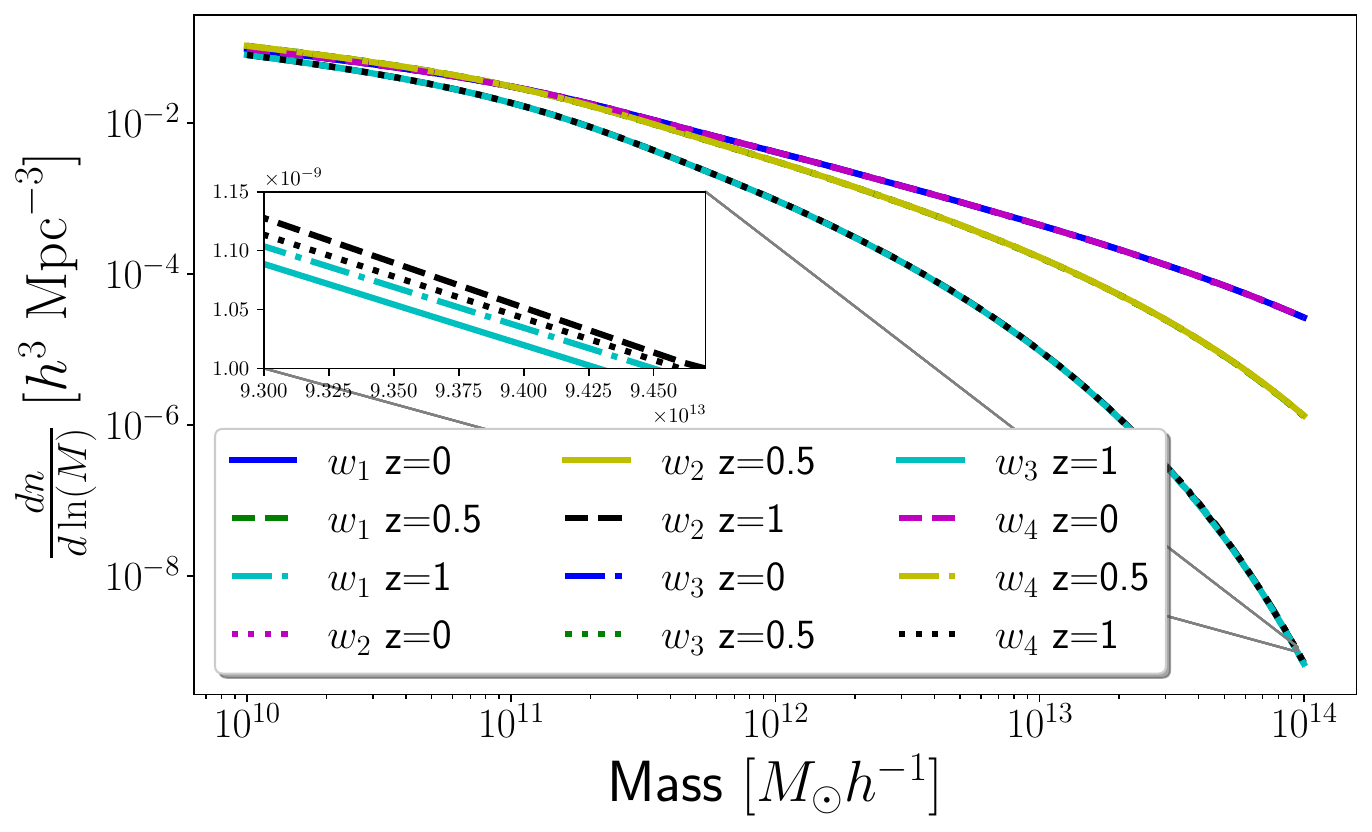}
 \caption{Comparison of the HMFs for model $w_1$, $w_2$, $w_3$, $w_4$ with $w_{\rm i}=-0.5$, $w_{\rm f}=-1$, $z_{\rm t}=0.5$, $\Gamma_1=10$, $\Gamma_2=6.5$, $\Gamma_3=0.07$, $\Gamma_4=0.033$. The models are indistinguishable.}
  \label{fig:hmf_comparison}
\end{figure}

Regarding the other models, they bear a strong resemblance to the model of Eq.~(\ref{eq:de_eos_1}). The main differences lie in the exact shape of the transition. As we have seen here and in previous sections, structure formation is quite insensitive to the exact form of the transition, thus reporting them would not add more insights on the physics. We just show in Fig.~\ref{fig:hmf_comparison} the HMF for the first four models, with fixed parameter values and different redshifts. The maximum deviations at redshift $z=1$ are $<5\%$ at high masses and $<0.5\%$ for an average mass galaxy.

\section{Outlooks and perspectives}\label{sec:discussion}

In this study, we analysed six fast transition DE models. Five of them are argued as effective quintessence fluids while all lie within the category of effective Horndeski fluids, i.e, these scenarios originally emerged from exploring the Horndeski parameter space. For those frameworks allowing a quintessence interpretation, we computed the quintessence field and potential. Our results are summarized below.

\begin{itemize}
    \item[-] The field potential exhibits a weak dependence on both the final value and the transition speed of the EOS. Notably, the dependence on the transition speed is particularly weak and we demonstrated that a Heaviside step transition can serve as a good approximation for fast transition models, with the benefit of having analytical $g(a)$ and $V(a)$.
\end{itemize}

We computed the effects of such models on the expansion history of the Universe and on structure formation. The performed analysis brought to light the most important aspects of the fast transition models. 
In the simplest analysis of the matter linear perturbations we found results consistent with the scalar field analysis.

\begin{itemize}
    \item[-] Linear matter perturbations are quite insensitive to the transition speed, suggesting that a Heaviside step transition can approximate well fast transition models also at perturbative level.
    \item[-] For all the nonphantom models we found that they result in lower density contrast respect to the standard model. We showed that this can be understood with the background evolution, by comparing the Hubble functions of the different models with the $\Lambda$CDM one.
    \item[-] Linear perturbations are insensitive to the redshift at which the transition happens ($z_{\rm t}$) if $z_{\rm t}\gtrapprox 2$. This is due to the fact that matter-DE equality happens between\footnote{The bounds are computed using the $w$CDM model with $w = -0.4$ and $w = -1$ for the first and second bounds, respectively.} $z=1$ and $z=0.3$. If the transition happens before matter-DE equality, when DE is sub-dominant, DE effects are not evident. On the other hand if the transition happens within DE domination, the effects are more pronounced.
    \item[-] We compared the case of $\delta_{\rm{de}}=0$ to $\delta_{\rm{de}}\neq0$ and found that when DE perturbations are included, the overall behavior remains qualitatively similar. The presence of DE perturbations results in damping of matter perturbations due to the interactions between the two fluids. If  DE has $w_{\rm de}<-1/3$, it tends to dilute matter perturbations, whereas if $w_{\rm de}>-1/3$, it tends to amplify them.
    \item[-] The models $w_1$, $w_2$, $w_3$ and $w_4$ are indistinguishable one from the other, because the maximum difference between them is less then $1\%$. The $w_6$ model, being phantom, results in perturbations higher than the standard model.
\end{itemize}

We also analysed matter perturbations in the nonlinear regime. We computed the virialization overdensity, $\zeta^*_{\rm{vir}}$, and the linearly extrapolated matter density contrast at collapse, $\delta_{\rm c}$. The latter quantity is very significant as we used it to compute the HMF. 

\begin{itemize}
    \item[-] We understood in a simple way the qualitative behavior of the virialization overdensity and its main features, such as the presence of a peak in the percentage difference with respect to $\Lambda$CDM and its position. In fact, we explained why the ratio between the Hubble function for a $w$CDM Universe to the standard model one, shows the same qualitative behavior as the ratio of the virialization overdensities. We analytically computed the location of the peak using the Hubble functions and obtained, for $w\rightarrow -1$, $z\approx 0.6$. 
    \item[-] The peaks in the virialization ratios are generally located at $z=0.4\pm 0.2$ and $z=0.8\pm 0.2$ which average at $z=0.6\pm 0.2$, i.e, very close to the transition redshift constraints \cite{Capozziello:2014zda,Capozziello:2021xjw,Muccino:2022rnd}, while far from the equivalence DE-dark matter \cite{Alfano:2023evg}. The Hubble function approximation does not work well in predicting the peaks when the $z_{\rm t}$ parameter is varied. In the case of $z_{\rm t}=2$ we found the peak at $z=2.9\pm 0.2$. 
    \item[-] We showed how the shape of the virialization overdensity can be understood by comparing the model's Hubble function to the EdS one. 
    \item[-] By varying the transition speed we found that also nonlinear perturbations are quite insensitive to the parameter, confirming again the previous results.
    \item[-] In the nonlinear regime, perturbations exhibit high sensitivity to the final EOS value in comparison to the linear regime. This is because in the nonlinear phase, the changes happen much faster\footnote{In the time it takes linear perturbations to reach $\delta=1.686$ the nonlinear ones reach infinity.} and thus even if DE has a short time after the transition in which it can act, the effects are quite large.
    \item[-] We compared the clustering case, $\delta_{\rm{de}}\neq 0$, to the smooth case, $\delta_{\rm{de}}= 0$. The main effect of DE perturbations is to dampen, on average, by $2\%$ the differences of the model with respect to the $\Lambda$CDM model. The dampening depends on the exact parameter value considered.
    \item[-] Consistently with the linear results, the models $w_1$, $w_2$, $w_3$ and $w_4$ are indistinguishable one from the other.
\end{itemize}
 
We then analysed statistical properties of the matter clustering, namely the linear and nonlinear power spectrum and the HMF. We found that:

\begin{itemize}
    \item[-] Freezing fast transition models can lower the present $\sigma_8$ value by around $8\%$, without affecting too much the background evolution. This suggests that it may be possible to relieve the $\sigma_8$ tension with such models. This possibility deserves a thorough investigation, and we intend to pursue it in a future work.
    \item[-] By comparing the nonlinear matter power spectrum for smooth and clustering DE, we found that clustering slightly suppresses the differences with respect to the LCDM model. In the smooth case, valleys at $k \approx 1\,h^{-1}\,{\rm Mpc}$ in the percentage difference are replaced by peaks at $k \approx 2\,h^{-1}\,{\rm Mpc}$ when clustering is allowed. This effect could be used to detect signatures of clustering DE.
    \item[-] Since perturbations are maximally different from the standard model at $z\approx 0.6$, we argue that this happens also for the HMF. The redshift range we have used is too narrow to prove this; thus, we plan to extend it in future work.
    \item[-] We showed that the differences between the HMF in the clustering and smooth case are small, approximately $1\%$ for average size galaxies, while they can become more important, approximately $5\%$ for massive galaxies.
    \item[-] We explained why the HMF shows a weak dependency on $w_{\rm f}$. The dependency of the HMF on the $z_{\rm t}$ parameter is very weak for $z_{\rm t}\gtrapprox 2$. For $z_{\rm t}=2$ at $z=0$, the model is indistinguishable from the standard one, while there are some signs of the transition at higher redshifts $z=0.5,1$.
    \item[-] All the models predict a lower number of massive galaxies, except Eq.~(\ref{eq:de_eos_5}) which is thawing, and predicts higher numbers of massive galaxies with respect to the $\Lambda$CDM model at $z=0$. This model with $\Gamma=-2.5$, predicts that from $z=0.5$ to $z=0$, the number of massive galaxies increases by $50\%$ with respect to the $\Lambda$CDM scenario, implying a period of fast galaxy merging which may be detectable. This is a common feature of the models: they predict a merging rate higher than in the $\Lambda$CDM Universe.
\end{itemize}

We conclude by presenting  perspectives and research challenges below.

\begin{itemize}
    \item[-] Can an evolving fast-transition DE influence gravitational lensing, and how? By changing the EOS of DE to a fast-transition model, we can study how a sudden period in which DE undergoes a transition affect this observable.
    \item[-] Determining whether freezing fast transition models can alleviate the $\sigma_8$ tension and to clarify their impact on the Hubble tension \citep[see also][]{Heisenberg:2022lob,Heisenberg:2022gqk,Lee:2022cyh}. Indeed, the presence of DE perturbations may influence baryon acoustic oscillations and the cosmic microwave background power spectrum. A dynamical DE also changes the expansion history of the Universe and thus the relation between the values of the Hubble parameter at different redshifts. Thus, we will explore the possibility of a model that can alleviate both tensions at the same time.
    \item[-] Investigating whether future data from the \textit{Euclid} mission will be accurate enough to discern differences in the nonlinear power spectrum between clustering and smooth DE cases.
\end{itemize}

\section*{Acknowledgements}

F.P. acknowledges partial support from the INFN grant InDark and from from the Italian Ministry of University and Research (\textsc{mur}), PRIN 2022 `EXSKALIBUR – Euclid-Cross-SKA: Likelihood Inference Building for Universe's Research', Grant No.\ 20222BBYB9, CUP C53D2300131 0006, and from the European Union -- Next Generation EU. F.P. also acknowledges support from the FCT project ``BEYLA -- BEYond LAmbda" with ref. number PTDC/FIS-AST/0054/2021. S.T. is grateful to Benjamin Bose for his invaluable assistance in running and understanding the R{\scriptsize E}A{\scriptsize CT} code.

\appendix

\section*{APPENDIX: ADDITIONAL MATERIAL}\label{sec:appendix}

In this Appendix we present some additional material which contributes to a better understanding of our results. In particular, we discuss how the model represented by the EOS $w_1$ depends on the free parameters and how similar or different the other models are, with respect to both background and perturbations.

\subsection{Background}

\begin{figure}
  \centering
  \includegraphics[width=1\columnwidth,clip]{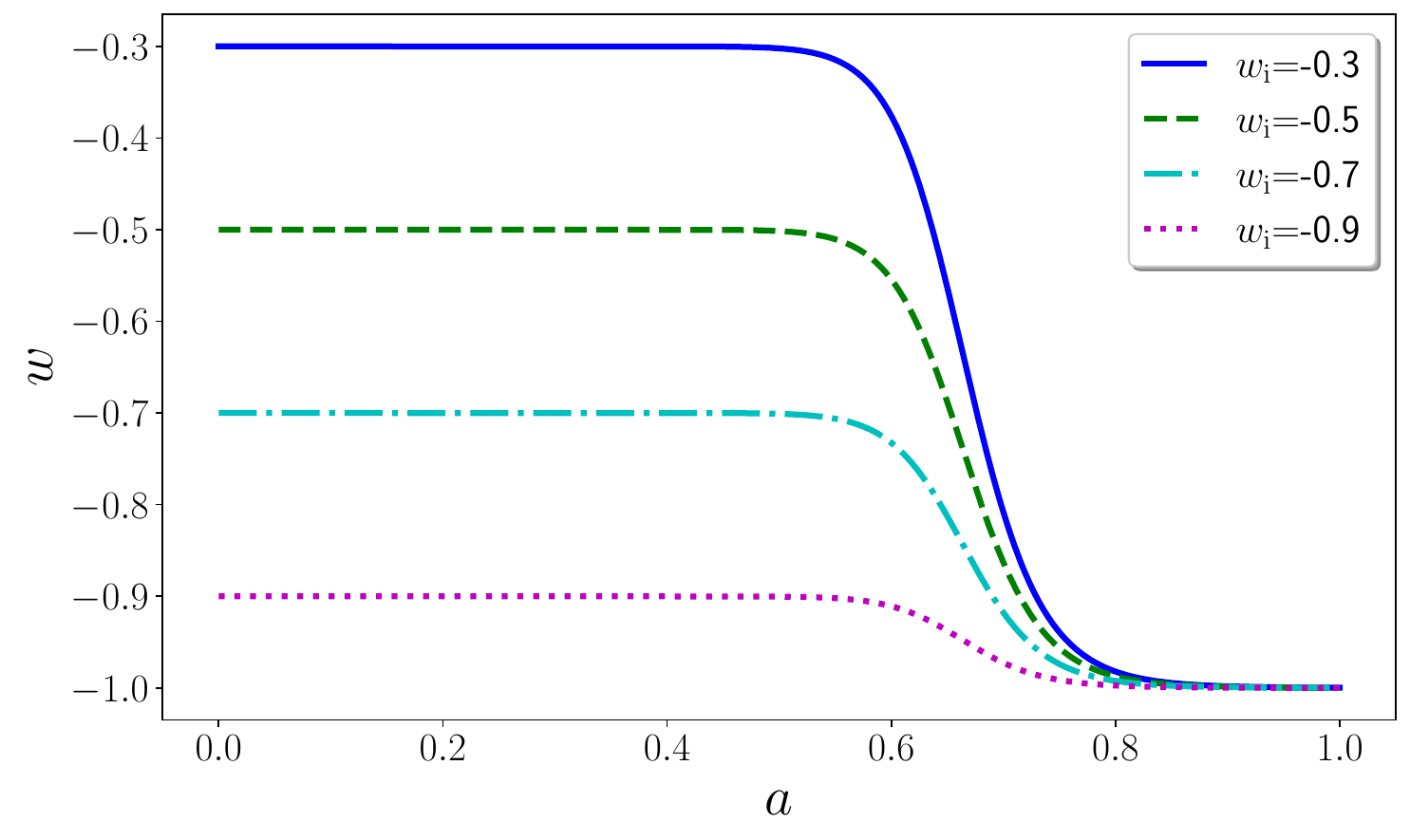}
  \caption{Behavior of the EOS for model $w_1$ with constants $w_{\rm f}=-1$, $\Gamma=10$, $z_{\rm t}=0.5$, when the parameter $w_{\rm i}$ is varied.}
  \label{fig:de_eos_1_w_i}
\end{figure}

\begin{figure}
    \centering
    \includegraphics[width=1\columnwidth,clip]{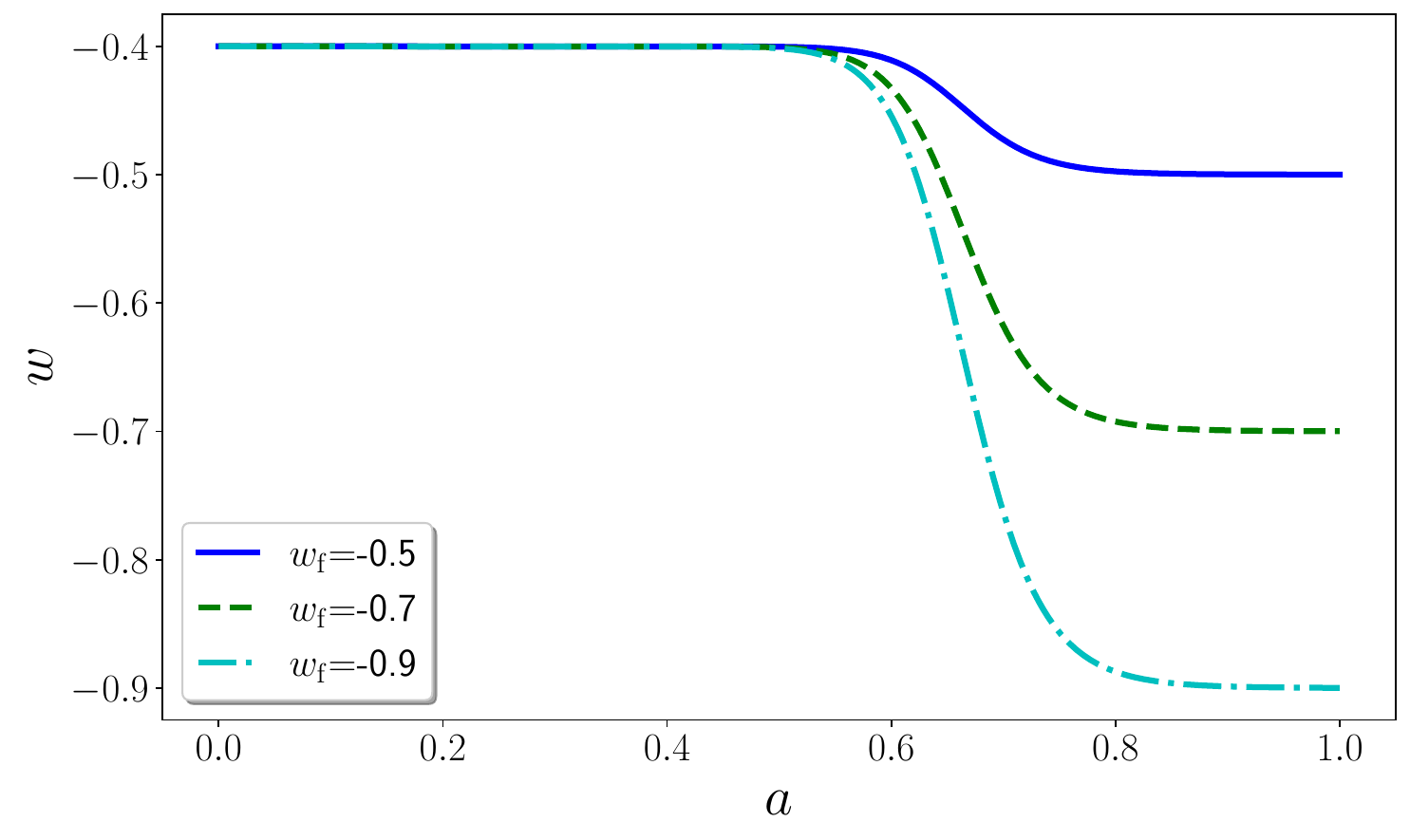}
    \includegraphics[width=1\columnwidth,clip]{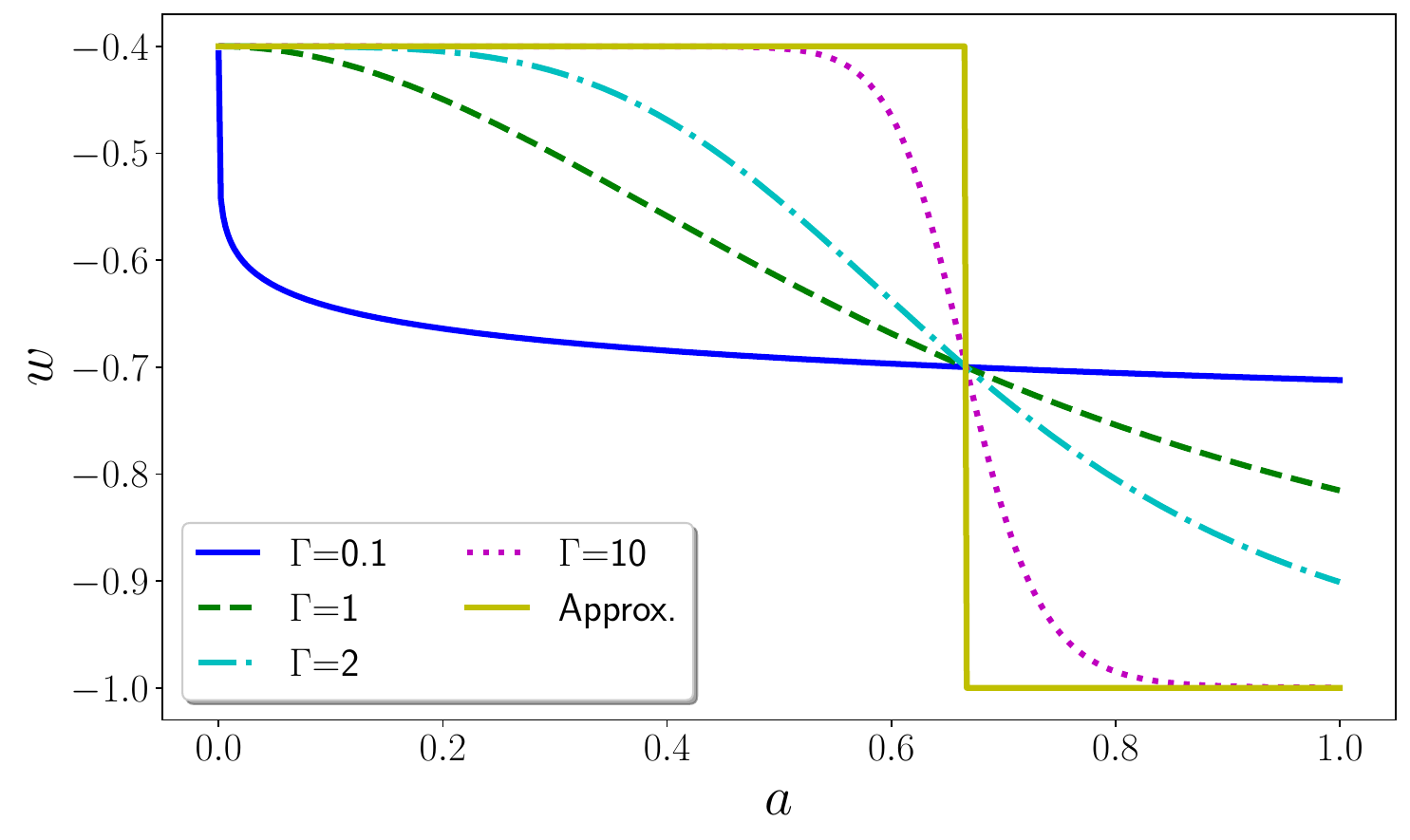}\vfil
    \caption{EOS of a model described by the $w_1$ functional form with fixed $w_{\rm i}=-0.4$ and $z_{\rm t}=0.5$. Top panel: we set $\Gamma=10$ and vary $w_{\rm f}$ parameter. Bottom panel: we fix $w_{\rm f}=-1$ and consider different values of the $\Gamma$ parameter. The yellow line is the Heaviside step function which approximates the model for $\Gamma\rightarrow\infty$. For small values of $\Gamma$, the model tends to a $w$CDM model with $w$ equal to the average of $w_{\rm i}$ and $w_{\rm f}$.}
    \label{fig:quint_1_trans_steepness_eos}
\end{figure}

In Fig.~\ref{fig:de_eos_1_w_i} we report the EOS for the chosen values of the $w_{\rm i}$ parameter, for model $w_1$, Eq.~(\ref{eq:de_eos_1}). For lower values of $w_{\rm i}$ the transition is more gentle, in the sense that the EOS derivative during the transition is lower. However, the time it takes to transition does not depend on $w_{\rm i}$. How the EOS $w_1$ depends on the $w_{\rm f}$ parameter is shown in the top panel of Fig.~\ref{fig:quint_1_trans_steepness_eos}. The potential and scalar field consequently vary but as the transition is recent, the potential and the field do not depend as strongly on the $w_{\rm f}$ parameter as they do on the $w_{\rm i}$ parameter, resulting in a narrow cluster of curves we calculated but decided not to show. In the bottom panel of Fig.~\ref{fig:quint_1_trans_steepness_eos} we present the EOS as we vary the transition speed parameter, $\Gamma$. The model exhibits a spectrum of behaviors, from approximating the $w$CDM with $w=(w_{\rm i} + w_{\rm f})/2$ when $\Gamma=0.1$, to resembling the Heaviside step transition when $\Gamma=10$. The true step transition is also shown and serves as an approximation for fast transition models.

\begin{figure}
    \centering
    \includegraphics[width=1\columnwidth,clip]{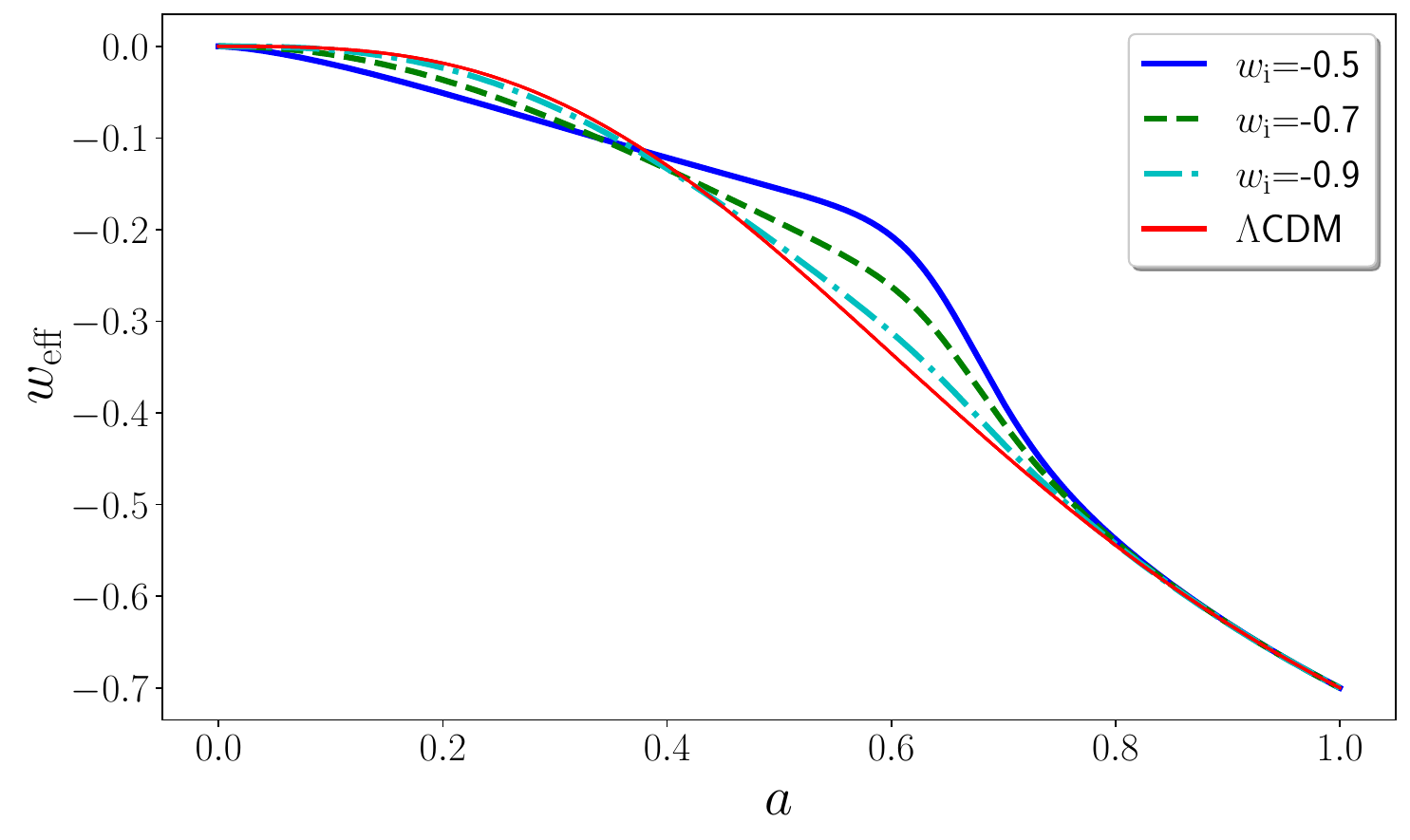}
    \includegraphics[width=1\columnwidth,clip]{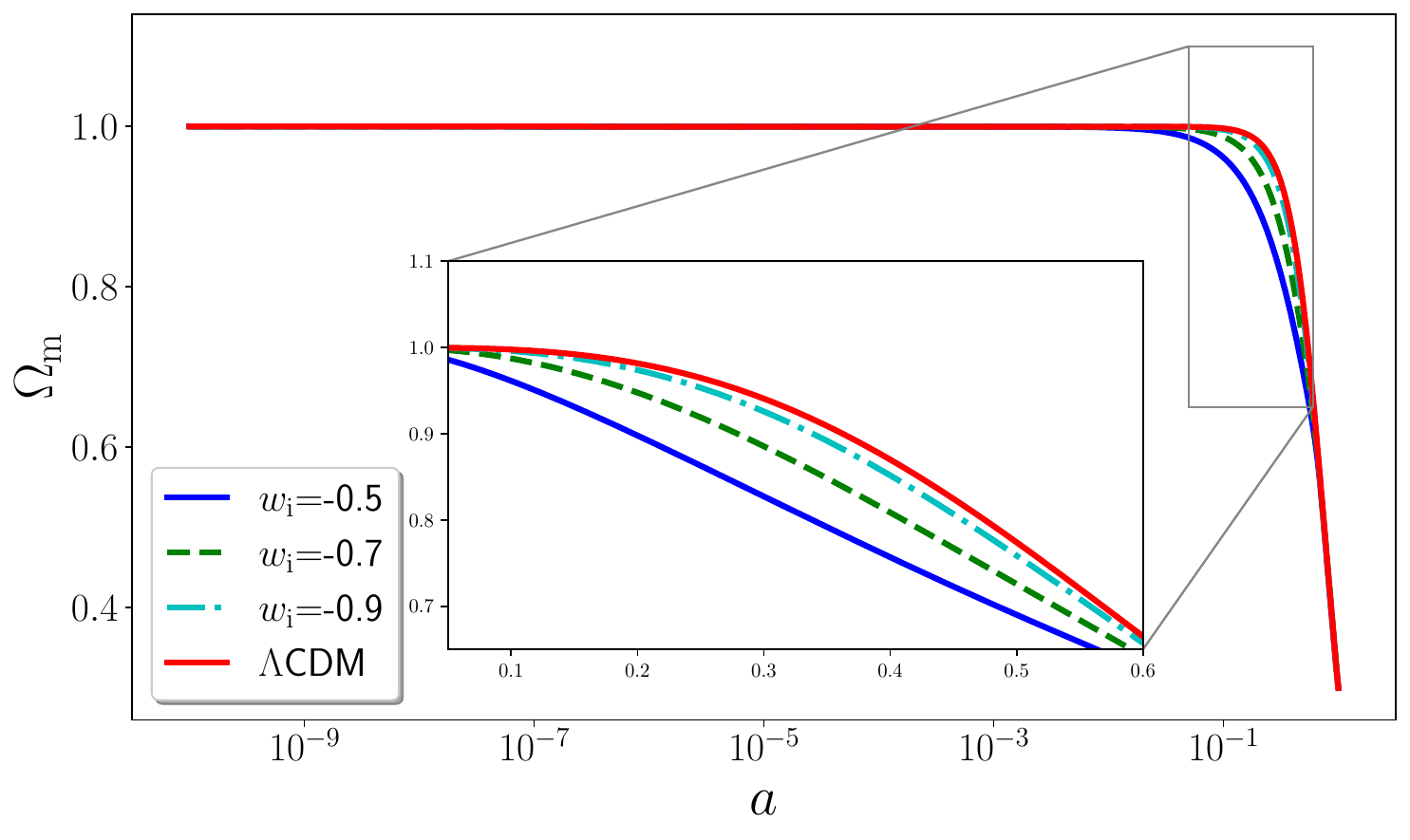}
    \caption{Plots for model $w_1$ with $w_{\rm f}=-1$, $\Gamma=10$, $z_{\rm t}=0.5$. Top panel: effective EOS. It shows that at early times, DE with $w_{\rm i}$ close to zero has an important impact on the background evolution of the Universe. Bottom panel: matter density parameters. Comparing the case $w_{\rm i}=-0.5$ to the $\Lambda$CDM curve shows that matter is not completely dominant at last scattering.}
    \label{fig:unpert_1_w_i_background}
\end{figure}

 To give the reader a more comprehensive understanding of the Universe evolution in the presence of the considered DE models, we present the evolution of some additional background quantities, namely the matter density parameter, the Hubble function and the model described by the EOS $w_5$. In Fig.~\ref{fig:unpert_1_w_i_background}, we show the effective EOS and the matter density parameter when DE is described by model $w_1$, with different values of the $w_{\rm i}$ parameter. The more $w_{\rm i}$ approaches zero, the more $\Omega_{\rm{m}}$, at early-times ($a\approx 10^{-3}$), differs from unity, even if slightly. In the limit of $w_{\rm i}=0$, at early times $\Omega_{\rm{m}}$ would be equal to $0.3$. Considering for a moment matter perturbations, this is useful to understand how the background affects clustering and remarks the importance of using the correct exponent for the growth factor in Eq.~(\ref{eq:deltaMatterLinear}).

\begin{figure}
    \centering
    \includegraphics[width=1\columnwidth,clip]{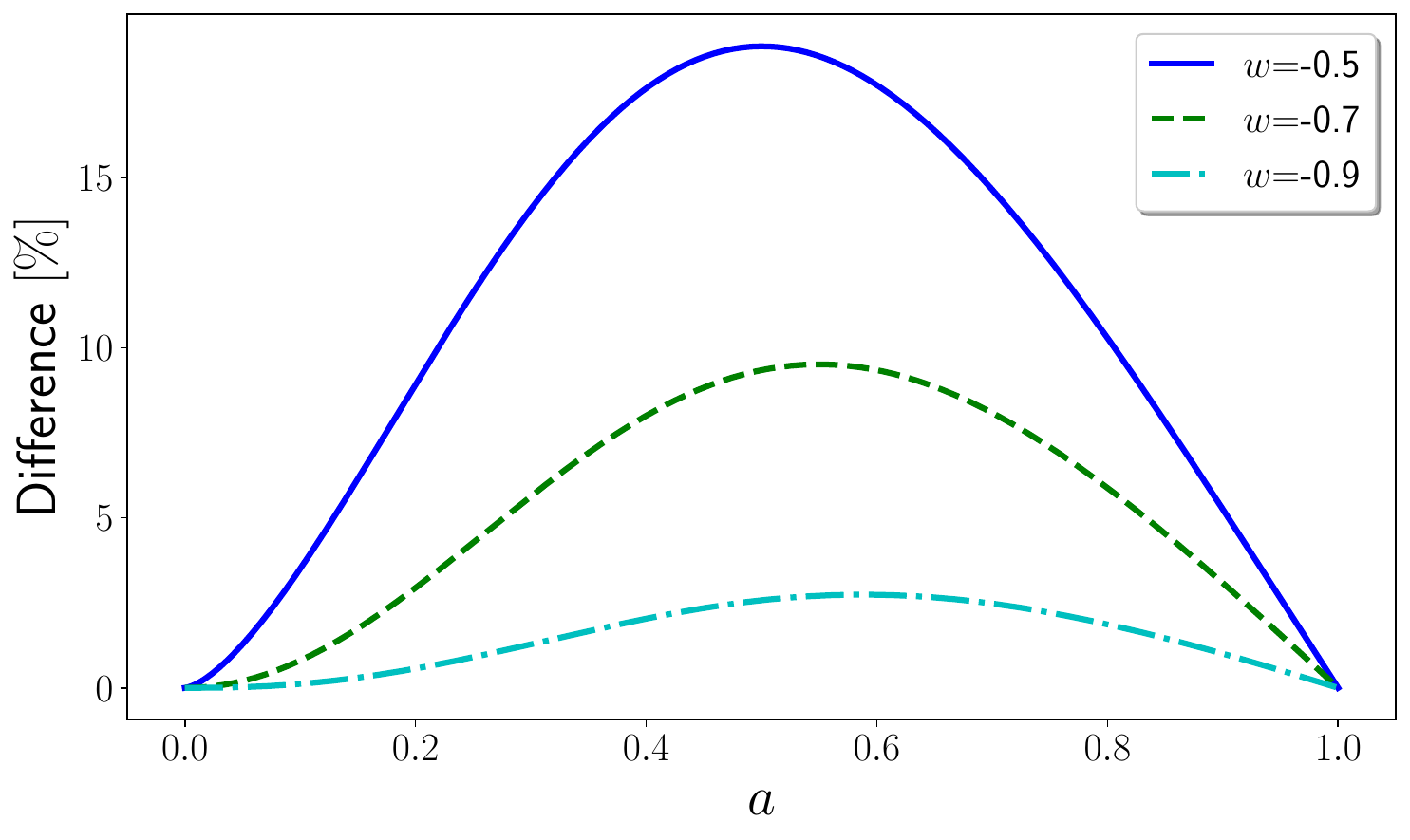}
    \caption{Percentage difference between the $w$CDM and $\Lambda$CDM Hubble functions, defined as $E_{w\rm{CDM}}=E_{\Lambda \rm{CDM}}(1+\%)$. This shows that if the late-time acceleration of the Universe increases, the Hubble function decreases. Even if we have shown this for the $w$CDM model, it qualitatively holds for fast transition models.}
    \label{fig:hubble_function}
\end{figure}

Figure ~\ref{fig:hubble_function} shows the percentage difference between the scaled Hubble function, $E(a)\equiv H(a)/H_0$, of the standard cosmological model and the $w$CDM model, which we use as a reference to better understand fast transition models. The main feature is that the $w$CDM model has an Hubble function always higher than the $\Lambda$CDM model, resulting in an augmented Hubble drag. We have shown in Sec. \ref{subsec:linearRegime} that this implies lower matter clustering in the $w$CDM Universe.

\begin{figure}
    \centering
    \includegraphics[width=1\columnwidth,clip]{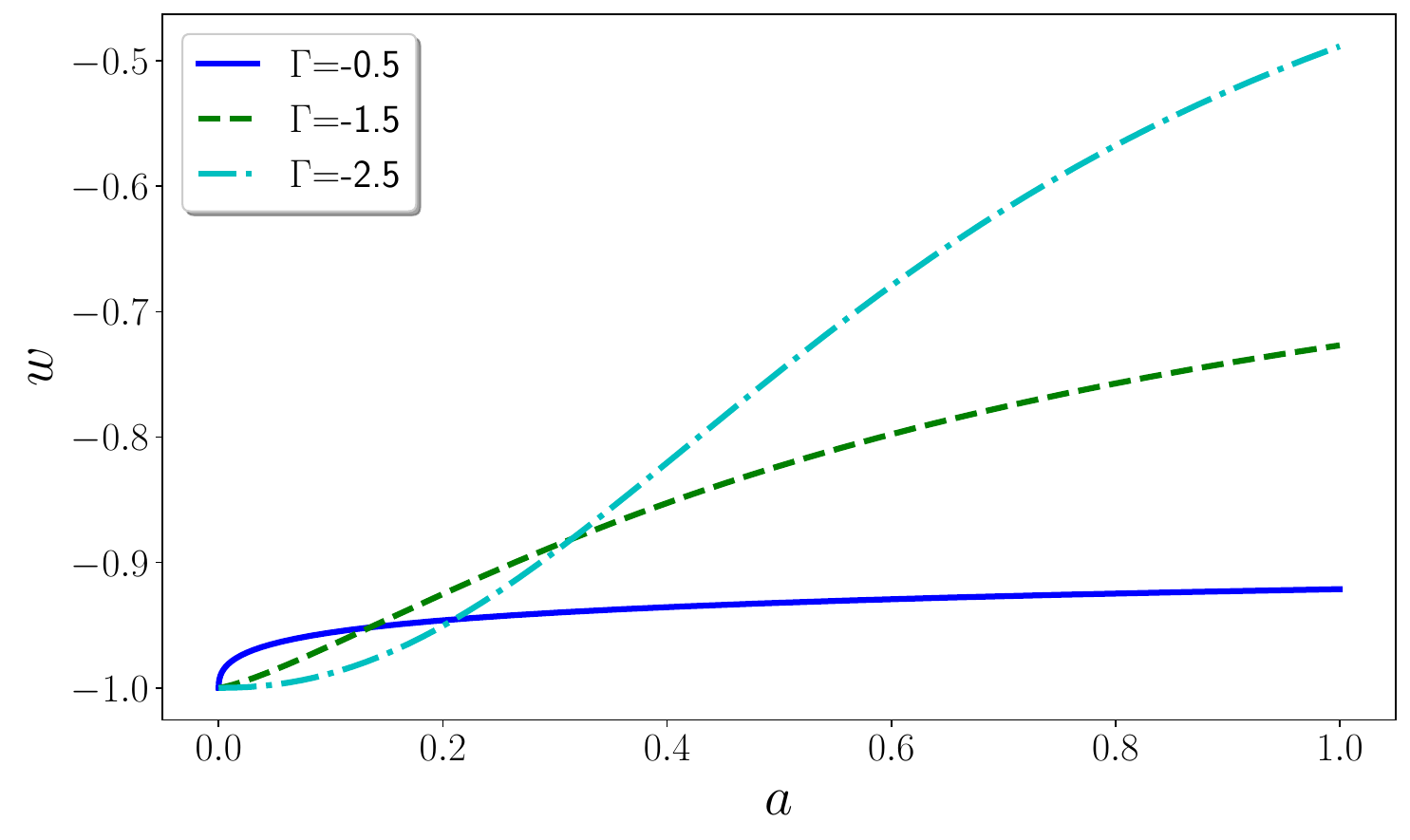}
    \includegraphics[width=1\columnwidth,clip]{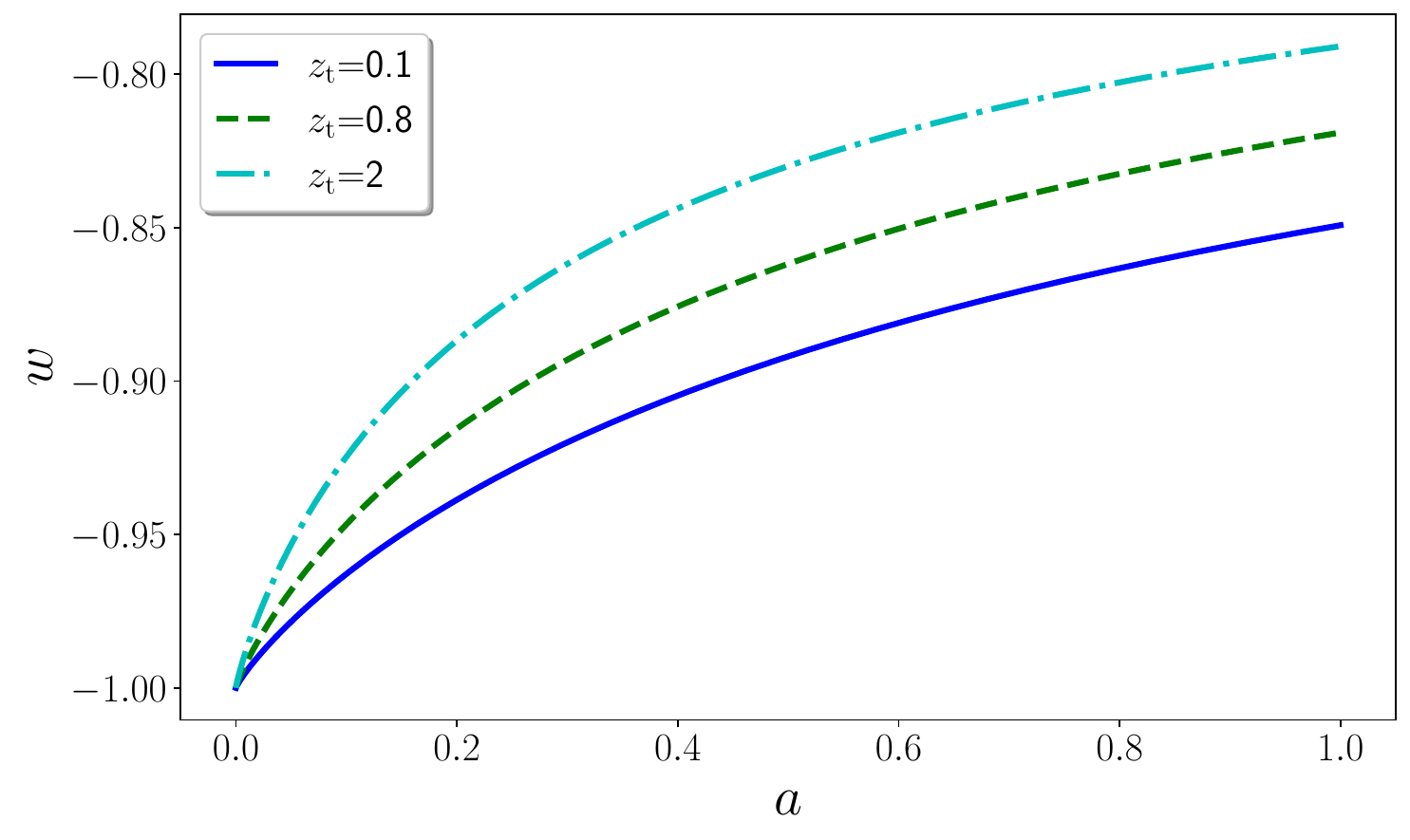}
    \caption{Plots for model $w_5$. Top panel: EOS with constant $z_{\rm t}=0.5$, for various values of $\Gamma$. Bottom panel: EOS with constant $\Gamma=-1$, for some values of $z_{\rm t}$.}
    \label{fig:pert_5}
\end{figure}

The last model we analyze is $w_5$, Eq.~(\ref{eq:de_eos_5}). Results are shown in Fig.~\ref{fig:pert_5}. In the top panel, we can see that at early times, the EOS is consistently close to $-1$, implying that matter dominates. In the bottom panel, instead, we show the effect of varying the parameter $z_{\rm t}$. We can note that the final EOS value and the $\Gamma$ and $z_{\rm t}$ parameters are highly dependent, while the initial value is fixed to $w_{\rm i}=-1$ making this model thawing.

\subsection{Linear and nonlinear perturbations}

\begin{figure}
    \centering
    \includegraphics[width=1\columnwidth,clip]{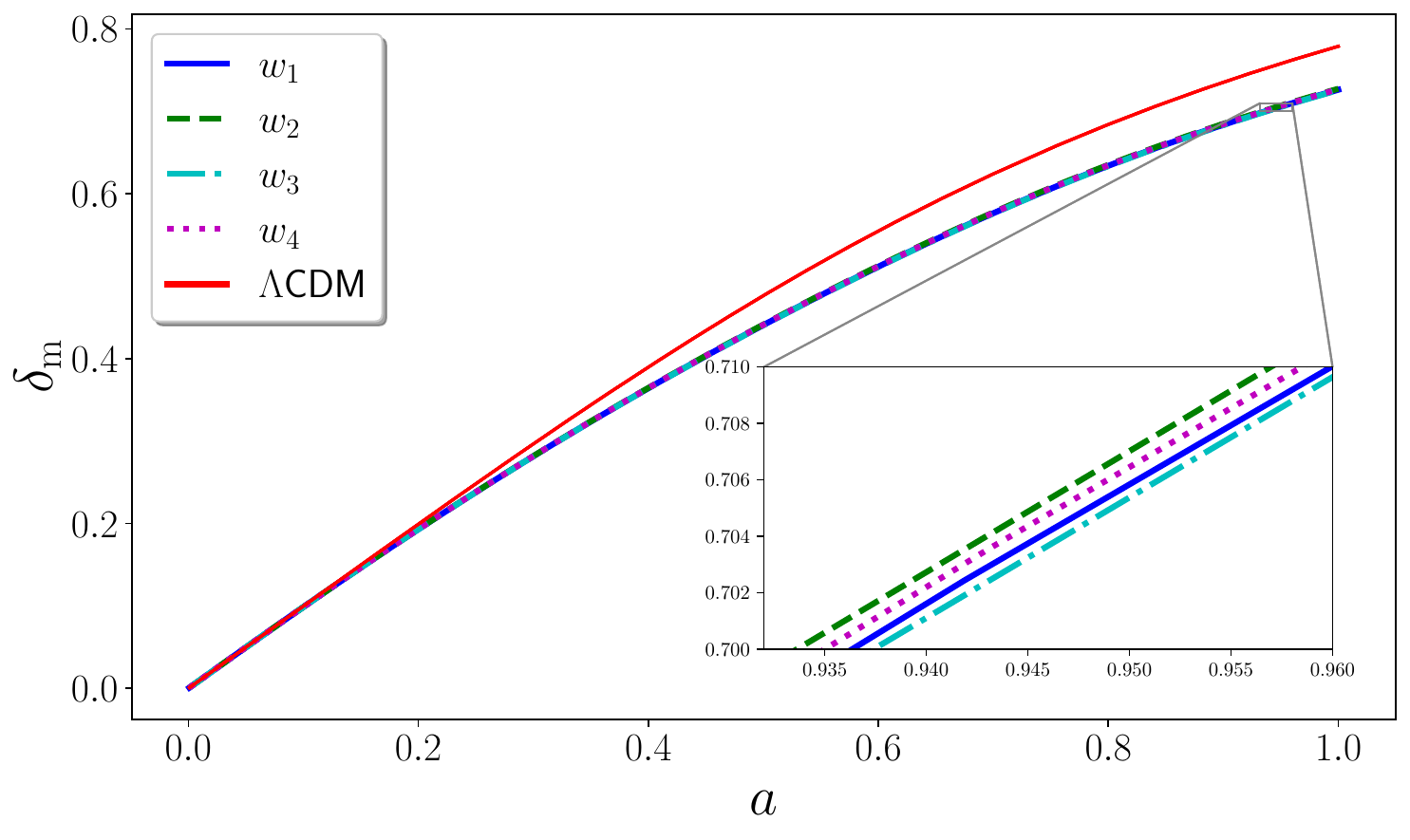}
    \caption{Plot for models $w_1$, $w_2$, $w_3$ and $w_4$, respectively, with $w_{\rm i}=-0.5$, $w_{\rm f}=-1$, $\Gamma_1=10$, $\Gamma_2=6.5$, $\Gamma_3=0.07$ and $\Gamma_4=0.033$. As we can see, the differences between these four models are negligible in the linear regime.}
    \label{fig:linear_model_diff}
\end{figure}

In the previous section, we should have considered all the other models, but we notice that among the six EOS considered, the first four, Eqs.~(\ref{eq:de_eos_1})--(\ref{eq:de_eos_4}), behave in the exactly same way, as shown in Fig.~\ref{fig:linear_model_diff}, justifying our choice to focus only on the model given in Eq.~(\ref{eq:de_eos_1}). To ensure a fair comparison, we carefully selected the transition speed parameters to align the models as closely as possible. The maximum deviation between the models is less than $1\%$.

\begin{figure}
    \centering
    \includegraphics[width=1\columnwidth,clip]{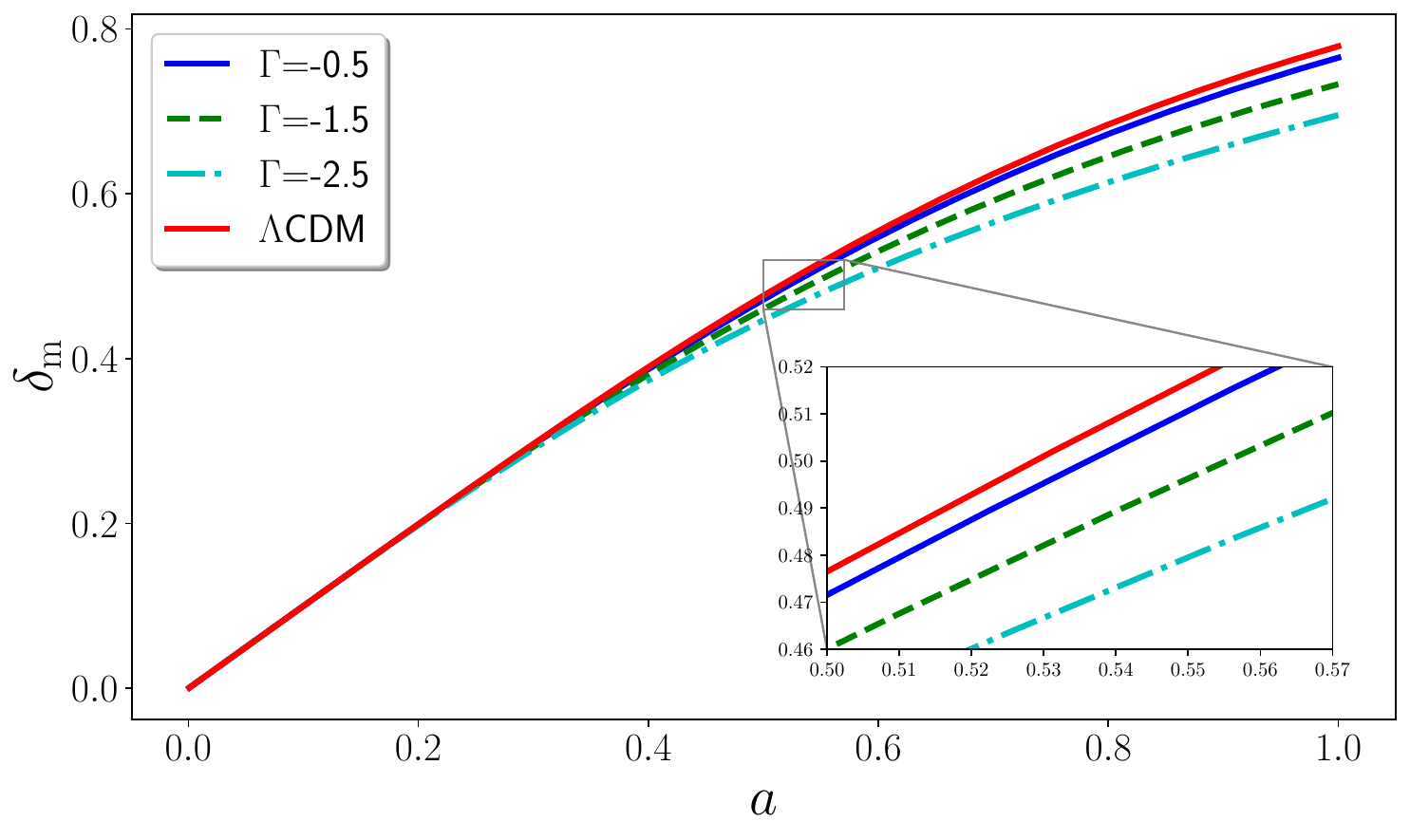}
    \caption{Plot for model $w_5$ with $z_{\rm t}=0.5$. Linear perturbations with $\delta_{\rm{de}}\neq 0$. }
    \label{fig:pert_5_trans_steepness_1}
\end{figure}

In view of the results reported in Fig.~\ref{fig:linear_model_diff}, we now focus on the fifth model, Eq.~(\ref{eq:de_eos_5}), which has only two parameters: $\Gamma$ and $z_{\rm t}$. These parameters are closely related, as their values determine the rate of the transition as well as the initial and final values of the EOS. We limit our analysis to the nonphantom regime, for which $\Gamma < 0$. The results for this model are presented in Fig.~\ref{fig:pert_5_trans_steepness_1}.

\begin{figure}
    \centering
    \includegraphics[width=1\columnwidth,clip]{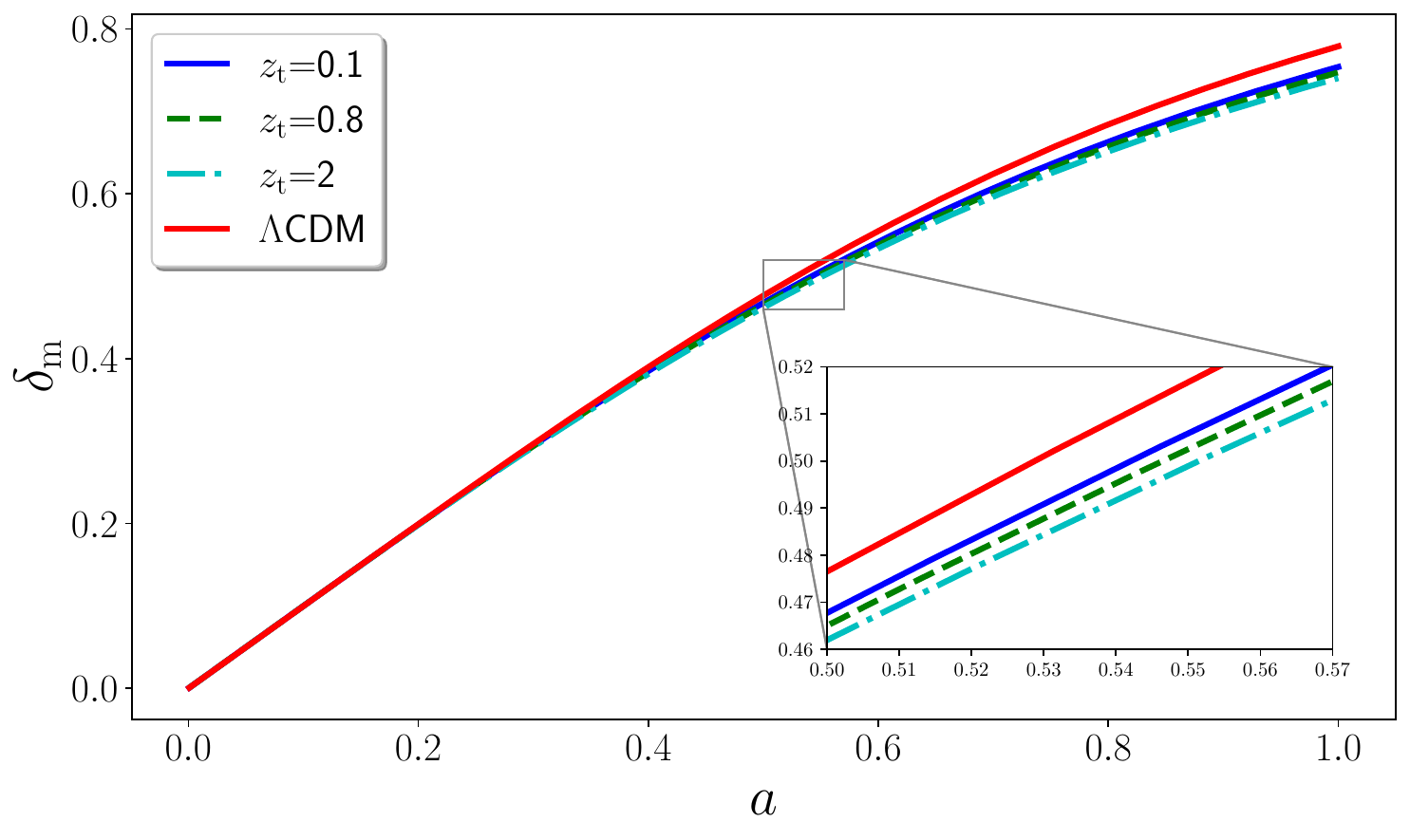}
    \caption{Plots for model $w_5$ with $\Gamma=-1$. Linear perturbations with $\delta_{\rm{de}}\neq 0$.}
    \label{fig:pert_5_z_t}
\end{figure}

The deviations from the $\Lambda$CDM model are constrained by the requirement that the acceleration parameter is negative today, which implies that $w(a=1) < -1/3$. The plot in Fig.~\ref{fig:pert_5_trans_steepness_1} covers almost the full range of admissible $\Gamma$ values. Varying the other free parameter of Eq.~(\ref{eq:de_eos_5}), $z_{\rm t}$, yields the linear perturbations shown in Fig.~\ref{fig:pert_5_z_t}. We notice that the parameter $\Gamma$ has a bigger impact than $z_{\rm t}$.

The last model, Eq.~(\ref{eq:de_eos_6}), has no free parameters and it is entirely in the phantom regime. Thus, as expected, it predicts larger perturbations than in the $\Lambda$CDM model. This happens because the DE component is subdominant until very late in the cosmic history. If we would normalise linear perturbations to be unity today, we would see that the growth factor has lower values than the $\Lambda$CDM model at early times, as structures need to grow more slowly to reach the same speed of the other models today.

\begin{figure}
    \centering
    \includegraphics[width=1\columnwidth,clip]{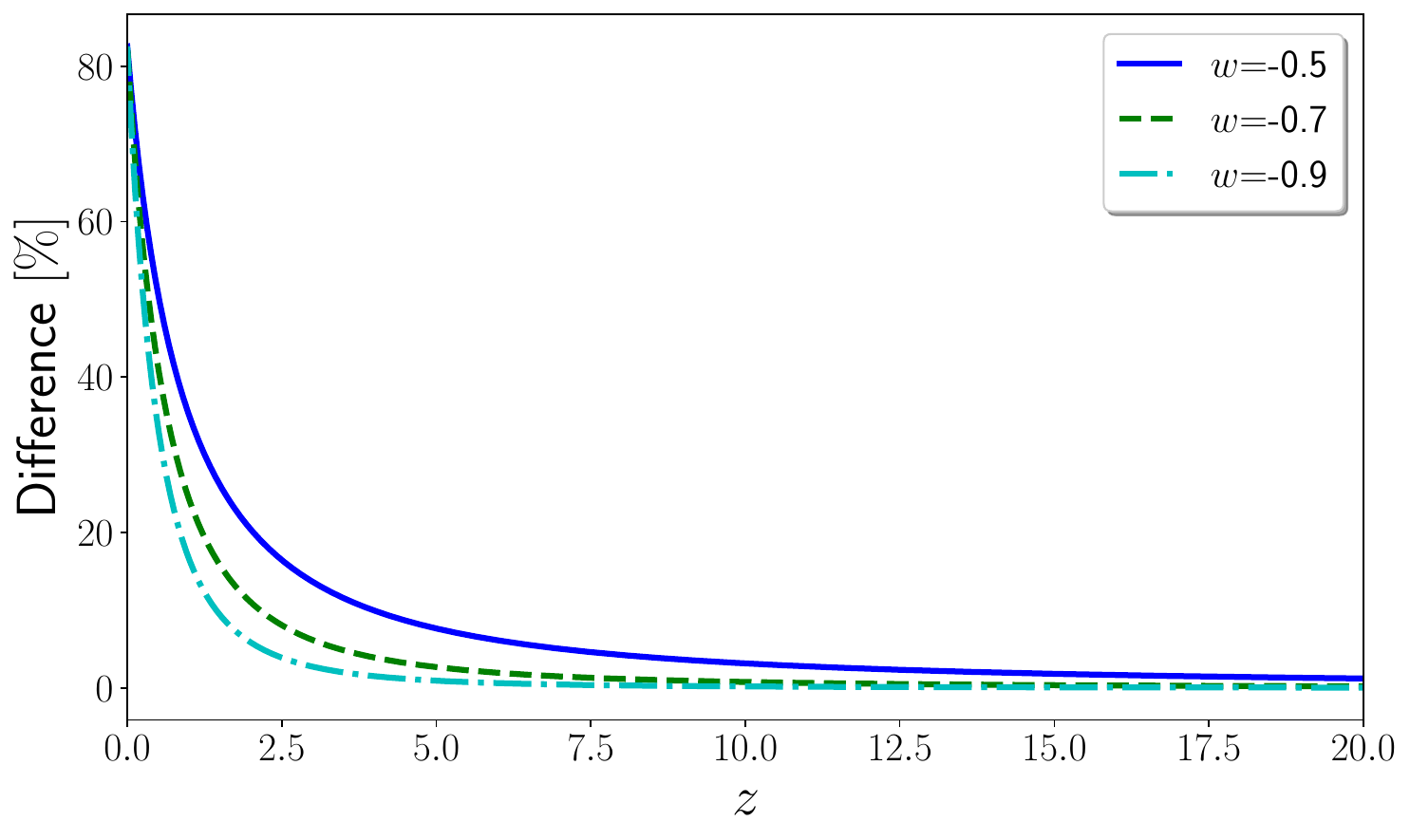}
    \caption{Percentage difference defined by $E_{w\rm{CDM}}(z)=\sqrt{\Omega_{\rm m}^0}E_{\rm{EdS}}(1+\%)$. Note that we scaled the $w$CDM Hubble function $E_{w{\rm CDM}}\rightarrow E_{w{\rm CDM}}/\sqrt{\Omega_{\rm m}^0}$ in order to make the $w$CDM model reduce to the EdS one at early times.}
    \label{fig:hubble_ratio_1}
\end{figure}

A Universe that, at late times, has a faster expansion rate, will exhibit lower values of the Hubble function, as shown in Figs.~\ref{fig:hubble_function} and~\ref{fig:hubble_ratio_1}. Given that the Hubble function dilutes perturbations, overdensities in a Universe expanding with a slower rate will face greater dilution while collapsing. Since the collapse time is fixed, to counterbalance the effect of a higher Hubble function, the virialization overdensity will consequently be higher.

\begin{figure}
    \centering
    \includegraphics[width=1\columnwidth,clip]{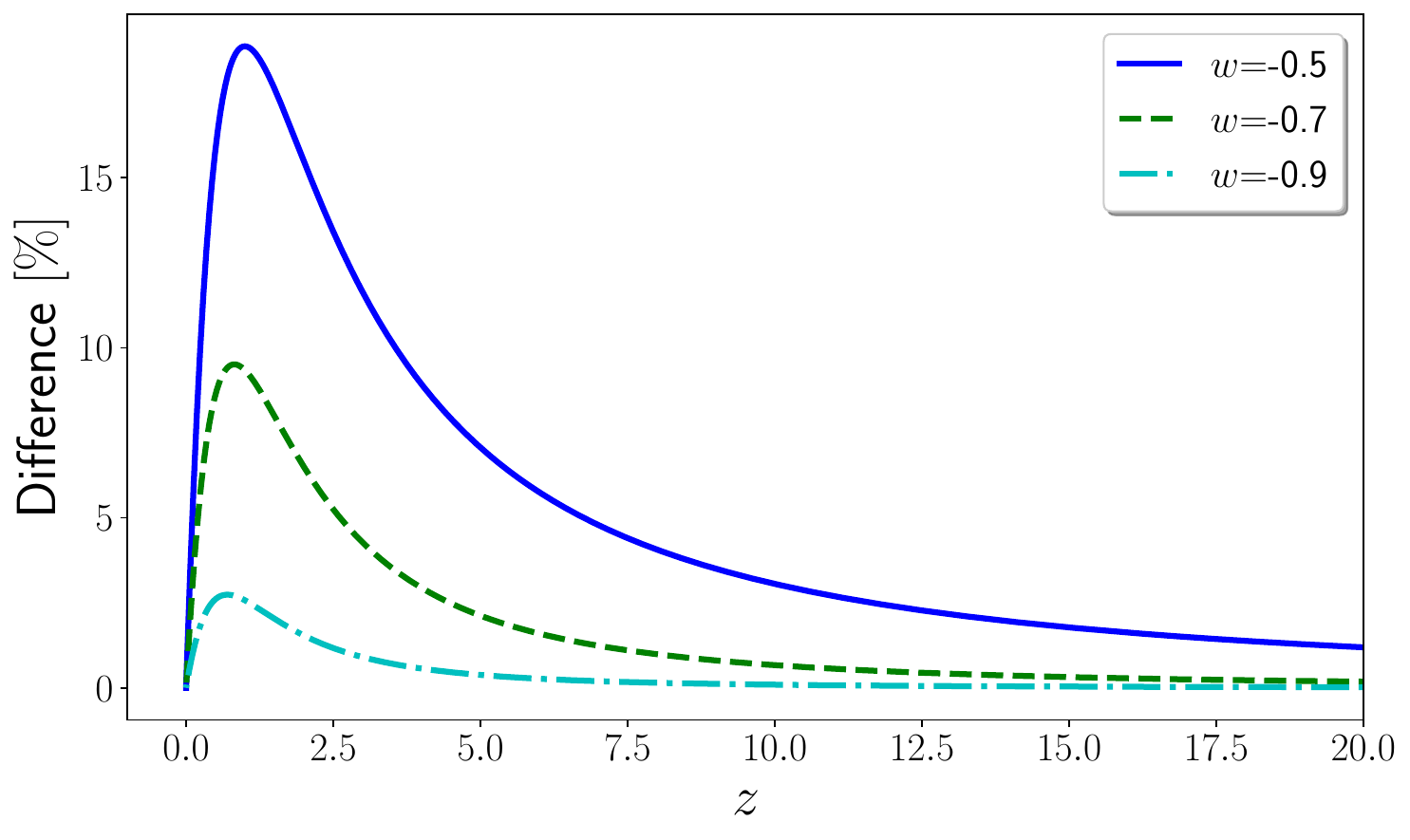}
    \includegraphics[width=1\columnwidth,clip]{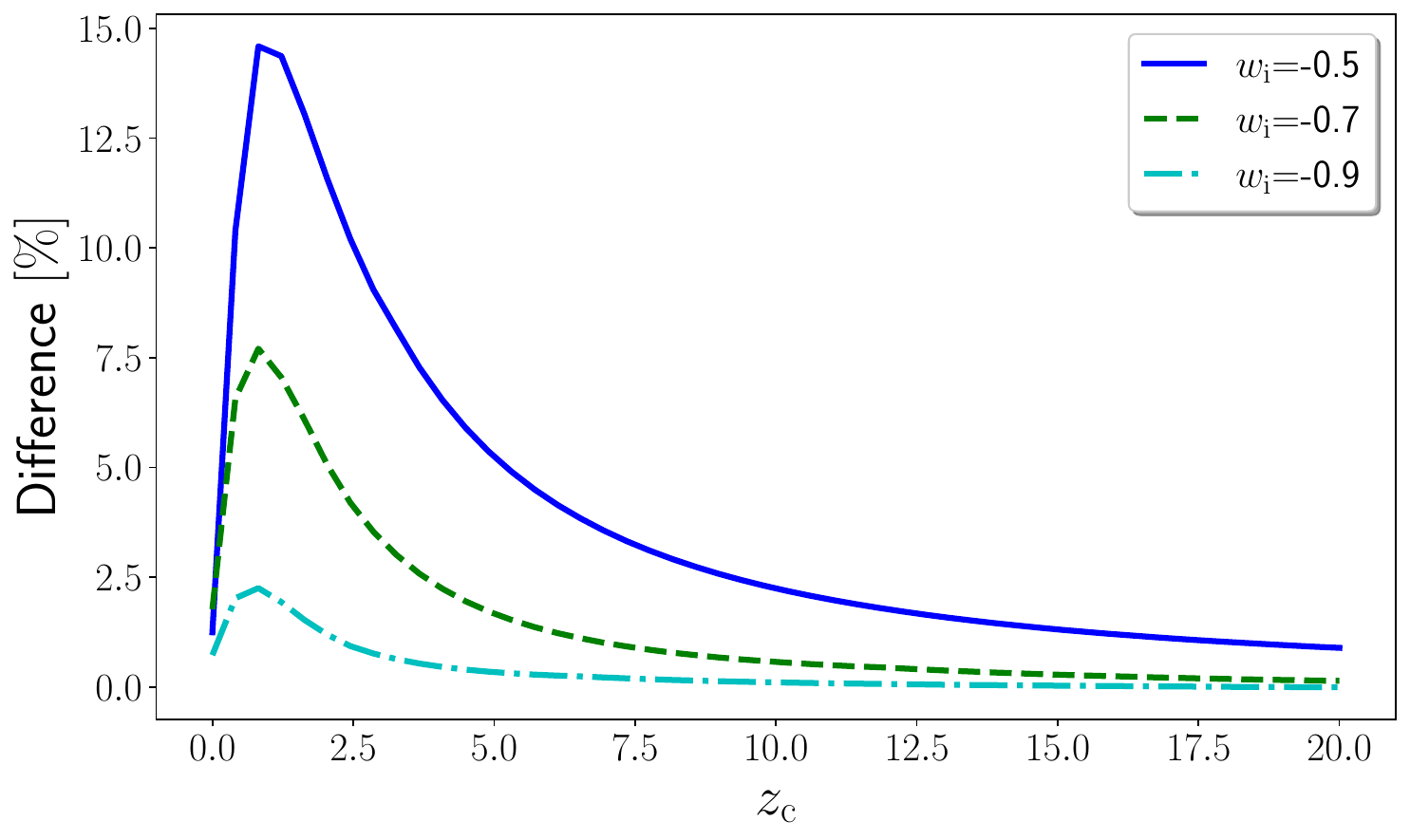}
    \includegraphics[width=1\columnwidth,clip]{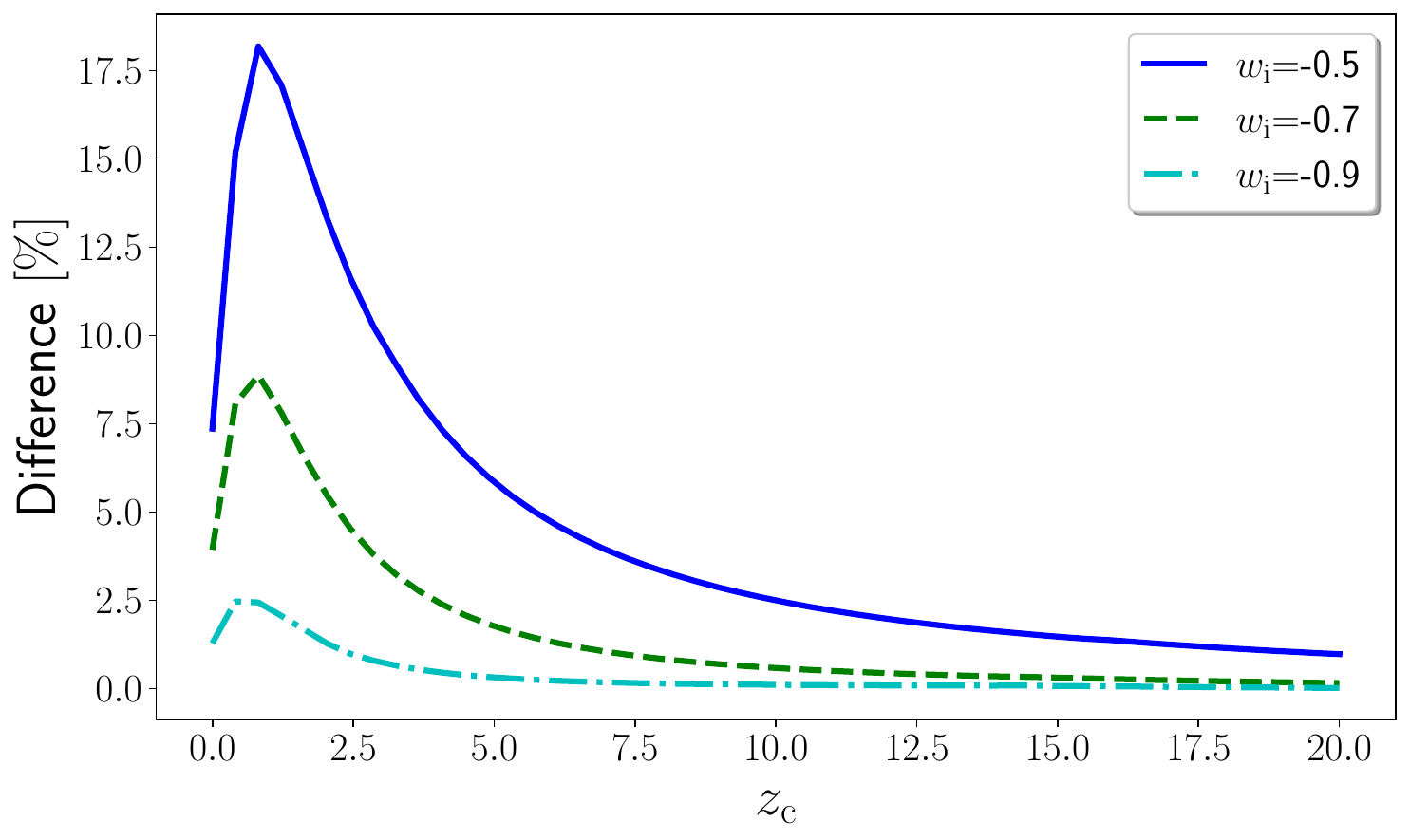}
    \caption{Top panel: percentage difference defined as $E_{w{\rm CDM}}(z)=E_{\Lambda{\rm CDM}}(z)(1+\%)$. This shows that DE models with $w_{\rm de}>-1$ result in Hubble functions higher than the $\Lambda$CDM model one and thus explain why those models result in higher virialization overdensities than the $\Lambda$CDM model. The maximum point of the ratio approximately corresponds to the time at which the difference between perturbations are most different from the $\Lambda$CDM model. For $w$ close to $-1$, the maximum point is approximately $z_{\rm{max}}\approx 0.6$. {\it Central panel}: percentage difference $\zeta_{\rm{m}}^*=\zeta_{\Lambda{\rm CDM}}^*(1+\%)$ where $\zeta_{\rm{m}}^*$ is the matter virialization overdensity for model $w_1$ and $\delta_{\rm{de}}\neq 0$. The behavior is similar to the ratio shown in the top plot. The redshift that maximizes the difference is $z_{\rm{max}}\approx 0.8$. Bottom panel: percentage difference defined by $\zeta_{\rm{m}}^*=\zeta_{\Lambda{\rm CDM}}^*(1+\%)$, where $\zeta_{\rm{m}}^*$ is the matter virialization overdensity for model $w_1$ with $\delta_{\rm{de}}=0$. The maximum points are located at respectively $z_{\rm{max}}=[0.4,0.8]$. We can therefore note that the estimate of $z_{\rm{max}}$ obtained from the $w$CDM Hubble function ratio in the top panel is precise within $0.2$ redshift units.}
    \label{fig:hubble_ratio_2}
\end{figure}

In order to understand the increase in the virialization overdensity at late times, it is important to note that the $\zeta^{*}_{\rm{vir}}$ remains constant in the EdS Universe. This means that if the Hubble function follows the same evolution as in the EdS Universe, the virialization overdensity will remain constant. However, if the Hubble function exceeds that of the EdS Universe, the virialization overdensity must increase to counterbalance the Hubble flow that attempts to dilute the perturbation. In Fig.~\ref{fig:hubble_ratio_1}, we present the percentage difference between $E_{w\text{CDM}}$ and $E_{\text{EdS}}$. Additionally, in the top panel of Fig.~\ref{fig:hubble_ratio_2}, we show the percentage difference between $E_{w\text{CDM}}$ and $E_{\Lambda \text{CDM}}$.

In Fig.~\ref{fig:hubble_ratio_2} we can see that for each choice of the parameter, there exists a time interval during which the virialization overdensity maximally deviates from the $\Lambda$CDM model. In the bottom panel of Fig.~\ref{fig:hubble_ratio_2} we show the percentage difference of $\zeta_{\rm{m}}^*$ with respect to the $\Lambda$CDM one, for the case of smooth DE while in the central panel DE is clustering. As we have pointed out in the linear perturbation analysis, for values of $w$ close to $-1$, allowing DE to cluster or not, does not have appreciable consequences on perturbations. The effect of imposing $\delta_{\rm{de}}=0$ is to enhance the difference between the fast transition model and the $\Lambda$CDM one. The increase in difference depends on the value of the $w_{\rm i}$ parameter. For $w_{\rm i}=-0.5$ the deviation from the $\Lambda$CDM model increases of about $3\%$, while for the other values of $w_{\rm i}$ the difference drops to $1\%$.

\begin{figure}
    \centering
    \includegraphics[width=1\columnwidth,clip]{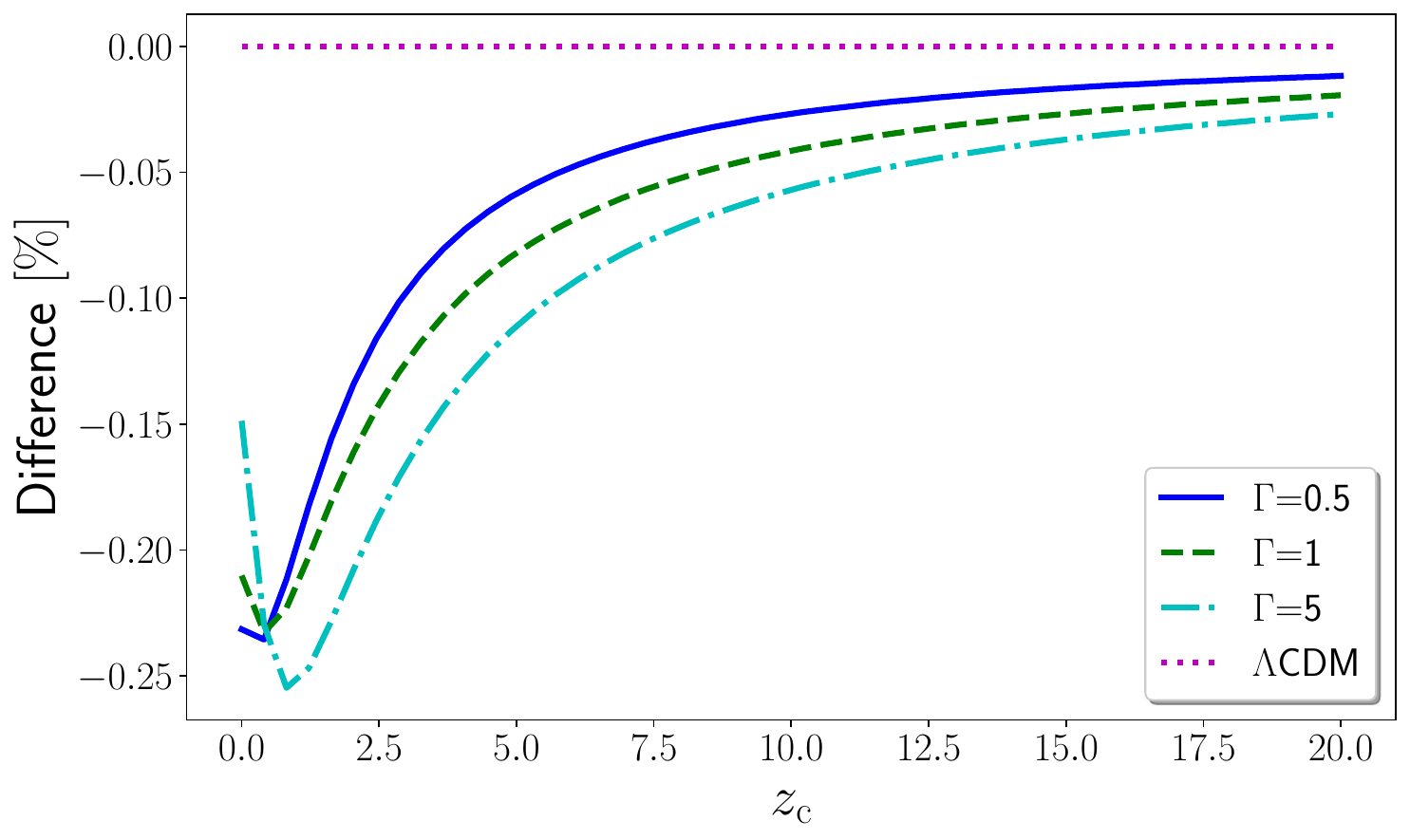}
    \includegraphics[width=1\columnwidth,clip]{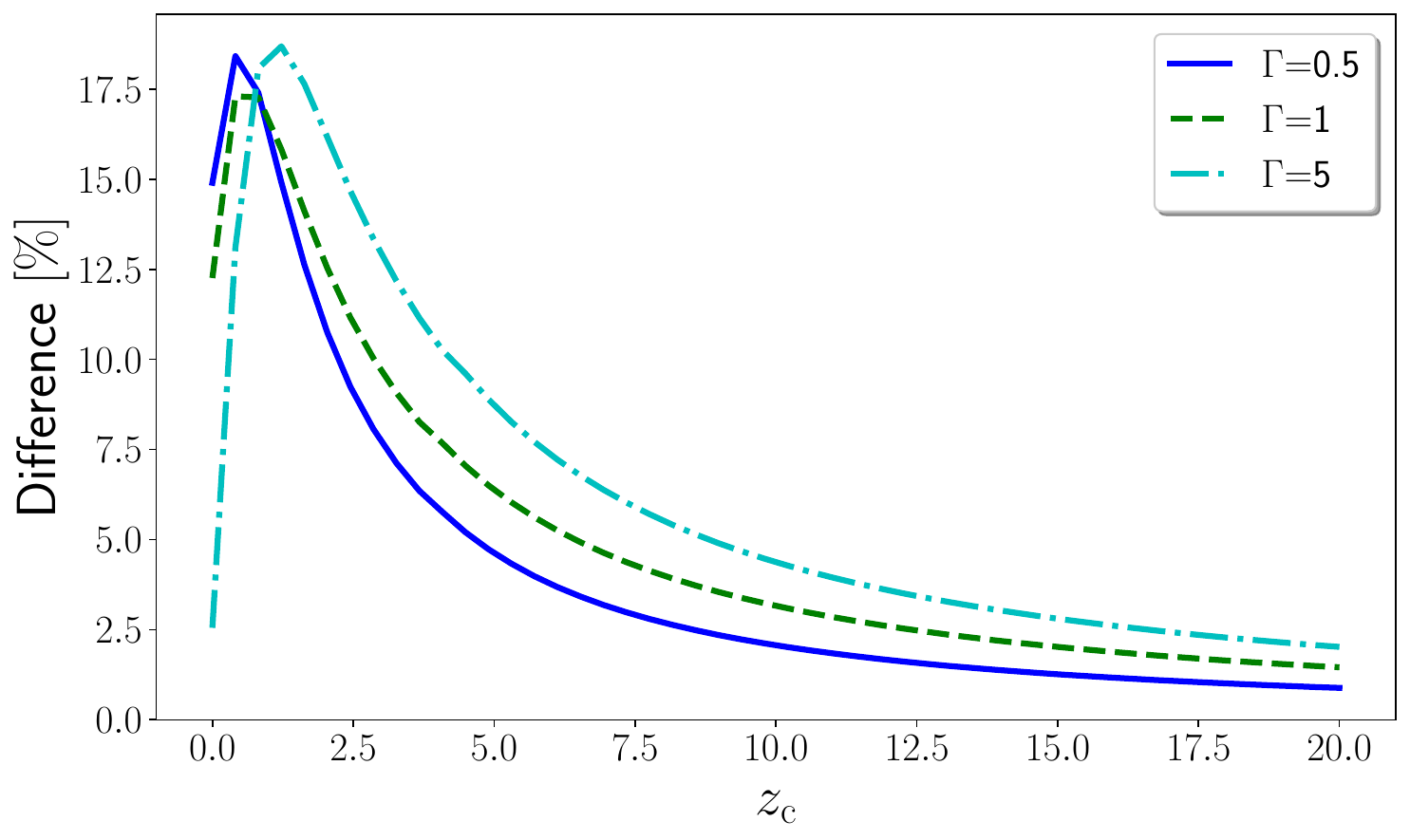}
    \caption{Plots for model $w_1$ with $w_{\rm i}=-0.4$, $z_{\rm t}=0.5$, $w_{\rm f}=-1$. Top panel: percentage difference of $\delta_{\rm c}$ with respect to the $\Lambda$CDM model. Bottom panel: percentage difference of $\zeta^{*}_{\rm{vir}}$ with respect to the $\Lambda$CDM model. The redshifts that maximize the differences with respect to the $\Lambda$CDM model are $z_{\rm{max}}=[0.4,1.2]$. }
    \label{fig:nonlin_pert_2_trans_steepness}
\end{figure}

We only consider clustering DE in the rest of this section. The results obtained when the $w_{\rm f}$ parameter is changed and DE is allowed to cluster are quite revealing. In the nonlinear regime, perturbations are much more sensitive to the final EOS value, $w_{\rm f}$, compared to the linear perturbations. Small variations in $w_{\rm f}$ can have a dramatic impact on the virialization overdensity and density contrast at $z_{\rm{c}}=0$.

\begin{figure}
    \centering
    \includegraphics[width=1\columnwidth,clip]{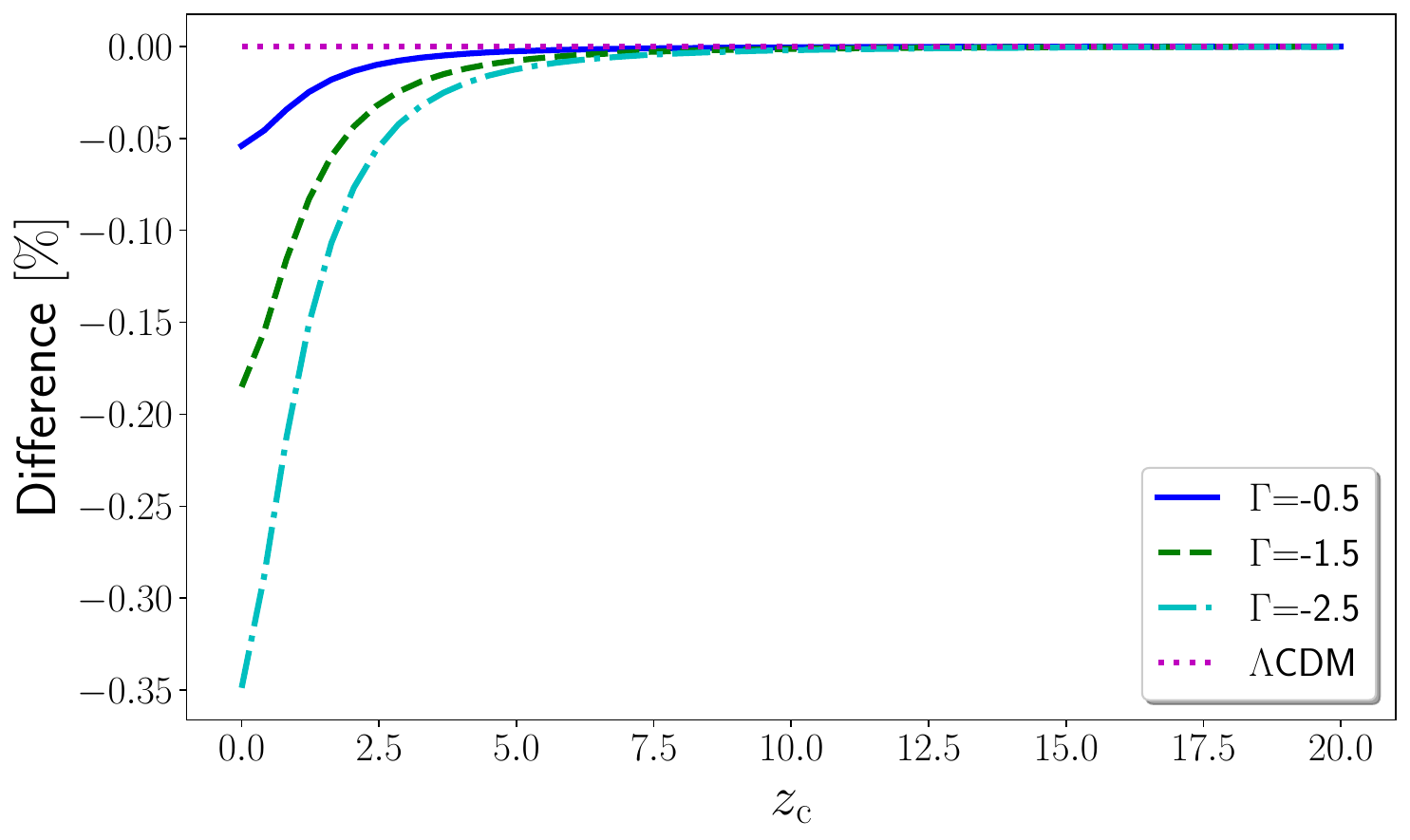}
    \includegraphics[width=1\columnwidth,clip]{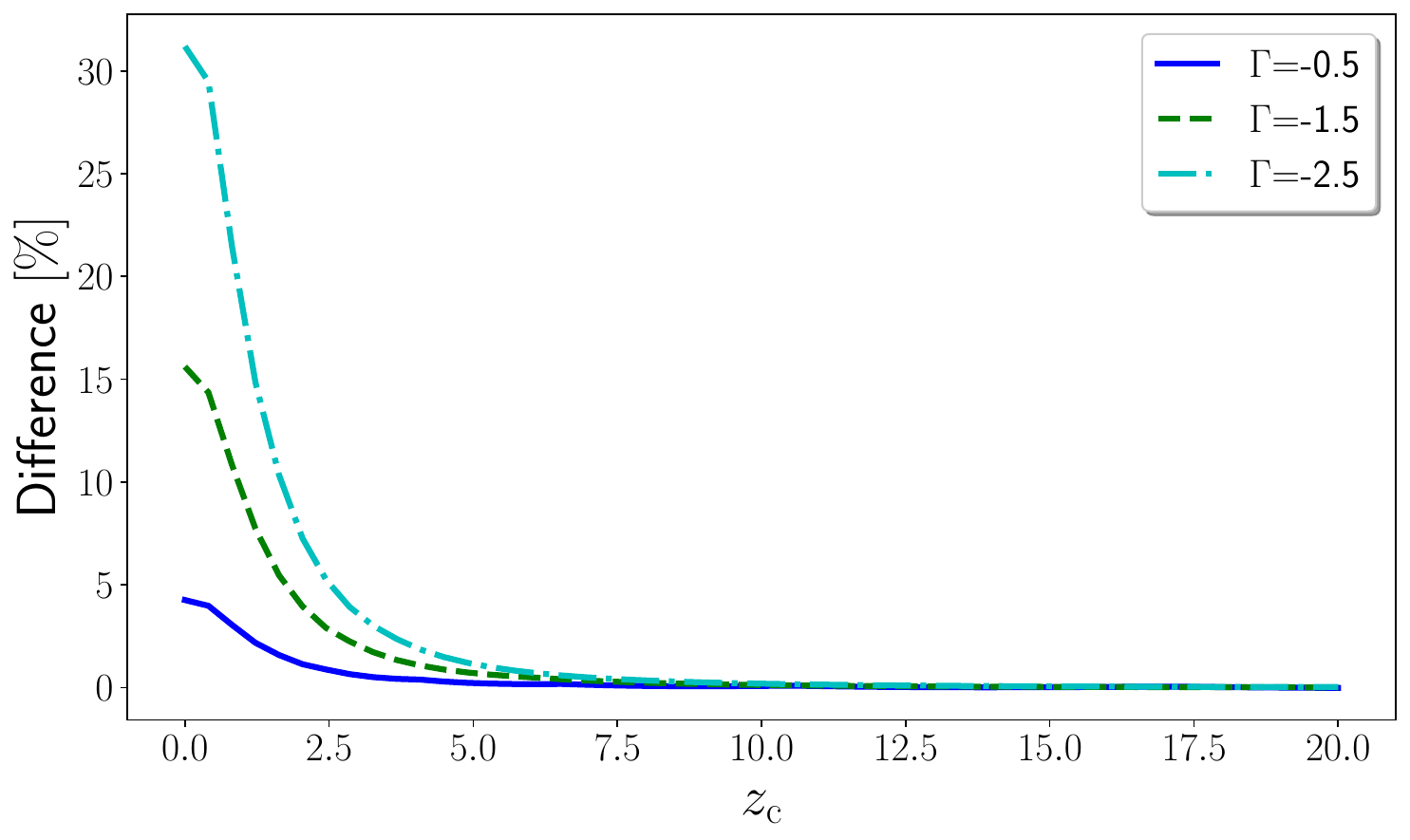}
    \caption{Plots for model $w_5$ with $z_{\rm t}=0.5$. Top panel: percentage difference of $\delta_{\rm c}$ with respect to the $\Lambda$CDM model. Bottom panel:  percentage difference of $\zeta^{*}_{\rm{vir}}$ with respect to the $\Lambda$CDM model.}
    \label{fig:nonlin_pert_5_trans_steepness}
\end{figure}

The time at which the transition occurs has considerable effects on structure formation only if the transition is recent. Only transitions with $z_{\rm t}\lesssim 2$ are distinguishable from the $\Lambda$CDM paradigm. This is because at early times matter dominates and thus an early-time transition is practically indistinguishable from the $\Lambda$CDM model. Interestingly, the effects are more visible in the approximate range of collapse times $z_{\rm c}\in[0.6,3]$, where we can observe deviations greater than $10\%$. The very recent transition with $z_{\rm t}=0.1$ has the strongest effect, since the transition happened after the DE with $w_{\rm i}=-0.4$ started to dominate.

\begin{figure}
    \centering
    \includegraphics[width=1\columnwidth,clip]{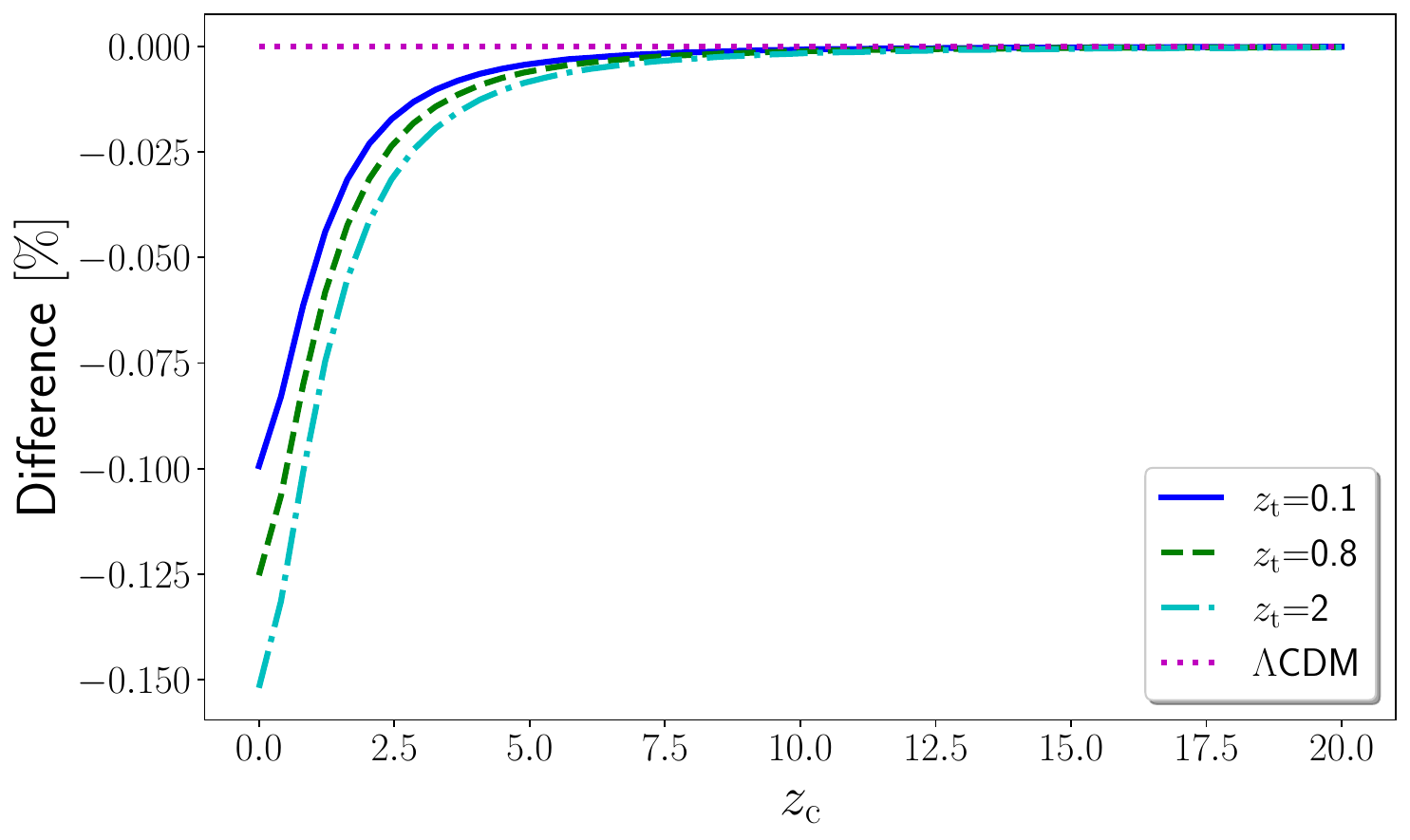}
    \includegraphics[width=1\columnwidth,clip]{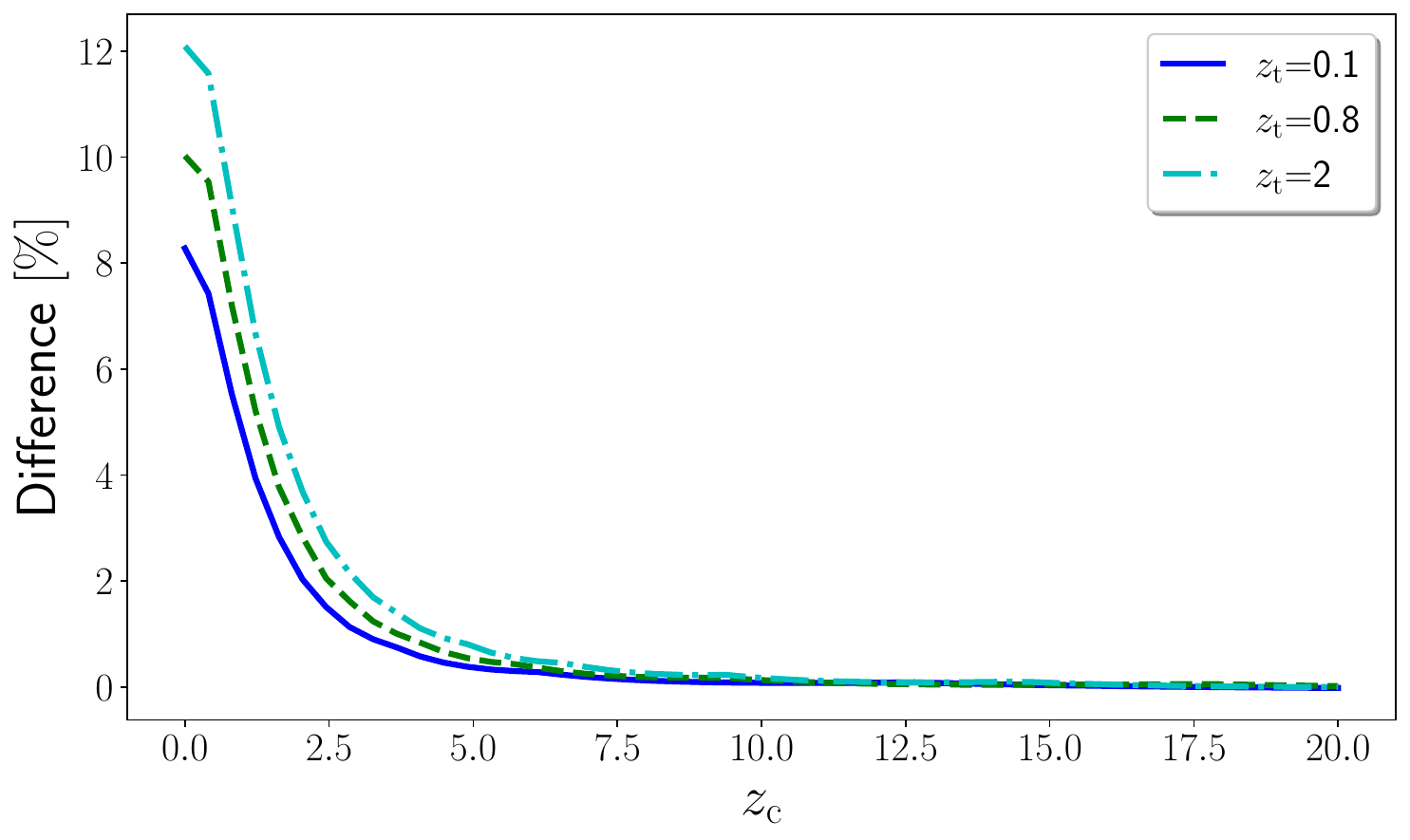}
    \caption{Plots for model $w_5$ with $\Gamma=-2$. Top panel: percentage difference of $\delta_{\rm c}$ with respect to the $\Lambda$CDM model. Bottom panel: percentage difference of $\zeta^{*}_{\rm{vir}}$ with respect to the $\Lambda$CDM model.}
    \label{fig:nonlin_pert_5_z_t}
\end{figure}

As one can see in Fig.~\ref{fig:nonlin_pert_2_trans_steepness}, variations in the $\Gamma$ parameter do not impact significantly the behavior of the nonlinear perturbations, strengthening the results obtained in Secs.~\ref{subsec:quintessencePotentials} and \ref{subsec:linearRegime}. Thus, a step transition is a very good approximation to transition models also in the nonlinear regime.

\begin{figure}
  \centering
  \includegraphics[width=1\columnwidth,clip]{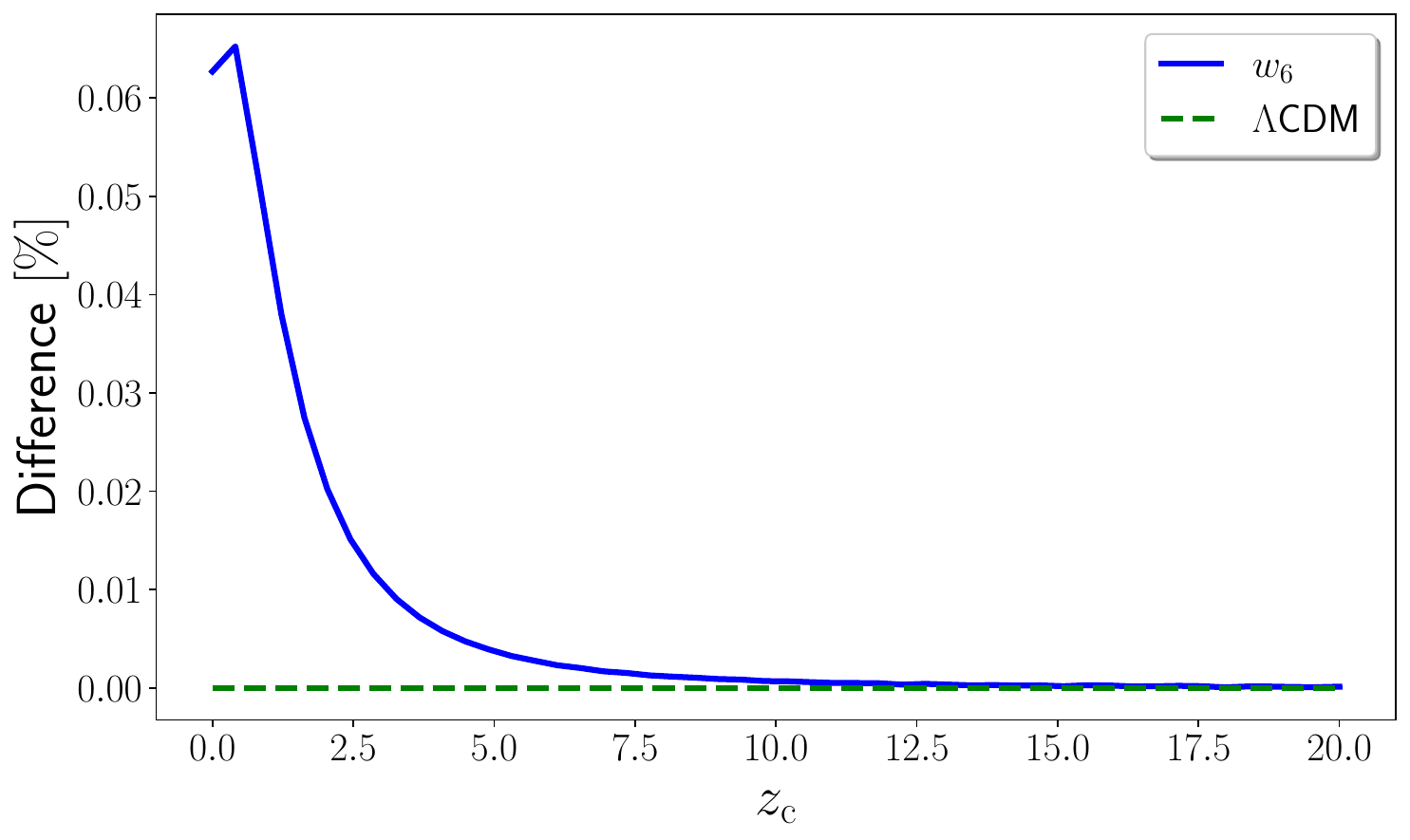}
  \includegraphics[width=1\columnwidth,clip]{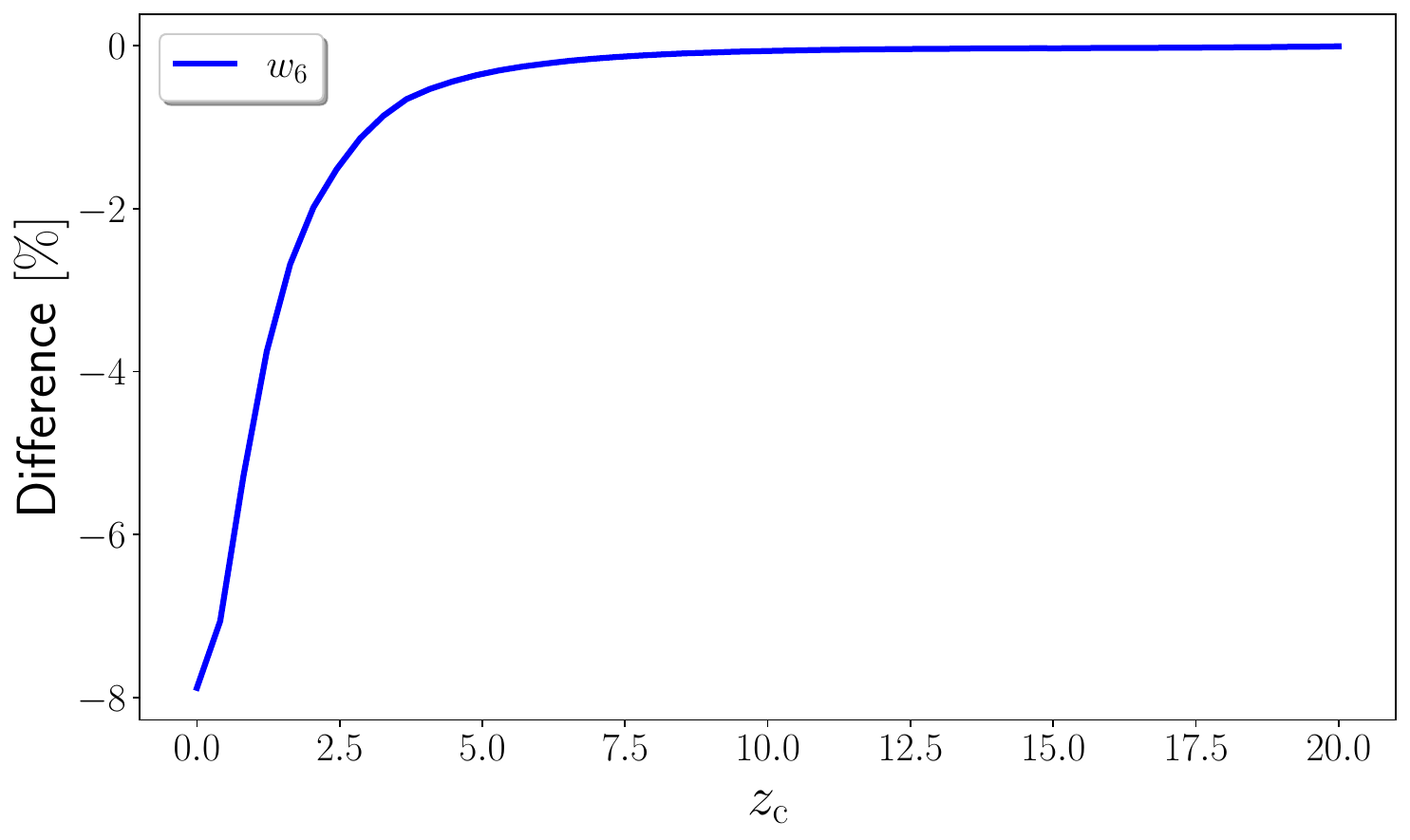}
  \caption{Plot for model $w_6$. Top panel: percentage difference of $\delta_{\rm c}$ with respect to the $\Lambda$CDM model. Bottom panel: percentage difference of $\zeta^{*}_{\rm{vir}}$ with respect to the $\Lambda$CDM model. The percentage difference with respect to the $\Lambda$CDM model peaks at $z=0$ at about $-8\%$.}
  \label{fig:nonlin_pert_6_z_t}
\end{figure}

We now shift our attention to the fifth model, Eq.~(\ref{eq:de_eos_5}), because it shows interesting differences with the previous ones. The only two free parameters are $\Gamma$ and $z_{\rm t}$. In the other models those parameters are almost independent: for example changing the transition redshift does not affect significantly the transition speed. In this model, the value of one parameter changes the behaviors that should be regulated by the other parameters. Results are shown in Fig.~\ref{fig:nonlin_pert_5_trans_steepness}. The DE parametrization in question has initial EOS values close to -1, causing matter to remain dominant at later times than the other models, and resulting in all curves converging quickly to the EdS model. We have varied the $\Gamma$ parameter in a range for which the EOS at $a=1$ is always less then $-1/3$.

\begin{figure}
\centering
\includegraphics[width=1\columnwidth,clip]{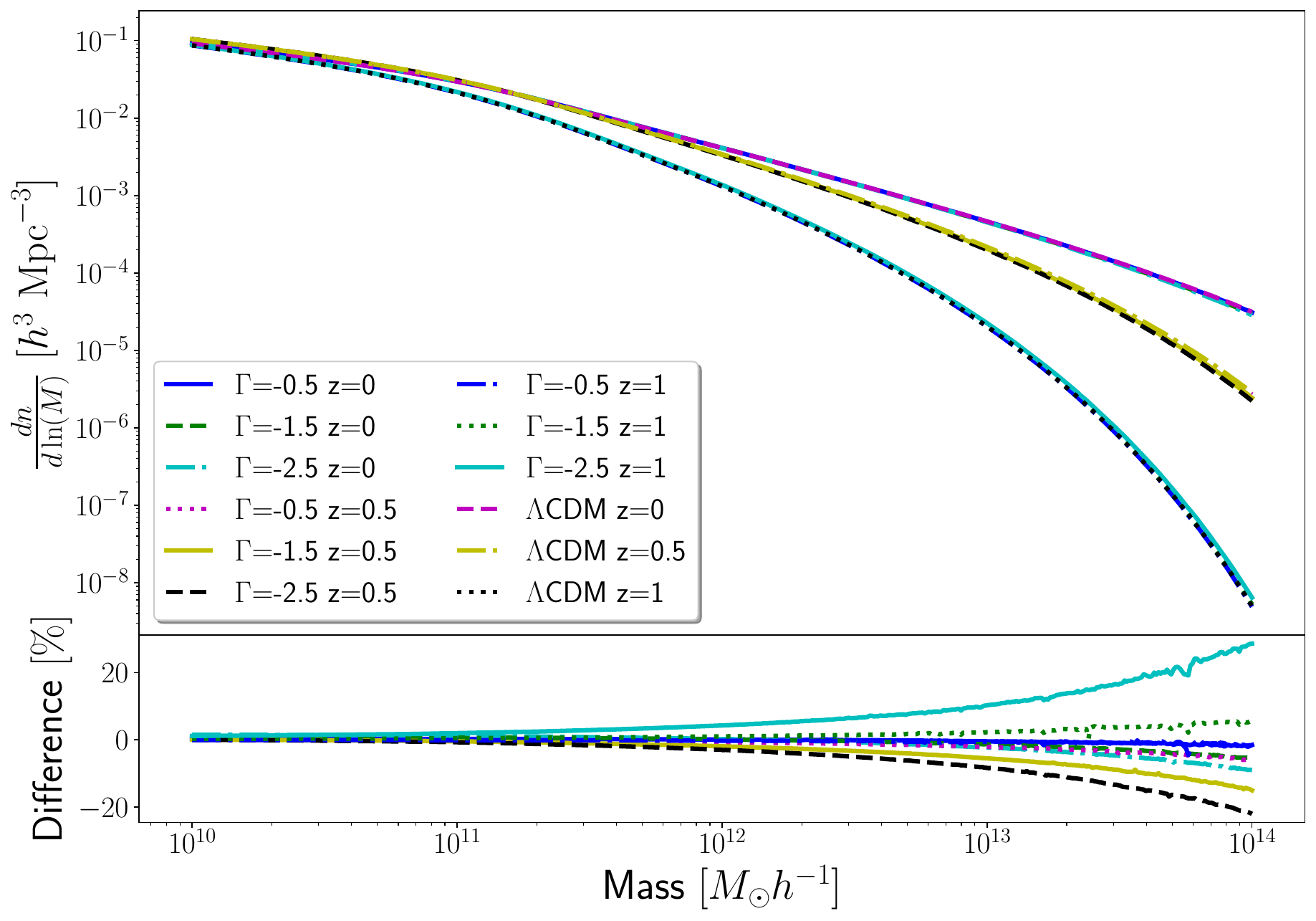}
 \caption{Plot for model $w_5$ with $z_{\rm t}=0.5$, $\delta_{\rm{de}}\neq0$. Top panel: ST mass function for various values of $\Gamma$ and redshift. Bottom panel: percentage difference with respect to the $\Lambda$CDM model defined through the relation $\Lambda$CDM $(1+\%)$.}
  \label{fig:mf_gamma_5}
\end{figure}

A notable difference compared to the model of Eq.~(\ref{eq:de_eos_1}), is that the redshift which maximizes the difference with respect to the $\Lambda$CDM model is consistently zero, rather than peaking around $z=0.8$. This characteristic remains true independently of the parameter we choose to vary and its value. This is probably due to the fact that this EOS, Eq.~(\ref{eq:de_eos_5}), does not behave as a fast transition, but more like a Chevallier--Polarski--Linder parametrization, as one can see in Fig.~\ref{fig:pert_5}.

In Fig.~\ref{fig:nonlin_pert_5_z_t} are shown the results obtained by varying the $z_{\rm t}$ parameter. Since the parameters are strongly intertwined, changing $z_{\rm t}$ also considerably changes the final EOS value and, in the end, the effect on perturbations is quite important.

The last model, Eq.~(\ref{eq:de_eos_6}), cannot be analysed in the same way since it results in $\delta_{\rm{de}}<-1$ that would correspond to a negative energy density. Thus we modified the code to impose $\delta_{\rm{de}}=-1$ when this happens. Results are shown in Fig.~\ref{fig:nonlin_pert_6_z_t}. This is the only intrinsically phantom model we analysed and the only case in which the density contrast is greater than that of the $\Lambda$CDM model while the virialization overdensity is lower.

While in the main body of this work we considered mainly the model described by Eq.~(\ref{eq:de_eos_1}), here we consider the HMF for the model of Eq.~(\ref{eq:de_eos_5}) and present our results in Fig.~\ref{fig:mf_gamma_5}. This model is unique due to its initial value of the EOS being fixed at $-1$, leading to a thawing behavior. Because of the initial condition $w_i=-1$, the curves do not cluster in groups with the same redshift, and the range of variation is narrower. For $\Gamma=-0.5$, the EOS is bounded between $-1$ and $-0.9$. For lower values (e.g., $\Gamma=-2.5$), the model shows a slow transition from $-1$ to $-0.5$. Focusing on the model with $\Gamma=-2.5$, we observe that at $z=0$, the model predicts a $30\%$ increase in massive galaxies, and at $z=0.5$, it predicts a $20\%$ decrease. This suggests a recent and brief period during which many smaller galaxies rapidly merged to form more massive ones. However, it is important to note that this effect is present in all the other models as well, with the difference being that the other models generally predict a lower number of massive galaxies than the $\Lambda$CDM model. This effect is particularly notable for this model since it predicts a higher number of massive galaxies compared to the standard model.

\bibliographystyle{unsrt}
\bibliography{bibliography}

\end{document}